\documentclass[pra,twocolumn]{revtex4-2}
\usepackage{graphicx}
\usepackage{amsmath}
\usepackage{amssymb}
\usepackage{bm}
\usepackage[T1]{fontenc}
\usepackage[latin9]{luainputenc}
\usepackage{stmaryrd}
\usepackage{subfigure}

\usepackage{float}

\newcommand{\beq}{\begin{eqnarray}}
\newcommand{\eeq}{\end{eqnarray}}

\usepackage{color}

\begin{document}

\title{Contour-time approach to the disordered Bose-Hubbard model in the
strong coupling regime}
\author{Ali Mokhtari-Jazi,$^1$ Matthew R. C. Fitzpatrick,$^2$ Malcolm P. Kennett,$^1$}

\affiliation{$^1$ Department of Physics, Simon Fraser University,\\
Burnaby, British Columbia V5A 1S6, Canada\\
$^2$Department of Physics, University of Victoria,\\
Victoria, British Columbia V8P 5C2, Canada}  
\date{\today}

\begin{abstract} 
There has been considerable interest in the disordered Bose Hubbard model (BHM) in recent years, particularly in the context of thermalization and many-body localization.  We develop a two-particle irreducible (2PI) strong-coupling approach to the disordered BHM that allows us to treat both equilibrium and out-of-equilibrium situations.  We obtain equations of motion for spatio-temporal correlations and explore their equilibrium solutions.  We study the equilibrium phase diagram as a function of disorder strength and discuss applications of the formalism to out-of-equilibrium situations. We also note that the disorder strengths where the emergence of non-ergodic dynamics was observed in a recent experiment [Choi {\it et al.} Science \textbf{352}, 1547 (2016)] appear to correspond to the Mott insulator -- Bose glass phase boundary.
\end{abstract}
\maketitle

\section{Introduction}

Isolated strongly interacting quantum systems with quenched disorder may fail to thermalize and enter a phase in which the entire spectrum is composed of localized states \cite{Anderson1958,Basko2006}.  Such many-body 
localized (MBL) states have been studied intensively in recent years \cite{Abanin2019,Oganesyan2007,Znidaric2008,Pal2010,Bardason2012,Vosk2013,Serbyn2013,Bauer2013,Kjall2014,Huse2014,Pekker2014,Nandkishore2015,Bera2015,Chandran2015,Ros2015,Vosk2015,Potter2015,Huse2016,Imbrie2016a,Imbrie2016b, Rademaker2016,Potter2016, Zhang2016, Dumitrescu2017, Imbrie2017, Wortis2017, Monthus2018, Goremykina2019,Leipner-Johns2019, Panda2019, Balasubramanian2020, Kiefer2020,Tarzia2020,Morningstar2020,Garratt2021,Tikhonov2021,Kiefer2021,Kiefer2022}.  In addition to their interest from a fundamental point of view, MBL states have also been suggested as having potential for use as quantum memories.

Aside from perturbative calculations \cite{Fleishman1980,Altshuler1997,Gornyi2005,Basko2006} most of the evidence for MBL states comes from numerical calculations in one dimension
\cite{Oganesyan2007,Znidaric2008,Pal2010,Bardason2012,Vosk2013,Serbyn2013,Kjall2014,Pekker2014,Bera2015,Chandran2015,Luitz2015,Ros2015,Vosk2015,Potter2015,Rademaker2016,Potter2016,Dumitrescu2017, Villalonga2018}.  
Imbrie \cite{Imbrie2016a,Imbrie2016b} has also provided rigorous arguments for the existence of MBL in one dimension under reasonable assumptions, although recent work has raised the question of
whether particles are fully localized \cite{Kiefer2020,Kiefer2021,Kiefer2022} or whether it is possible to reach large enough system sizes to study the MBL phase \cite{Panda2019}.  In dimensions higher than one, which is beyond the reach of many numerically exact methods, the 
situation is less clear and there is theoretical evidence and arguments for and against MBL \cite{Chandran2016,Lev2016,Agarwal2017,DeRoeck2017a,DeRoeck2017b,Thomson2018,Abanin2019,Gopalakrishnan2019,Wahl2019,Doggen2020,Theveniaut2020,Kshetrimayum2020,Chertkov2021,Pietracaprina2021}.

Experimentally, there are indications of localization in disordered, interacting many body cold atom systems in optical lattices in two and higher dimensions \cite{Kondov2015,Schreiber2015,Choi2016,Bordia2017,Luschen2017}.
Of particular interest for our work is the experiment by Choi {\it et al.} \cite{Choi2016}, in which the relaxation dynamics of disordered bosons in a two dimensional optical lattice were studied. 
Starting from an initial condition in which all of the atoms were localized on one side of a trap, Choi {\it et al.} observed the imbalance as a function of time and found that beyond a critical 
disorder strength, their system failed to thermalize in the time window of their experiment.  Yan {\it et al.} \cite{Yan2017} applied Gutzwiller mean-field theory (GMFT) to the two-dimensional disordered Bose Hubbard model and were able to reproduce the main experimental results, even though GMFT is unable to capture MBL, raising the possibility that the experiments probe glassy dynamics 
rather than MBL.  This highlights the need to develop theoretical methods to investigate the out-of-equilibrium dynamics of the disordered Bose Hubbard model in dimensions greater than one.

There has been considerable study of the out-of-equilibrium dynamics of the Bose-Hubbard model realized in optical lattices 
\cite{Greiner2002,Bloch2005,Jaksch2005,Lewenstein2007,Bloch2008,Hung2010,Bakr2010,Kennett2013,Gross2017}.  In order to obtain
spatial as well as temporal information, correlations are of particular interest, and a variety of methods, such as exact diagonalization (ED) and time-dependent
density-matrix renormalization-group methods (t-DMRG) have been used
in one dimension \cite{Clark2004,Kollath2007,Lauchli2008,Bernier2011,Cheneau2012,Barmettler2012,Trotzky2012,Bernier2012,Cevolani2018,Despres2019}. 
In two dimensions, where many of these approaches become less effective, methods for calculating correlations include perturbative corrections to 
Gutzwiller mean-field theory \cite{Navez2010,Trefzger2011,Krutitsky2014,Queisser2014}, time-dependent variational Monte Carlo \cite{Carleo2014}, doublon-holon pair theories \cite{Yanay2016} 
and tensor network methods \cite{Koneko2022}.

An alternative approach has been developed by two of us that is based on a two particle irreducible (2PI) out-of-equilibrium strong coupling approach to the BHM (2PISC)
\cite{Kennett2011,Fitzpatrick2018a,Fitzpatrick2018b,Fitzpatrick2019,Kennett2020}. This approach allows the treatment of the dynamics of the order parameter and 
correlation functions on an equal footing and we have previously used it to demonstrate excellent agreement \cite{Mokhtari-Jazi2021} with experiments investigating
the spreading of correlations for bosons in optical lattices in one and two dimensions \cite{Cheneau2012,Takasu2020}.  It also has the attractive feature that 
it allows for the inclusion of disorder averaging, which we make use of to study the disordered Bose-Hubbard model.

The presence of disorder in the BHM can lead to an additional phase in between the superfluid and Mott insulator, the Bose glass \cite{Fisher1989}, and can reduce the size of the Mott lobes
\cite{Fisher1989,Freericks1996}.  The experiments by Choi {\it et al.} \cite{Choi2016} have focused attention on the out-of-equilibrium dynamics of the disordered Bose Hubbard model and 
both many-body localization and glassiness for bosons in one \cite{Rispoli2019,Yao2020,Kim2021,Chen2021,Villa2021} and two or more dimensions \cite{Lin2012,Thomson2016,Meldgin2016,Yan2017,Wahl2019,Bertoli2019,Geissler2020,Geissler2021,Kim2021,Chertkov2021,Souza2021}.  This activity motivates our extension of the 2PISC formalism for 
the BHM to include disorder to provide an additional route to investigate the out-of-equilibrium dynamics of the disordered BHM.

The main result of this paper is that we develop a 2PI framework that allows us to treat both the equilibrium and out-of-equilibrium behaviour of the disordered Bose Hubbard model. This allows
us to obtain equations of motion for the superfluid order parameter and spatio-temporal correlations.  We obtain solutions of these equations in the equilibrium case and investigate the 
Mott insulator phase boundary as a function of disorder strength and calculate the collective excitation spectrum both in and outside the Mott phase.  We find that our results compare 
favourably with quantum Monte Carlo (QMC) simulations in two \cite{Soyler2011} and three \cite{Gurarie2009} dimensions.  We also note that the disorder strengths at which Ref.~\cite{Choi2016}
found the emergence of non-ergodic dynamics appear to correspond to the Mott insulator -- Bose glass phase boundary.

This paper is structured as follows: in Sec.~\ref{sec:Model} we introduce the disordered Bose Hubbard model and formalism, deriving an effective theory.  We use our effective theory to obtain 2PI
equations of motion which we then solve for equilibrium properties of the model in Sec.~\ref{sec:EoM}. We conclude and discuss our results in Sec.~\ref{sec:Conc}

\section{Model and formalism} \label{sec:Model}

In this section, we introduce the disordered Bose Hubbard model and
discuss the generalization of the strong-coupling approach developed
in Refs.~\citep{Sengupta2005,Kennett2011,Fitzpatrick2018a} for the standard BHM to the disordered
case allowing for both equilibrium and out-of-equilibrium behaviour. The Hamiltonian for the disordered BHM is

\begin{equation}
\hat{H}_{\text{BHM}}^{\text{dis}}\left(t;\epsilon\right)=\hat{H}_{J}\left(t\right)+\hat{H}_{0}+\hat{H}_{\epsilon},\label{eq:BHM Hamiltonian - 1}
\end{equation}
\noindent where
\begin{equation}
\hat{H}_{J}=-\sum_{\left\langle \vec{r}_{1},\vec{r}_{2}\right\rangle }J_{\vec{r}_{1}\vec{r}_{2}}\left(\hat{a}_{\vec{r}_{1}}^{\dagger}\hat{a}^{\phantom{\dagger}}_{\vec{r}_{2}}+\hat{a}^{\phantom{\dagger}}_{\vec{r}_{2}}\hat{a}_{\vec{r}_{1}}^{\dagger}\right),\label{eq:H_J defined - 1}
\end{equation}
\begin{equation}
\hat{H}_{0}=\frac{U}{2}\sum_{\vec{r}}\hat{n}_{\vec{r}}\left(\hat{n}_{\vec{r}}-1\right)+\sum_{\vec{r}}\left(V_{\vec{r}}-\mu\right)\hat{n}_{\vec{r}},\label{eq:H_0 defined - 1}
\end{equation}
\begin{equation}
\hat{H}_{\epsilon}=\sum_{\vec{r}}\epsilon_{\vec{r}} \: \hat{n}_{\overline{r}},\label{eq:H_e defined - 1}
\end{equation}
\noindent with $\hat{a}_{\vec{r}}^{\dagger}$ and $\hat{a}_{\vec{r}}$
annihilation and creation operators for bosons on lattice site $\vec{r}$
respectively, $\hat{n}_{\vec{r}}\equiv\hat{a}_{\vec{r}}^{\dagger}\hat{a}^{\phantom{\dagger}}_{\vec{r}}$
the number operator, $U$ the interaction strength, $V_{\vec{r}}$
a harmonic trapping potential, $\mu$ the chemical potential, and
$\epsilon_{r}$ an on-site disorder potential. The disorder potential is drawn from a Gaussian
distribution

\begin{equation}
\mathcal{P}\left[\epsilon_{\vec{r}}\right]=\sqrt{\frac{4\ln2}{\pi\Delta_{\epsilon}^{2}}}e^{-\frac{4\left(\text{ln}2\right)\epsilon_{\vec{r}}^{2}}{\Delta_{\epsilon}^{2}}},\label{eq: Disorder Distribution - 1}
\end{equation}

\noindent with  $\Delta_{\epsilon}$ being the full-width at half maximum for the distribution. The notation $\left\langle \vec{r}_{1},\vec{r}_{2}\right\rangle $
indicates a sum over nearest neighbours only.

\subsection{Contour-time formalism\label{subsec:Contour formalism}}
\begin{figure}[t]
\subfigure{
    \includegraphics[width=0.8\columnwidth]{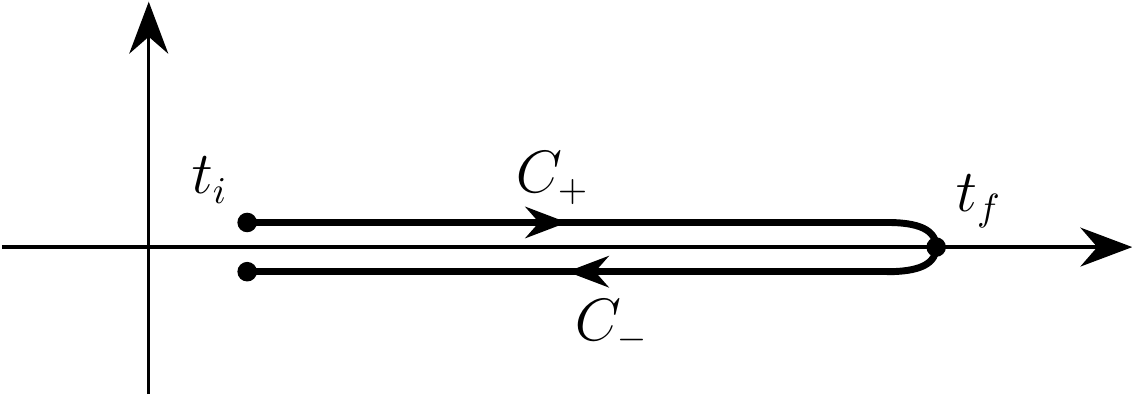}
}\caption{Contour for a system initially prepared at time $t_{i}$. $t_{f}$
is the maximum real-time considered in the problem, which may be set
to $t_{f}\to\infty$ without loss of generality.}
\label{fig:fig1}
\end{figure}

The general formalism that we discuss and adopt in this paper was
developed in a previous paper by two of us; we refer the reader to Ref.~\citep{Fitzpatrick2018a}
for further details on the formalism.
We use the contour-time formalism \citep{Schwinger1961,Keldysh1964,Rammer1986,Semenoff1984,Landsman1987,Chou1985},
which replaces the notion of real time along the real line with contour time, a complex valued time on a contour in the complex plane. Furthermore, an appropriate choice of contour is particularly attractive for
studying disordered systems as it eliminates the need to use replicas
in carrying out the average over the quenched disorder \citep{Kamenev1999,Chamon1999}.
For systems initially prepared in out-of-equilibrium states, one can work with a contour $C$ of the form illustrated
in Fig.~\ref{fig:fig1}. A popular alternative to this contour is
the Schwinger-Keldysh (SK) closed-time path \citep{Schwinger1961,Keldysh1964}
which is also suitable for certain out-of-equilibrium problems. However,
unlike contour $C$, the SK contour ignores transient phenomena and,
more importantly, information about the initial state. Given that
we are interested in comparing long-time density profiles with that
of the initial state, contour $C$ is a more appropriate choice. 

 \vspace{1cm}

\subsection{Contour-ordered Green's functions\label{subsec:Contour-ordered Green's functions}}

In deriving our effective theory of the disordered BHM, we calculate
various contour-ordered Green's functions (COGFs). We define the $n$-point
COGF as \citep{Chou1985}
\begin{widetext}
\begin{eqnarray}
G_{\vec{r}_{1}\ldots\vec{r}_{n}}^{a_{1}\ldots a_{n}}\left(\tau_{1},\ldots,\tau_{n};\epsilon\right) & \equiv\left(-i\right)^{n-1}\text{Tr}\left\{ \hat{\rho}_{i}T_{C}\left[\hat{a}_{\vec{r}_{1}}^{a_{1}}\left(\tau_{1};\epsilon\right)\ldots\hat{a}_{\vec{r}_{n}}^{a_{n}}\left(\tau_{n};\epsilon\right)\right]\right\} \nonumber \\
 & \equiv\left(-i\right)^{n-1}\left\langle T_{C}\left[\hat{a}_{\vec{r}_{1}}^{a_{1}}\left(\tau_{1};\epsilon\right)\ldots\hat{a}_{\vec{r}_{n}}^{a_{n}}\left(\tau_{n};\epsilon\right)\right]\right\rangle _{\hat{\rho}_{i}},\label{eq:COGFs defined - 1}
\end{eqnarray}
\end{widetext}

\noindent where $\hat{\rho}_{i}$ is the state
operator representing the initial state of the system and the $a_{i}$
upper indices are defined such that

\begin{equation}
\hat{a}_{\vec{r}}^{1}\equiv\hat{a}^{\phantom{1}}_{\vec{r}},\quad\hat{a}_{\vec{r}}^{2}\equiv\hat{a}_{\vec{r}}^{\dagger},\label{eq:introducing nambu indices - 1}
\end{equation}

\noindent and $\hat{a}_{\vec{r}}^{a}\left(\tau;\epsilon\right)$ are
the bosonic fields in the Heisenberg picture with respect to $\hat{H}_{\text{BHM}}\left(\tau;\epsilon\right)$ [Eq.~(\ref{eq:BHM Hamiltonian - 1})]

\begin{widetext}
\begin{align}
\hat{a}_{\vec{r}}^{a}\left(\tau;\epsilon\right) & =U_{C}\left(\tau_{i},\tau;\epsilon\right)\hat{a}_{\vec{r}}^{a}~U_{C}\left(\tau,\tau_{i};\epsilon\right),\label{eq:Heisenberg fields - 1}\\
U_{C}\left(\tau_{i},\tau;\epsilon\right)&
=\begin{cases}
T_{C}\left[e^{-i\int_{C\left(\tau,\tau^{\prime}\right)}d\tau^{\prime\prime}\hat{H}_{\text{BHM}}^{\text{dis}}\left(\tau^{\prime\prime};\epsilon\right)}\right], & \text{\text{if $\tau$ later than $\tau'$}},\\[4mm]
T_{C}\left[e^{i\int_{C\left(\tau,\tau^{\prime}\right)}d\tau^{\prime\prime}\hat{H}_{\text{BHM}}^{\text{dis}}\left(\tau^{\prime\prime};\epsilon\right)}\right], & \text{if $\tau'$ later than $\tau$}.
\end{cases} \label{eq:evolution operator - 1}
\end{align}
\end{widetext}

\noindent Here we have introduced explicitly the complex contour time
argument $\tau$, the sub-contour $C\left(\tau,\tau^{\prime}\right)$
which goes from $\tau$ to $\tau^{\prime}$ along the contour $C$,
and the contour time ordering operator $T_{C}$, which orders strings
of operators according to their position on the contour, with operators
at earlier contour times placed to the right.

Given the somewhat cumbersome notation in expressions such as that
in Eq.~(\ref{eq:COGFs defined - 1}), we make extensive use of a
compact notation where we write an arbitrary function $X$ as

\begin{equation}
X_{\vec{r}_{1}\ldots\vec{r}_{n},\tau_{1}\ldots\tau_{n};\epsilon}^{a_{1}\ldots a_{n}}\equiv X_{\vec{r}_{1}\ldots\vec{r}_{n}}^{a_{1}\ldots a_{n}}\left(\tau_{1}\ldots\tau_{n};\epsilon\right),\label{eq:compact notation for 2PI calculations - 1}
\end{equation}

\noindent and introduce the following implicit summation convention

\begin{eqnarray}
X_{\tau;\epsilon}^{a}Y_{\tau;\epsilon}^{\overline{a}} & = & \sum_{a_{1}a_{2}}\int_{C}d\tau\sigma_{1}^{a_{1}a_{2}}X^{a_{1}}\left(\tau;\epsilon\right)Y^{a_{2}}\left(\tau;\epsilon\right),\label{eq:Einstein summation convention for tau - 1}
\end{eqnarray}

\noindent where $\sigma_{i}$ is the $i^{\text{th}}$ Pauli matrix, $\overline{1}=2$ and $\overline{2}=1$. Note that we only include
the $\epsilon$ parameter in Eq.~(\ref{eq:compact notation for 2PI calculations - 1})
if the function $X$ depends on the disorder configuration.

\subsection{Generating functional $\mathcal{Z}\left[f;\epsilon\right]$\label{subsec:Generating functional Z}}

The COGFs above can be derived from a generating functional $\mathcal{Z}\left[f;\epsilon\right]$,
which can be cast in the following path integral form \citep{Semenoff1984,Kennett2011,Fitzpatrick2018a,Fitzpatrick2019}:
\begin{align}
\mathcal{Z}\left[f;\epsilon\right]&=\int\left[\mathcal{D}a\right]\left\langle \mathbf{a}\left(\tau_i\right)\left|\hat{\rho}_{i}\left(\epsilon\right)\right|\mathbf{a}\left(\tau_f\right)\right\rangle\nonumber\\ 	&\quad\times e^{-\frac{1}{2}\left\{\mathbf{a}\left(\tau_i\right).\mathbf{a}\left(\tau_i\right)+ \mathbf{a}\left(\tau_f \right). \mathbf{a}\left(\tau_f\right)\right \}}\nonumber\\ 	  &\quad\times e^{iS\left[a;\epsilon\right]+iS_{f}\left[a\right]},\label{eq:Path integral form of Z - 1}
\end{align}

\noindent where $S\left[a; \epsilon\right]$ is the action for the disordered BHM
\begin{equation}
    S\left[a; \epsilon\right]=S_{J}\left[a\right]+S_{0}\left[a\right]+S_{\epsilon}\left[a\right],\label{eq:S_BHM^dis - 1}
\end{equation}
\noindent with
\begin{align}
S_{J}\left[a\right]&=\frac{1}{2!}\sum_{\vec{r}_{1}\vec{r}_{2}} \left\{2J_{\vec{r}_{1}\vec{r}_{2},\tau_{1}\tau_{2}}^{a_{1}a_{2}}\right\} a_{\vec{r}_{1},\tau_{1}}^{\overline{a_{1}}}a_{\vec{r}_{2},\tau_{2}}^{\overline{a_{2}}},\label{eq:S_J - 1}\\
S_{0}\left[a\right] & =\frac{1}{2!}\sum_{\vec{r}}a_{\vec{r},\tau_{1}}^{\overline{a_{1}}} \left\{-\partial_{\tau_{1}\tau_{2}}^{a_{1}a_{2}}-\left\{ V_{\vec{r}}-\mu\right\}\zeta_{\tau_{1}\tau_{2}}^{a_{1}a_{2}}\right\}a_{\vec{r},\tau_{2}}^{\overline{a_{2}}}\nonumber\\
& \quad-\frac{1}{4!}\sum_{\vec{r}}\left\{U\zeta_{\tau_{1}\tau_{2}\tau_{3}\tau_{4}}^{a_{1}a_{2}a_{3}a_{4}}\right\}a_{\vec{r},\tau_{1}}^{\overline{a_{1}}}a_{\vec{r},\tau_{2}}^{\overline{a_{2}}}a_{\vec{r},\tau_{3}}^{\overline{a_{3}}}a_{\vec{r},\tau_{4}}^{\overline{a_{4}}}\label{eq:S_U - 1},\\
S_{\epsilon}\left[a\right]&=\frac{1}{2!}\sum_{\vec{r}}\left\{-\epsilon_{\vec{r}}\:\zeta_{\tau_{1}\tau_{2}}^{a_{1}a_{2}}\right\}a_{\vec{r},\tau_{1}}^{\overline{a_{1}}}a_{\vec{r},\tau_{2}}^{\overline{a_{2}}},\label{eq:S_epsilon - 1}
\end{align}
\noindent with 
\begin{align}
    J_{\vec{r}_{1}\vec{r}_{2},\tau_{1}\tau_{2}}^{a_{1}a_{2}}&=J_{\vec{r}_{1}\vec{r}_{2}}\zeta_{\tau_{1}\tau_{2}}^{a_{1}a_{2}},\label{eq:J tensor defined - 1}\\
    \zeta_{\tau_{1}\tau_{2}}^{a_{1}a_{2}}&=\delta_{\tau_{1}\tau_{2}}\sigma_{1}^{a_{1}a_{2}},\label{eq:zeta^(2) defined - 1}
\end{align}
\noindent and
\begin{equation}
\zeta_{\tau_{1}\tau_{2}\tau_{3}\tau_{4}}^{a_{1}a_{2}a_{3}a_{4}}=2\delta_{\tau_{1}\tau_{2}}\delta_{\tau_{2}\tau_{3}}\delta_{\tau_{3}\tau_{4}}\sigma^{a_{1}a_{2}a_{3}a_{4}},\label{eq:zeta^(4) defined - 1}
\end{equation}
\noindent where
\begin{eqnarray}
\sigma^{a_{1}a_{2}a_{3}a_{4}}=\left\{\begin{array}{cc}
1, & \text{if }\left\{ a_{m}\right\} _{m=1}^{4}\in P\left(\left\{ 1,1,2,2\right\} \right) ,\\
0, & \text{otherwise.}
\end{array}
\right. 
\label{eq:generalized sigma tensor defined - 1}
\end{eqnarray}

\noindent $S_{f}\left[a\right]$ is the
source term

\begin{equation}
S_{f}\left[a\right]=\sum_{\vec{r}}f_{\vec{r},\tau}^{a}a_{\vec{r},\tau}^{\overline{a}},\label{eq:f source term - 1}
\end{equation}

\noindent and $\int\left[\mathcal{D}a\right]$ is the coherent-state
measure. Note that in the path-integral formalism $a_{\vec{r}}^{1}=a^{\phantom{1}}_{\vec{r}}$
and $a_{\vec{r}}^{2}=a_{\vec{r}}^{*}$. In this formalism, we can
rewrite averages of the form $\left\langle T_{C}\left[\ldots\right]\right\rangle _{\hat{\rho}_{i}}$
as follows

\begin{equation}
\left\langle T_{C}\left[\hat{a}_{\vec{r}_{1},\tau_{1};\epsilon}^{a_{1}}\ldots\hat{a}_{\vec{r}_{n},\tau_{n};\epsilon}^{a_{n}}\right]\right\rangle _{\hat{\rho}_{i}}\equiv\left\langle a_{\vec{r}_{1},\tau_{1}}^{a_{1}}\ldots a_{\vec{r}_{n},\tau_{n}}^{a_{n}}\right\rangle _{S},\label{eq:Rewriting averages - 1}
\end{equation}

\noindent where contour ordering is now implicit in the path integral
representation \citep{Negele}. Occasionally, we drop the action subscript
$\left\langle \ldots\right\rangle _{S}\to\left\langle \ldots\right\rangle $
for brevity.

To derive the COGFs in Eq.~(\ref{eq:COGFs defined - 1}) from $\mathcal{Z}\left[f;\epsilon\right]$,
we take appropriate functional derivatives with respect to the sources
and set the sources to zero afterwards
\begin{equation}
    G_{\vec{r}_{1}\ldots\vec{r}_{n},\tau_{1}\ldots\tau_{n;\epsilon}}^{a_{1}\ldots a_{n}}=i\left(-1\right)^{n}\left.\frac{\delta^{n}\mathcal{Z}\left[f;\epsilon\right]}{\delta f_{\vec{r}_{1},\tau_{1}}^{\overline{a_{1}}}\ldots\delta f_{\vec{r}_{n},\tau_{n}}^{\overline{a_{n}}}}\right|_{f\rightarrow0}.\label{eq:COGFs from Z - 1}
\end{equation}

\subsection{Disorder averaging\label{subsec:Disorder averaging}}

We are ultimately interested in calculating disorder averaged COGFs

\begin{equation}
    \check{G}_{\vec{r}_{1}\ldots\vec{r}_{n},\tau_{1}\ldots\tau_{n}}^{a_{1}\ldots a_{n}}=\left(\prod_{\vec{r}}\int_{-\infty}^{\infty}d\mathcal{\epsilon}_{\vec{r}}\mathcal{P}\left[\mathcal{\epsilon}_{\vec{r}}\right]\right)G_{\vec{r}_{1}\ldots\vec{r}_{n},\tau_{1}\ldots\tau_{n;\epsilon}}^{a_{1}\ldots a_{n}},
    \label{eq:avg COGF defined - 1}
\end{equation}
\noindent where for a quantity $\theta$ we denote the disorder average $\check{\theta}$ with a carat. Using Eq.~\eqref{eq:COGFs from Z - 1} we can determine an expression for calculating the disorder-averaged COGFs:

\begin{equation}
    \check{G}_{\vec{r}_{1}\ldots\vec{r}_{n},\tau_{1}\ldots\tau_{n}}^{a_{1}\ldots a_{n}}=i\left(-1\right)^{n}\left.\frac{\delta^{n}\check{\mathcal{Z}}\left[f\right]}{\delta f_{\vec{r}_{1},\tau_{1}}^{\overline{a_{1}}}\ldots\delta f_{\vec{r}_{n},\tau_{n}}^{\overline{a_{n}}}}\right|_{f\rightarrow0},
    \label{eq:avg COGF from avg Z - 1}
\end{equation}
\noindent where $\check{\mathcal{Z}}\left[f\right]$ is the disorder-average of $\mathcal{Z}\left[f;\epsilon\right]$.

\subsection{Effective theory of the disordered BHM\label{subsec:Effective theory of disorder BHM}}

We develop an effective theory that is suitable for studying the dynamics
of the disordered BHM in the strong coupling regime. The approach
can be outlined as follows: first we calculate the disorder average
of $\mathcal{Z}\left[f;\epsilon\right]$, which gives us an effective
theory $S_{\text{eff}}^{\text{dis}}$ in terms of the original $a$-fields,
then we apply various Hubbard-Stratonovich transformations such that
we can obtain a strong coupling expansion of the theory. The resulting
effective strong coupling theory introduces two auxiliary fields $z$
and $\mathcal{Q}$. We derive identities relating the correlators
of these two auxiliary fields to those of the original $a$-fields.
One can then apply a two-particle irreducible effective action
approach \citep{Cornwall1974} to the effective theory to obtain equations
of motion for the correlation functions.

We begin by performing the disorder average of $\mathcal{Z}\left[f;\epsilon\right]$

\begin{align}
\mathcal{\check{Z}}\left[f\right]	&=\int\left[\mathcal{D}a\right]\left\langle \mathbf{a}\left(\tau_{i}\right)\left|\hat{\rho}_{i}\right|\mathbf{a}\left(\tau_{f}\right)\right\rangle \nonumber\\
	&\quad\quad\times e^{-\frac{1}{2}\left\{ \mathbf{a}\left(\tau_{i}\right).\mathbf{a}\left(\tau_{i}\right)+\mathbf{a}\left(\tau_{f}\right).\mathbf{a}\left(\tau_{f}\right)\right\} }\nonumber\\
	&\quad\quad\times e^{iS_{\Delta}\left[a\right]}e^{iS_{J}\left[a\right]+iS_{0}\left[a\right]+iS_{f}\left[a\right]},\label{eq:avg Z - 2}
\end{align}

\noindent where 
\begin{equation}
    e^{iS_{\Delta}\left[a\right]}=\left\{ \prod_{\vec{r}}\int_{-\infty}^{\infty}d\mathcal{\epsilon}_{\vec{r}}\mathcal{P}\left[\mathcal{\epsilon}_{\vec{r}}\right]\right\} e^{iS_{\mathcal{\epsilon}}\left[a\right]}. \label{eq:exp-action-Delta}
\end{equation}

\noindent Next, using Eqs.~\eqref{eq: Disorder Distribution - 1} and \eqref{eq:S_epsilon - 1} we calculate $e^{iS_{\Delta}\left[a\right]}$:

\begin{align}e^{iS_{\Delta}\left[a\right]} & =\prod_{\vec{r}}\int_{-\infty}^{\infty}d\epsilon_{\vec{r}}\mathcal{P}\left[\epsilon_{\vec{r}}\right]\exp\left\{ \frac{i}{2!}\left(-\epsilon_{\vec{r}}\:\zeta_{\tau_{1}\tau_{2}}^{a_{1}a_{2}}\right)a_{\vec{r},\tau_{1}}^{\overline{a_{1}}}a_{\vec{r},\tau_{2}}^{\overline{a_{2}}}\right\} \nonumber\\
 & =\exp\left\{ \frac{i}{2!}\sum_{\vec{r}}\left(-\frac{1}{4}M_{\left\llbracket \tau_{1}\tau_{2}\right\rrbracket \left\llbracket \tau_{3}\tau_{4}\right\rrbracket }^{\left\llbracket a_{1}a_{2}\right\rrbracket \left\llbracket a_{3}a_{4}\right\rrbracket }\right)\right.\nonumber\\
 & \quad\quad\quad\quad\quad\quad\quad\quad\left.\times a_{\vec{r},\tau_{1}}^{\overline{a_{1}}}a_{\vec{r},\tau_{2}}^{\overline{a_{2}}}a_{\vec{r},\tau_{3}}^{\overline{a_{3}}}a_{\vec{r},\tau_{4}}^{\overline{a_{4}}}\right\}, \label{eq:expS_{D}elta-1}
\end{align}
\noindent where
\begin{equation}
M_{\left\llbracket \tau_{1}\tau_{2}\right\rrbracket \left\llbracket \tau_{3}\tau_{4}\right\rrbracket }^{\left\llbracket a_{1}a_{2}\right\rrbracket \left\llbracket a_{3}a_{4}\right\rrbracket }\equiv -i\widetilde{\Delta}_{\mathcal{\epsilon}}^{2}\zeta_{\tau_{1}\tau_{3}}^{a_{1}a_{3}}\zeta_{\tau_{2}\tau_{4}}^{a_{2}a_{4}},\label{eq: M matrix - 1}
\end{equation}
\noindent and
\begin{equation}
\widetilde{\Delta}_{\mathcal{\epsilon}}^{2}=\frac{\Delta_{\epsilon}^{2}}{8\ln2}.\label{eq:Delta defined - 1}
\end{equation}

\noindent To decouple the quartic $a$-field term in Eq.~\eqref{eq:Delta defined - 1},
we perform a Hubbard-Stratonovich transformation (similarly to e.g. Ref.~\citep{Kennett2001})
\begin{widetext}
\begin{align}
e^{iS_{\Delta}\left[a\right]} & =\int\left[\mathcal{DQ}\right]\exp\left\{ \frac{i}{2!}\sum_{\vec{r}}\left[M^{-1}\right]_{\left\llbracket \tau_{1}\tau_{2}\right\rrbracket \left\llbracket \tau_{3}\tau_{4}\right\rrbracket }^{\left\llbracket a_{1}a_{2}\right\rrbracket \left\llbracket a_{3}a_{4}\right\rrbracket }\mathcal{Q}_{\vec{r}\vec{r},\tau_{1}\tau_{2}}^{\overline{a_{1}}\overline{a_{2}}}\mathcal{Q}_{\vec{r}\vec{r},\tau_{3}\tau_{4}}^{\overline{a_{3}}\overline{a_{4}}}+\frac{i}{2!}\sum_{\vec{r}}\mathcal{Q}_{\vec{r}\vec{r},\tau_{1}\tau_{2}}^{a_{1}a_{2}}a_{\vec{r},\tau_{1}}^{\overline{a_{1}}}a_{\vec{r},\tau_{2}}^{\overline{a_{2}}}\right\} \nonumber\\
 & \equiv\int\left[\mathcal{DQ}\right]e^{iS_{M^{-1}}\left[\mathcal{Q}\right]+iS_{\mathcal{Q}}\left[a\right]},
\end{align}
\end{widetext}

\noindent where

\begin{equation}
\left[M^{-1}\right]_{\left\llbracket \tau_{1}\tau_{2}\right\rrbracket \left\llbracket \tau_{3}\tau_{4}\right\rrbracket }^{\left\llbracket a_{1}a_{2}\right\rrbracket \left\llbracket a_{3}a_{4}\right\rrbracket }\equiv \frac{i}{\widetilde{\Delta}_{\mathcal{\epsilon}}^{2}}\zeta_{\tau_{1}\tau_{3}}^{a_{1}a_{3}}\zeta_{\tau_{2}\tau_{4}}^{a_{2}a_{4}},\label{eq: M matrix inv - 1}
\end{equation}

\begin{equation}
S_{M^{-1}}\left[\mathcal{Q}\right]=\frac{1}{2!}\sum_{\vec{r}}\left[M^{-1}\right]_{\left\llbracket \tau_{1}\tau_{2}\right\rrbracket \left\llbracket \tau_{3}\tau_{4}\right\rrbracket }^{\left\llbracket a_{1}a_{2}\right\rrbracket \left\llbracket a_{3}a_{4}\right\rrbracket }\mathcal{Q}_{\vec{r}\vec{r},\tau_{1}\tau_{2}}^{\overline{a_{1}}\overline{a_{2}}}\mathcal{Q}_{\vec{r}\vec{r},\tau_{3}\tau_{4}}^{\overline{a_{3}}\overline{a_{4}}},\label{eq: S_M_inv - 1}
\end{equation}
\noindent and
\begin{equation}
S_{\mathcal{Q}}\left[a\right]=\frac{1}{2!}\sum_{\vec{r}}\mathcal{Q}_{\vec{r}\vec{r},\tau_{1}\tau_{2}}^{a_{1}a_{2}}a_{\vec{r},\tau_{1}}^{\overline{a_{1}}}a_{\vec{r},\tau_{2}}^{\overline{a_{2}}},\label{eq: S_Q - 1}
\end{equation}

\noindent and $\mathcal{Q}$ is an auxiliary field introduced by the
transformation.

At this point, the generating functional $\mathcal{Z}\left[f,K\right]$
can be written as

\begin{align}
\mathcal{\check{Z}}\left[f\right]&=\int\left[\mathcal{D}a\right]\left\langle \mathbf{a}\left(\tau_{i}\right)\left|\hat{\rho}_{i}\right|\mathbf{a}\left(\tau_{f}\right)\right\rangle e^{-\frac{1}{2}\left\{ \mathbf{a}\left(\tau_{i}\right).\mathbf{a}\left(\tau_{i}\right)+\mathbf{a}\left(\tau_{f}\right).\mathbf{a}\left(\tau_{f}\right)\right\} }\nonumber\\
&\quad\times\int\left[\mathcal{D}Q\right]e^{i\left(S_{J}\left[a\right]+S_{0}\left[a\right]+S_{M^{-1}}\left[\mathcal{Q}\right]+S_{f}\left[a\right]+S_{\mathcal{Q}}\left[a\right]\right)} \:\: .\label{eq:avg Z - 3}
\end{align}

\noindent Next, following Refs.~\citep{Dupuis2001,Sengupta2005,Kennett2011,Fitzpatrick2018a},
we decouple the hopping term by performing another Hubbard-Stratonovich transformation

\begin{widetext}
\begin{align}
\mathcal{Z}\left[f\right] & =\int\left[\mathcal{D}a\right]\left\langle \mathbf{a}\left(\tau_{i}\right)\left|\hat{\rho}_{i}\right|\mathbf{a}\left(\tau_{f}\right)\right\rangle e^{-\frac{1}{2}\left\{ \mathbf{a}\left(\tau_{i}\right).\mathbf{a}\left(\tau_{i}\right)+\mathbf{a}\left(\tau_{f}\right).\mathbf{a}\left(\tau_{f}\right)\right\} } \nonumber \\
 & \quad\times\int\left[\mathcal{DQ}\right]\int\left[\mathcal{D}\psi\right]e^{i\left(-S_{J^{-1}}\left[\psi\right]+S_{0}\left[a\right]+S_{M^{-1}}\left[\mathcal{Q}\right]-S_{\psi}\left[a\right]+S_{f}\left[a\right]+S_{\mathcal{Q}}\left[a\right]\right)},\label{eq:avg Z - 4}
\end{align}
\end{widetext}

\noindent where

\begin{equation}
S_{J^{-1}}\left[\psi\right]=\frac{1}{2!}\sum_{\vec{r}_{1}\vec{r}_{2}}\left(\frac{1}{2}\left[J^{-1}\right]_{\vec{r}_{1}\vec{r}_{2},\tau_{1}\tau_{2}}^{a_{1}a_{2}}\right)\psi_{\vec{r}_{1},\tau_{1}}^{\overline{a_{1}}}\psi_{\vec{r}_{2},\tau_{2}}^{\overline{a_{2}}},\label{eq: S_J_inv - 1}
\end{equation}
\noindent with
\begin{equation}
S_{\psi}\left[a\right]=\sum_{\vec{r}}\psi_{\vec{r},\tau}^{a}a_{\vec{r},\tau}^{\overline{a}},\label{eq: S_f_plus_psi - 1}
\end{equation}

\noindent and $\psi$ is another auxiliary field. By making a field
substitution, $\psi_{\vec{r},\tau}^{a}\to-\psi_{\vec{r},\tau}^{a}+f_{\vec{r},\tau}^{a}$,
and rearranging terms in Eq.~\eqref{eq:avg Z - 4} we get

\begin{align}
\mathcal{\check{Z}}\left[f\right]	=\int\left[\mathcal{D}Q\right]\left[\mathcal{D}\psi\right]e^{i\left(-S_{J^{-1}}\left[\psi-f\right]+S_{M^{-1}}\left[\mathcal{Q}\right]\right)}\mathcal{Z}_{0}\left[\psi,\mathcal{Q}\right],\label{eq:avg Z - 5}
\end{align}
\noindent where

\begin{align}
    \mathcal{Z}_{0}\left[\psi,\mathcal{Q}\right] &\equiv  e^{iW_{0}\left[\psi,\mathcal{Q}\right]}\nonumber\\
	&=\int\left[\mathcal{D}a\right]\left\langle \mathbf{a}\left(\tau_{i}\right)\left|\hat{\rho}_{i}\right|\mathbf{a}\left(\tau_{f}\right)\right\rangle \nonumber\\
	&\quad\quad\times e^{-\frac{1}{2}\left\{ \mathbf{a}\left(\tau_{i}\right).\mathbf{a}\left(\tau_{i}\right)+\mathbf{a}\left(\tau_{f}\right).\mathbf{a}\left(\tau_{f}\right)\right\} } \nonumber\\
	&\quad\quad\times e^{i\left(S_{0}\left[a\right]+S_{\psi}\left[a\right]+S_{\mathcal{Q}}\left[a\right]\right)}.
	\label{eq:Z0 - 1}
\end{align}

\noindent In this context $\psi$ and $\mathcal{Q}$ take the same form as $f$ and $K$ in the 2PI generating functionals introduced in Ref.~\cite{Fitzpatrick2018a}, i.e. by taking functional derivatives of $\mathcal{Z}_{0}\left[\psi,\mathcal{Q}\right]$ (or $W_{0}\left[\psi,\mathcal{Q}\right]$) with respect to $\psi$ and $\mathcal{Q}$ one can generate all n-point COGFs (or CCOGFs). In this case the generating functionals $\mathcal{Z}_{0}\left[\psi,\mathcal{Q}\right]$ and $W_{0}\left[\psi,\mathcal{Q}\right]$ are governed by a different theory than that introduced in Ref.~\cite{Fitzpatrick2018a}.

$W_{0}\left[\psi,\mathcal{Q}\right]$ generates all the $n$-point CCOGFs in the limit of zero disorder and hopping
for a system prepared in the initial state $\hat{\rho}_{i}$
\begin{equation}
    \mathcal{G}_{\vec{r}_{1}\ldots\vec{r}_{n},\tau_{1}\ldots\tau_{n}}^{a_{1}\ldots a_{n},c}=\left(-1\right)^{n-1}\left.\frac{\delta^{n}W_{0}\left[\psi,\mathcal{Q}\right]}{\delta\psi_{\vec{r}_{1},\tau_{1}}^{\overline{a_{1}}}\ldots\delta\psi_{\vec{r}_{n},\tau_{n}}^{\overline{a_{n}}}}\right|_{\psi,\mathcal{Q}\rightarrow0},
    \label{eq:n-point CCOGFs - 1}
\end{equation}
\noindent as well as a set of generalized CCOGFs defined by:
\begin{widetext}
\begin{align}
   & \mathcal{G}_{\vec{r}_{1}\ldots\vec{r}_{n_{1}}\left\llbracket \vec{r}_{1}^{'}\vec{r}_{1}^{''}\right\rrbracket \ldots\left\llbracket \vec{r}_{n_{2}}^{'}\vec{r}_{n_{2}}^{''}\right\rrbracket ,\tau_{1}\ldots\tau_{n_{1}}\left\llbracket \tau_{1}^{'}\tau_{1}^{''}\right\rrbracket \ldots\left\llbracket \tau_{n_{2}}^{'}\tau_{n_{2}}^{''}\right\rrbracket }^{a_{1}\ldots a_{n_{1}}\left\llbracket a_{1}^{'}a_{1}^{''}\right\rrbracket \ldots\left\llbracket a_{n_{2}}^{'}a_{n_{2}}^{''}\right\rrbracket ,c}\nonumber\\
   &\quad \quad \equiv-\left(-1\right)^{n_{1}}\left(2i\right)^{n_{2}}\left.\frac{\delta^{n_{1}+n_{2}}W_{0}\left[\psi,\mathcal{Q}\right]}{\delta\psi_{\vec{r}_{1},\tau_{1}}^{\overline{a_{1}}}\ldots\delta\psi_{\vec{r}_{n_{1}},\tau_{n_{1}}}^{\overline{a_{n_{1}}}}\delta\mathcal{Q}_{\vec{r}_{1}^{'}\vec{r}_{1}^{''},\tau_{1}^{'}\tau_{1}^{''}}^{\overline{a_{1}^{'}a_{1}^{''}}}\ldots\delta Q_{\vec{r}_{n_{2}}^{'}\vec{r}_{n_{2}}^{''},\tau_{n_{2}}^{'}\tau_{n_{2}}^{''}}^{\overline{a_{n_{2}}^{'}a_{n_{2}}^{''}}}}\right|_{\psi,\mathcal{Q}\rightarrow0}.
   \label{eq:n-point CCOGFs generalized - 1}
\end{align}

\noindent These functions are connected in a particular sense: indices that are paired inside a pair of brackets $\left\llbracket \ldots\right\rrbracket $ should be thought of as indices belonging to a single field. If we assume an initial state of the form
\begin{equation}
    \hat{\rho}_{i}=\otimes_{\vec{r}}\left| n_{i,\vec{r}}\right\rangle \left\langle n_{i,\vec{r}} \right|,
    \label{eq:rho-density - 1}
\end{equation}

\noindent the CCOGFs defined in Eqs.~\eqref{eq:n-point CCOGFs - 1} and~\eqref{eq:n-point CCOGFs generalized - 1} vanish unless all site indices are equal. Moreover, when Eq.~\eqref{eq:rho-density - 1} holds, correlators of the form $\left\langle a_{\vec{r}_1, \tau_1}^{a_1} \ldots a_{\vec{r}_n, \tau_n}^{a_n}\right\rangle_{S_0}$ vanish unless the number of $a^{*}$-fields equals the number of $a$-fields. By inverting Eq.~\eqref{eq:n-point CCOGFs generalized - 1} we may rewrite $W_0$ as

\begin{align}
W_{0}\left[\psi,\mathcal{Q}\right] & =-\Theta\left(n_1 +n_2 -1/2\right)\sum_{n_{1}=0}^{\infty}\sum_{n_{2}=0}^{\infty}\frac{\left(-i\right)^{n_{2}}}{\left(2n_{1}\right)!n_{2}! 
2^{n_{2}}}\prod_{m_{1}=1}^{2n_{1}}\left(\sum_{r_{m_{1}}}\right)\prod_{m_{2}=1}^{n_{2}}\left(\sum_{r_{m_{2}}^{\prime}r_{m_{2}}^{\prime\prime}}\right)\nonumber \\
 & \quad\quad\quad\times\mathcal{G}_{\vec{r}_{1}\ldots\vec{r}_{2n_{1}}\left\llbracket \vec{r}_{1}^{\prime}\vec{r}_{1}^{\prime\prime}\right\rrbracket \ldots\left\llbracket \vec{r}_{n_{2}}^{\prime}\vec{r}_{n_{2}}^{\prime\prime}\right\rrbracket ,\tau_{1}\ldots\tau_{2n_{1}}\left\llbracket \tau_{1}^{\prime}\tau_{1}^{\prime\prime}\right\rrbracket \ldots\left\llbracket \tau_{n_{2}}^{\prime}\tau_{n_{2}}^{\prime\prime}\right\rrbracket }^{a_{1}\ldots a_{2n_{1}}\left\llbracket a_{1}^{\prime}a_{1}^{\prime\prime}\right\rrbracket \ldots\left\llbracket a_{n_{2}}^{\prime}a_{n_{2}}^{\prime\prime}\right\rrbracket ,c}\nonumber \\
 &\quad\quad \quad\times\psi_{r_{1},\tau_{1}}^{\overline{a_{1}}}\ldots\psi_{r_{2n_{1}},\tau_{2n_{1}}}^{\overline{a_{2n_{1}}}}\mathcal{Q}_{r_{1}^{\prime}r_{1}^{\prime\prime},\tau_{1}^{\prime}\tau_{1}^{\prime\prime}}^{\overline{a_{1}^{\prime}}\overline{a_{1}^{\prime\prime}}}\ldots\mathcal{Q}_{r_{n_{2}}^{\prime}r_{n_{2}}^{\prime\prime},\tau_{n_{2}}^{\prime}\tau_{n_{2}}^{\prime\prime}}^{\overline{a_{n_{2}}^{\prime}}\overline{a_{n_{2}}^{\prime\prime}}},\label{eq: W_0 - 1}
\end{align}
\noindent where $\Theta\left(x\right)$ is the Heaviside function. Once again, following Refs.~\citep{Dupuis2001,Sengupta2005,Kennett2011,Fitzpatrick2018a},
we perform another Hubbard-Stratonovich transformation to decouple
the inverse hopping term such that

\begin{equation}
\mathcal{\check{Z}}\left[f\right]=\int\left[\mathcal{D}z\right]\int\left[\mathcal{DQ}\right]e^{i\left(S_{J}\left[z\right]+S_{M^{-1}}\left[\mathcal{Q}\right]+\widetilde{W}_{0}\left[z,\mathcal{Q}\right]+S_{f}\left[z\right]\right)},\label{eq:avg Z - 6}
\end{equation}
\end{widetext}

\noindent where

\begin{equation}
e^{i\widetilde{W}_{0}\left[z,\mathcal{Q}\right]}=\int\left[\mathcal{D}\psi\right]e^{i\left(W_{0}\left[\psi,\mathcal{Q}\right]+S_{z}\left[\psi\right]\right)},\label{eq:W_0 tilde - 1}
\end{equation}
\noindent with
\begin{equation}
S_{z}\left[\psi\right]=\sum_{\vec{r}}z_{\vec{r},\tau}^{a}\psi_{\vec{r},\tau}^{\overline{a}}.\label{eq:S_z - 1}
\end{equation}

\noindent The effective theory is obtained by adding all of the
action terms excluding sources

\begin{equation}
S_{\text{eff}}^{\text{dis}}\left[z,\mathcal{Q}\right]=S_{J}\left[z\right]+S_{M^{-1}}\left[\mathcal{Q}\right]+\widetilde{W}_{0}\left[z,\mathcal{Q}\right].\label{eq: S_eff^dis - 1}
\end{equation}

\noindent We next perform a cumulant expansion of $\widetilde{W}_{0}$,
similar to that found in Refs.~\citep{Dupuis2001,Sengupta2005,Kennett2011,Fitzpatrick2018a}
although the calculation is more complicated in the disordered case.

Even in the compact notation we introduced in Sec.~\ref{subsec:Contour-ordered Green's functions},
the resulting expression for the effective theory is quite cumbersome
to write out. We therefore condense the notation further such that

\begin{align}
X^{\chi_{i}} & =\left(\begin{array}{c}
X^{z_{i}}\\
X^{\mathcal{Q}_{i^{\prime}i^{\prime\prime}}}
\end{array}\right)\nonumber \\
 & =\left(\begin{array}{c}
X_{\vec{r}_{i},\tau_{i}}^{a_{i}} \\[2mm]
X_{\vec{r}_{i}^{\prime}\vec{r}_{i}^{\prime\prime},\tau_{i}^{\prime}\tau_{i}^{\prime\prime}}^{a_{i}^{\prime}a_{i}^{\prime\prime}}
\end{array}\right),\label{eq:further condensed notation - 1}
\end{align}

\begin{align}
X^{\chi_{i}}Y^{\chi_{i}} & =X^{z_{i}}Y^{z_{i}}+X^{\mathcal{Q}_{i^{\prime}i^{\prime\prime}}}Y^{\mathcal{Q}_{i^{\prime}i^{\prime\prime}}}\nonumber \\
 & =\sum_{\vec{r}_{i}}X_{\vec{r}_{i},\tau_{i}}^{a_{i}}Y_{\vec{r}_{i},\tau_{i}}^{\overline{a_{i}}}\nonumber \\
 & \quad+\sum_{\vec{r}_{i}^{\prime}\vec{r}_{i}^{\prime\prime}}X_{\vec{r}_{i}^{\prime}\vec{r}_{i}^{\prime\prime},\tau_{i}^{\prime}\tau_{i}^{\prime\prime}}^{a_{i}^{\prime}a_{i}^{\prime\prime}}Y_{\vec{r}_{i}^{\prime}\vec{r}_{i}^{\prime\prime},\tau_{i}^{\prime}\tau_{i}^{\prime\prime}}^{\overline{a_{i}^{\prime}}\overline{a_{i}^{\prime\prime}}}.\label{eq:further condensed notation - 2}
\end{align}

\noindent Using the above shorthand notation, the effective theory
can be expressed as follows

\begin{align}
S_{\text{eff}}^{\text{dis}}\left[\Phi\right] & =\left(\frac{1}{2!}\left[g_{0}^{-1}\right]^{\chi_{1}\chi_{2}}\right)\Phi^{\chi_{1}}\Phi^{\chi_{2}}\nonumber \\
 & \quad+\sum_{n=1}^{\infty}\frac{1}{n!}g^{\chi_{1}\ldots\chi_{n}}\Phi^{\chi_{1}}\ldots\Phi^{\chi_{n}},\label{eq: S_eff^dis - 2}
\end{align}

\noindent where

\begin{eqnarray}
\Phi^{\chi_{i}} & = & \left(\begin{array}{c}
z_{\vec{r}_{i},\tau_{i}}^{a_{i}}\\[2mm]
\mathcal{Q}_{\vec{r}_{i}^{\prime}\vec{r}_{i}^{\prime\prime},\tau_{i}^{\prime}\tau_{i}^{\prime\prime}}^{a_{i}^{\prime}a_{i}^{\prime\prime}}
\end{array}\right).\label{eq:Phi field - 1}
\end{eqnarray}
\noindent The couplings for quadratic terms in the theory are:
\begin{align}
\left[g_{0}^{-1}\right]^{z_{1}z_{2}} & =\left[\left(\mathcal{G}^{c}\right)^{-1}\right]_{\vec{r}_{1}\vec{r}_{2},\tau_{1}\tau_{2}}^{a_{1}a_{2}},\label{eq: g_0^psi^psi inv - 1}\\
\left[g_{0}^{-1}\right]^{z_{1}\mathcal{Q}_{23}} & =0,\label{eq: g_0^psi^Q inv - 1}\\
\left[g_{0}^{-1}\right]^{\mathcal{Q}_{12}z_{3}} & =0,\label{eq: g_0^Q^psi inv - 1}\\
\left[g_{0}^{-1}\right]^{\mathcal{Q}_{12}\mathcal{Q}_{34}} & =\delta_{\left\llbracket \vec{r}_{1}\vec{r}_{2}\right\rrbracket \left\llbracket \vec{r}_{3}\vec{r}_{4}\right\rrbracket }\left[M^{-1}\right]_{\left\llbracket \tau_{1}\tau_{2}\right\rrbracket \left\llbracket \tau_{3}\tau_{4}\right\rrbracket }^{\left\llbracket a_{1}a_{2}\right\rrbracket \left\llbracket a_{3}a_{4}\right\rrbracket },\label{eq: g_0^Q^Q inv - 1}
\end{align}
\noindent with
\begin{equation}
\delta_{\left\llbracket \vec{r}_{1}\vec{r}_{2}\right\rrbracket \left\llbracket \vec{r}_{3}\vec{r}_{4}\right\rrbracket }=\begin{cases}
1, & \text{if }\vec{r}_{1}=\vec{r}_{2}=\vec{r}_{3}=\vec{r}_{4}\\
0, & \text{otherwise}
\end{cases},\label{eq:generalized delta_r - 1}
\end{equation}

\noindent and the vertices $g^{\chi_{1}\ldots\chi_{n}}$ are combinations of the CCOGFs generated from $W_{0}$. The presence of
the contour ordering operator $T_{C}$ in the CCOGFs leads to symmetry
under permutations $\left\{ p_{1},\ldots,p_{n}\right\} $ of the sequence
$\left\{ 1,\ldots,n\right\} $:

\begin{align}
\left[g_{0}^{-1}\right]^{\chi_{1}\chi_{2}} & =\left[g_{0}^{-1}\right]^{\chi_{p_{1}}\chi_{p_{2}}},\label{eq:g_0 inv symmetry relation - 1}\\
g^{\chi_{1}\ldots\chi_{n}} & =g^{\chi_{p_{1}}\ldots\chi_{p_{n}}}.\label{eq:g symmetry relation - 1}
\end{align}

The action in Eq.~\eqref{eq: S_eff^dis - 2} contains an infinite
sum, therefore for practical calculations we truncate the action keeping only terms to order $\mathcal{O}\left[\psi^{n}\mathcal{Q}^{m}\right]$
where $4-2m\ge n\ge0$ and $2\ge m\ge0$. Out of these terms, the
only non-vanishing terms (modulo index-permutations) are

\begin{align}
g^{\mathcal{Q}_{12}}&=\frac{i}{2}\mathcal{G}_{\left\llbracket \vec{r}_{1}\vec{r}_{2}\right\rrbracket ,\left\llbracket \tau_{1}\tau_{2}\right\rrbracket }^{\left\llbracket a_{1}a_{2}\right\rrbracket ,c}\nonumber\\
 & \quad + \frac{1}{4}\mathcal{G}_{\left\llbracket \vec{r}_{1}\vec{r}_{2}\right\rrbracket \vec{r}_{3}\vec{r}_{4} ,\left\llbracket \tau_{1}\tau_{2}\right\rrbracket  \tau_{3}\tau_{4}}^{\left\llbracket a_{1}a_{2}\right\rrbracket a_3 a_4,c} \left[\left(\mathcal{G}^{c}\right)^{-1}\right]_{\vec{r}_{3}\vec{r}_{4},\tau_{3}\tau_{4}}^{\overline{a_{3}a_{4}}},
\label{eq: g^Q - 1}
\end{align}

\begin{equation}
g^{z_{1}z_{2}}=2J_{\vec{r}_{1}\vec{r}_{2},\tau_{1}\tau_{2}}^{a_{1}a_{2}}+\tilde{u}_{\vec{r}_{1}\vec{r}_{2},\tau_{1}\tau_{2}}^{a_{1}a_{2}},\label{eq:g^psi^psi - 1}
\end{equation}
\begin{align}
    g^{Q_{12}Q_{34}}&=\frac{1}{4}\mathcal{G}_{\left\llbracket r_{1}r_{2}\right\rrbracket \left\llbracket r_{3}r_{4}\right\rrbracket ,\left\llbracket \tau_{1}\tau_{2}\right\rrbracket \left\llbracket \tau_{3}\tau_{4}\right\rrbracket }^{\left\llbracket a_{1}a_{2}\right\rrbracket \left\llbracket a_{3}a_{4}\right\rrbracket ,c}, \label{eq:g^Q^Q - 1}
\end{align}

\begin{align}
g^{z_{1}z_{2}\mathcal{Q}_{34}} =\frac{1}{3}u_{\vec{r}_{1}\vec{r}_{2}\left\llbracket \vec{r}_{3}\vec{r}_{4}\right\rrbracket ,\tau_{1}\tau_{2}\left\llbracket \tau_{3}\tau_{4}\right\rrbracket }^{a_{1}a_{2}\left\llbracket a_{3}a_{4}\right\rrbracket },\label{eq:g^psi^psi^Q - 1}
\end{align}

\begin{equation}
g^{z_{1}z_{2}z_{3}z_{4}}=u_{\vec{r}_{1}\vec{r}_{2}\vec{r}_{3}\vec{r}_{4},\tau_{1}\tau_{2}\tau_{3}\tau_{4}}^{a_{1}a_{2}a_{3}a_{4}},\label{eq:g^psi^psi^psi^psi - 1}
\end{equation}
\noindent where

\begin{align}
    \tilde{u}_{\vec{r}_{1}\vec{r}_{2},\tau_{1}\tau_{2}}^{a_{1}a_{2}}=-\frac{1}{2!}\sum_{\vec{r}_{3}\vec{r}_{4}}u_{\vec{r}_{1}\vec{r}_{2}\vec{r}_{3}\vec{r}_{4},\tau_{1}\tau_{2}\tau_{3}\tau_{4}}^{a_{1}a_{2}a_{3}a_{4}}\left\{ i\mathcal{G}_{\vec{r}_{3}\vec{r}_{4},\tau_{3}\tau_{4}}^{\overline{a_{3}}\overline{a_{4}},c}\right\} ,
\end{align}

\begin{align}
u_{\vec{r}_{1}\vec{r}_{2}\left\llbracket \vec{r}_{3}\vec{r}_{4}\right\rrbracket ,\tau_{1}\tau_{2}\left\llbracket \tau_{3}\tau_{4}\right\rrbracket }^{a_{1}a_{2}\left\llbracket a_{3}a_{4}\right\rrbracket } & \equiv \frac{3i}{2}\prod_{m=1}^{2}\left\{ \sum_{\vec{r}_{m}^{\prime}}\left[\left(\mathcal{G}^{c}\right)^{-1}\right]_{\vec{r}_{m}\vec{r}_{m}^{\prime},\tau_{m}\tau_{m}^{\prime}}^{a_{m}a_{m}^{\prime}}\right\} \nonumber \\
 & \phantom{\frac{3i}{2}\prod_{m=1}^{2}}\times\mathcal{G}_{\vec{r}_{1}^{\prime}\vec{r}_{2}^{\prime}\left\llbracket \vec{r}_{3}\vec{r}_{4}\right\rrbracket ,\tau_{1}^{\prime}\tau_{2}^{\prime}\left\llbracket \tau_{3}\tau_{4}\right\rrbracket }^{\overline{a_{1}^{\prime}}\overline{a_{2}^{\prime}}\left\llbracket a_{3}a_{4}\right\rrbracket ,c},
\end{align}
 
\noindent and

\begin{align}
u_{\vec{r}_{1}\vec{r}_{2}\vec{r}_{3}\vec{r}_{4},\tau_{1}\tau_{2}\tau_{3}\tau_{4}}^{a_{1}a_{2}a_{3}a_{4}} & =-\prod_{m=1}^{4}\left\{ \sum_{\vec{r}_{m}^{\prime}}\left[\left(\mathcal{G}^{c}\right)^{-1}\right]_{\vec{r}_{m}\vec{r}_{m}^{\prime},\tau_{m}\tau_{m}^{\prime}}^{a_{m}a_{m}^{\prime}}\right\} \nonumber \\
 & \quad\phantom{-\prod_{m=1}^{4}}\times\mathcal{G}_{\vec{r}_{1}^{\prime}\vec{r}_{2}^{\prime}\vec{r}_{3}^{\prime}\vec{r}_{4}^{\prime},\tau_{1}^{\prime}\tau_{2}^{\prime}\tau_{3}^{\prime}\tau_{4}^{\prime}}^{\overline{a_{1}^{\prime}}\overline{a_{2}^{\prime}}\overline{a_{3}^{\prime}}\overline{a_{4}^{\prime}},c}.
\end{align}

In this work, we do not consider a trapping potential for simplicity. Therefore, local quantities such as $n$-point CCOGFs in the atomic limit have no spatial dependency, and henceforth we will drop the $\vec{r}$ index for these quantities.

\section{Equations of motion}
\label{sec:EoM}

We now apply a 2PI approach to our effective theory to obtain equations
of motion for the mean-field and the full two-point CCOGF for the $\Phi$-fields (the
``full propagator'' from now on). In a related work~\cite{Fitzpatrick2018a}, we applied this approach to the homogeneous BHM in the strong coupling limit. We follow the same general procedure in this work with a few modifications to account for the additional vertices and auxiliary fields that appear in our effective theory of the disordered BHM as compared to that of the homogeneous BHM. Here we only briefly outline the 2PI calculation. For a more detailed exposition of the 2PI approach, see Ref.~\cite{Fitzpatrick2018a}.

We define the mean-field $\mathcal{V}^{\chi_{1}}$ and full propagator
$\mathcal{V}^{\chi_{1}\chi_{2},c}$ as follows 
\begin{widetext}
\begin{equation}
\mathcal{V}^{\chi_{1}}=\left\langle \Phi^{\chi_{1}}\right\rangle =\left(\begin{array}{c}
\left\langle z_{r_{1},\tau_{1}}^{a_{1}}\right\rangle \\
\left\langle \mathcal{Q}_{r_{1}^{\prime}r_{1}^{\prime\prime},\tau_{1}^{\prime}\tau_{1}^{\prime\prime}}^{a_{1}^{\prime}a_{1}^{\prime\prime}}\right\rangle 
\end{array}\right),\label{eq:V^(1) - 1}
\end{equation}

\begin{equation}
i\mathcal{V}^{\chi_{1}\chi_{2},c}=\left\langle \Phi^{\chi_{1}}\Phi^{\chi_{2}}\right\rangle =\left(\begin{array}{cc}
\left\langle z_{r_{1},\tau_{1}}^{a_{1}}z_{r_{2},\tau_{2}}^{a_{2}}\right\rangle ^{c} & \left\langle z_{r_{1},\tau_{1}}^{a_{1}}\mathcal{Q}_{r_{2}^{\prime}r_{2}^{\prime\prime},\tau_{2}^{\prime}\tau_{2}^{\prime\prime}}^{a_{2}^{\prime}a_{2}^{\prime\prime}}\right\rangle ^{c}\\
\left\langle \mathcal{Q}_{r_{1}^{\prime}r_{1}^{\prime\prime},\tau_{1}^{\prime}\tau_{1}^{\prime\prime}}^{a_{1}^{\prime}a_{1}^{\prime\prime}}\psi_{r_{2},\tau_{2}}^{a_{2}}\right\rangle ^{c} & \left\langle \mathcal{Q}_{r_{1}^{\prime}r_{1}^{\prime\prime},\tau_{1}^{\prime}\tau_{1}^{\prime\prime}}^{a_{1}^{\prime}a_{1}^{\prime\prime}}\mathcal{Q}_{r_{2}^{\prime}r_{2}^{\prime\prime},\tau_{2}^{\prime}\tau_{2}^{\prime\prime}}^{a_{2}^{\prime}a_{2}^{\prime\prime}}\right\rangle ^{c}
\end{array}\right).\label{eq:V^(2) - 1}
\end{equation}
\end{widetext}

\noindent Next, we consider the 2PI Dyson's equation

\begin{equation}
\left[\mathcal{V}^{-1}\right]^{\chi_{1}\chi_{2},c}=\left[D^{-1}\right]^{\chi_{1}\chi_{2}}-\left[\Sigma^{\left( \text{2PI}\right)}\right]^{\chi_{1}\chi_{2}},\label{eq:Dyson's equation - 1}
\end{equation}

\noindent where
\begin{align}
\left[D^{-1}\right]^{\chi_{1}\chi_{2}} & \equiv\frac{\delta^{2}S\left[\mathcal{V}^{\chi}\right]}{\delta\mathcal{V}^{\chi_{1}}\delta\mathcal{V}^{\chi_{2}}}\nonumber \\
 & =\left[g_{0}^{-1}\right]^{\chi_{1}\chi_{2}}-\left[\Sigma^{\left( 1\right)}\right]^{\chi_{1}\chi_{2}}.\label{eq:D^inv - 1}
\end{align}

\noindent $\left[\Sigma^{\left(1\right)}\right]^{\chi_{1}\chi_{2}}$ is the ``1-loop'' self-energy

\begin{align}
\left[\Sigma^{\left(1\right)}\right]^{\chi_{1}\chi_{2}} & =-g^{\chi_{1}\chi_{2}}-g^{\chi_{1}\chi_{2}\chi_{3}}\mathcal{V}^{\chi_{3}}\nonumber \\
 & \quad-\frac{1}{2}g^{\chi_{1}\chi_{2}\chi_{3}\chi_{4}}\mathcal{V}^{\chi_{3}}\mathcal{V}^{\chi_{4}}.\label{eq:Sigma^1-loop - 1}
\end{align}

\noindent $\left[\Sigma^{\left(\text{2PI}\right)}\right]^{\chi_{1}\chi_{2}}$ is
the 2PI self-energy

\begin{equation}
\left[\Sigma^{\left(\text{2PI}\right)}\right]^{\chi_{1}\chi_{2}}=2i\frac{\delta\Gamma_{2}\left[\mathcal{V}^{\chi},\mathcal{V}^{\chi\chi^{\prime},c}\right]}{\delta\mathcal{V}^{\chi_{1}\chi_{2},c}},\label{eq: Self energy - 1}
\end{equation}

\noindent and $\Gamma_{2}\left[\mathcal{V}^{\left(1\right)},\mathcal{V}^{\left(2\right)}\right]$
is the sum of all 2PI connected vacuum diagrams in the theory with
vertices determined by the action

\begin{align}
S_{\text{int}}\left[\Phi;\mathcal{V}^{\left(1\right)}\right] & =\frac{1}{3!}g^{\chi_{1}\chi_{2}\chi_{3}}\Phi^{\chi_{1}}\Phi^{\chi_{2}}\Phi^{\chi_{3}}\nonumber \\
 & \quad+\frac{1}{3!}g^{\chi_{1}\chi_{2}\chi_{3}\chi_{4}}\Phi^{\chi_{1}}\Phi^{\chi_{2}}\Phi^{\chi_{3}}\mathcal{V}^{\chi_{4}}\nonumber \\
 & \quad+\frac{1}{4!}g^{\chi_{1}\chi_{2}\chi_{3}\chi_{4}}\Phi^{\chi_{1}}\Phi^{\chi_{2}}\Phi^{\chi_{3}}\Phi^{\chi_{4}}.\label{eq: S_int - 1}
\end{align}

\noindent To first order in the vertices we have

\begin{equation}
\left[\Sigma^{\left(\text{2PI}\right)}\right]^{\chi_{1}\chi_{2}}=-\frac{i}{2}g^{\chi_{1}\chi_{2}\chi_{3}\chi_{4}}\mathcal{V}^{\chi_{3}\chi_{4},c}.\label{eq:Sigma^(D.B.) - 1}
\end{equation}

\noindent If we define the full self-energy $\Sigma^{\chi_{1}\chi_{2}}$
to be

\begin{equation}
\Sigma^{\chi_{1}\chi_{2}}=\left[\Sigma^{\left(1\right)}\right]^{\chi_{1}\chi_{2}}+\left[\Sigma^{\left(\text{2PI}\right)}\right]^{\chi_{1}\chi_{2}},\label{eq:Full self energy - 1}
\end{equation}

\noindent then we may rearrange the Dyson's equation as follows

\begin{equation}
\mathcal{V}^{\chi_{1}\chi_{2},c}=\left[g_{0}\right]^{\chi_{1}\chi_{2}}+\left[g_{0}\right]^{\chi_{1}\chi_{3}}\Sigma^{\chi_{3}\chi_{4}}\mathcal{V}^{\chi_{4}\chi_{2},c}.\label{eq:Dyson's equation - 2}
\end{equation}
Additionally, the equation of motion for the mean field $\mathcal{V}^{\chi_{1}}$ is 

\begin{equation}
\frac{\delta S}{\delta\mathcal{V}^{\chi_{1}}}+\frac{i}{2}\left[\frac{\delta\left[D^{-1}\right]^{\chi_{2}\chi_{3}}}{\delta\mathcal{V}^{\chi_{1}}}\mathcal{V}^{\chi_{2}\chi_{3},c}\right]=0. \label{eq:Mean-field EOM-1}
\end{equation}
\noindent As will be clear shortly, we are mostly interested
in finding $\mathcal{V}^{z_{1}}$ and $\mathcal{V}^{z_{1}z_{2},c}$, that can be calculated from Eqs.~\eqref{eq:Mean-field EOM-1} and~\eqref{eq:Dyson's equation - 2} respectively, for the equilibrium and out-of-equilibrium scenarios . First, we start with $\mathcal{V}^{z_{1}z_{2},c}$. From Dyson's equation~\eqref{eq:Dyson's equation - 2} we can write
\begin{align}
\mathcal{V}^{z_{1}z_{2},c} & =\left[g_{0}\right]^{z_{1}z_{2}}\nonumber \\
 & \quad+\left[g_{0}\right]^{z_{1}z_{3}}\Sigma^{z_{3}z_{4}}\mathcal{V}^{z_{4}z_{2},c}\nonumber \\
 & \quad+\left[g_{0}\right]^{z_{1}z_{3}}\Sigma^{z_{3}\mathcal{Q}_{45}}\mathcal{V}^{\mathcal{Q}_{45}z_{2},c},\label{eq:Dyson's equation - 3}
\end{align}
\noindent where
\begin{align}
\Sigma^{z_{1}z_{2}} & =-g^{z_{1}z_{2}}-g^{z_{1}z_{2}\mathcal{Q}_{34}}\mathcal{V}^{\mathcal{Q}_{34}}\nonumber \\
 & \quad-\frac{1}{2}g^{z_{1}z_{2}z_{3}z_{4}}\mathcal{V}^{z_{3}}\mathcal{V}^{z_{4}}\nonumber \\
 & \quad-\frac{i}{2}g^{z_{1}z_{2}z_{3}z_{4}}\mathcal{V}^{z_{3}z_{4},c},\label{eq:Sigma^psi^psi - 1}\\
\Sigma^{z_{1}\mathcal{Q}_{23}} & =-g^{z_{1}\mathcal{Q}_{23}z_{4}}\mathcal{V}^{z_{4}}.\label{eq:Sigma^psi^Q - 1}
\end{align}

\noindent To make further progress, we  cast Eq.~\eqref{eq:Dyson's equation - 3}
in terms of the correlation functions of the original $a$-fields.
By inspection, one can see from Eq.~\eqref{eq:avg Z - 6} that $\mathcal{Z}\left[f,K\right]$
is the generator of COGFs of the $z$-fields in addition to the $a$-fields.
This implies that
\begin{align}
&\mathcal{V}^{z_{1}}=\check{G}_{\vec{r}_{1},\tau_{1}}^{a_{1}}\equiv \check{\phi}_{\vec{r}_{1},\tau_{1}}^{a_{1}},\label{eq:V^z - 1}\\
&\quad\mathcal{V}^{z_{1}z_{2},c}=\check{G}_{\vec{r}_{1}\vec{r}_{2},\tau_{1}\tau_{2}}^{a_{1}a_{2},c},\label{eq:V^z^z - 1}
\end{align}

\noindent where $\check{\phi}$ denotes the superfluid order parameter. Since
we consider hopping strengths below the critical value of the Mott
insulator to superfluid transition and assume that our initial state
is of the form given in Eq.~\eqref{eq:rho-density - 1}, we may safely assume that $\mathcal{V}^{z_{1}}=0$. Note that this assumption is not valid when the system is in the superfluid phase as will be discussed in Sec.~\ref{subsec:Super-Fluid}. For $\mathcal{V}^{\mathcal{Q}_{12}}$,
we follow a similar calculation to that in Ref\@.~\citep{Pairault2000}:
using Eqs.~\eqref{eq:avg COGF from avg Z - 1} and \eqref{eq:avg Z - 5}
we can write
\begin{widetext}
\begin{align}
\check{G}_{\vec{r}_{1}\vec{r}_{2},\tau_{1}\tau_{2}}^{a_{1}a_{2},c}	&=\check{G}_{\vec{r}_{1}\vec{r}_{2},\tau_{1}\tau_{2}}^{a_{1}a_{2}}\nonumber\\
	&=i\underset{f\rightarrow0}{\text{lim}}\frac{\delta^{2}\check{\mathcal{Z}}\left[f\right]}{\delta f_{\vec{r}_{1},\tau_{1}}^{\overline{a_{1}}}\delta f_{\vec{r}_{2},\tau_{2}}^{\overline{a_{2}}}}\nonumber\\
	&=-2\underset{f\rightarrow0}{\text{lim}}\int\left[\mathcal{D}a\right]\left\langle \mathbf{a}\left(\tau_{i}\right)\left|\hat{\rho}_{i}\right|\mathbf{a}\left(\tau_{f}\right)\right\rangle e^{-\frac{1}{2}\left\{ \mathbf{a}\left(\tau_{i}\right).\mathbf{a}\left(\tau_{i}\right)+\mathbf{a}\left(\tau_{f}\right).\mathbf{a}\left(\tau_{f}\right)\right\} }\nonumber\\
	&\quad\quad\quad\quad\quad\quad\times\int\left[\mathcal{D}Q\right]e^{i\left(S_{J}\left[a\right]+S_{0}\left[a\right]+S_{M^{-1}}\left[\mathcal{Q}\right]\right)}\left(\frac{\delta\left\{ e^{i\left(S_{f}\left[a\right]+S_{\mathcal{Q}}\left[a\right]\right)}\right\} }{\mathcal{Q}_{\vec{r}_{1}\vec{r}_{2},\tau_{1}\tau_{2}}^{\overline{a_{1}a_{2}}}}\right),\label{eq:V^Q - 1}
\end{align}

\noindent then integrate by parts to get

\begin{align}
\check{G}_{\vec{r}_{1}\vec{r}_{2},\tau_{1}\tau_{2}}^{a_{1}a_{2},c}	&=2\underset{f\rightarrow0}{\text{lim}}\int\left[\mathcal{D}a\right]\left\langle \mathbf{a}\left(\tau_{i}\right)\left|\hat{\rho}_{i}\right|\mathbf{a}\left(\tau_{f}\right)\right\rangle e^{-\frac{1}{2}\left\{ \mathbf{a}\left(\tau_{i}\right).\mathbf{a}\left(\tau_{i}\right)+\mathbf{a}\left(\tau_{f}\right).\mathbf{a}\left(\tau_{f}\right)\right\} }\nonumber\\
	&\quad\quad\quad\quad\quad\quad\times\int\left[\mathcal{D}Q\right]e^{i\left(S_{J}\left[a\right]+S_{0}\left[a\right]+S_{f}\left[a\right]+S_{\mathcal{Q}}\left[a\right]\right)}\left(\frac{\delta\left\{ e^{i\left(S_{M^{-1}}\left[\mathcal{Q}\right]\right)}\right\} }{\mathcal{Q}_{\vec{r}_{1}\vec{r}_{2},\tau_{1}\tau_{2}}^{\overline{a_{1}a_{2}}}}\right)\nonumber\\
	&=2i \left[M^{-1}\right]_{\left\llbracket \tau_{1}\tau_{2}\right\rrbracket \left\llbracket \tau_{3}\tau_{4}\right\rrbracket }^{\left\llbracket a_{1}a_{2}\right\rrbracket \left\llbracket a_{3}a_{4}\right\rrbracket }\int\left[\mathcal{D}Q\right]\mathcal{Q}_{\vec{r}_{3}\vec{r}_{4},\tau_{3}\tau_{4}}^{\overline{a_{3}a_{4}}}e^{i\tilde{S}\left[\mathcal{Q}\right]},\label{eq:V^Q - 2}
\end{align}
where
\begin{align}
    e^{i\tilde{S}\left[\mathcal{Q}\right]}	&=\int\left[\mathcal{D}a\right]\left\langle \mathbf{a}\left(\tau_{i}\right)\left|\hat{\rho}_{i}\right|\mathbf{a}\left(\tau_{f}\right)\right\rangle e^{-\frac{1}{2}\left\{ \mathbf{a}\left(\tau_{i}\right).\mathbf{a}\left(\tau_{i}\right)+\mathbf{a}\left(\tau_{f}\right).\mathbf{a}\left(\tau_{f}\right)\right\} }\nonumber\\
	&\quad\quad\quad\times e^{i\left(S_{J}\left[a\right]+S_{0}\left[a\right]+S_{M^{-1}}\left[\mathcal{Q}\right]+S_{f}\left[a\right]+S_{\mathcal{Q}}\left[a\right]\right)}.
	\label{eq:S-bar-Q-action}
\end{align}
\end{widetext}
Continuing with Eq.~\eqref{eq:V^Q - 2}, we have
\begin{align}
    \check{G}_{\vec{r}_{1}\vec{r}_{2},\tau_{1}\tau_{2}}^{a_{1}a_{2},c}	&=2i\left[M^{-1}\right]_{\left\llbracket \tau_{1}\tau_{2}\right\rrbracket \left\llbracket \tau_{3}\tau_{4}\right\rrbracket }^{\left\llbracket a_{1}a_{2}\right\rrbracket \left\llbracket a_{3}a_{4}\right\rrbracket }\left\langle \mathcal{Q}_{\vec{r}_{3}\vec{r}_{4},\tau_{3}\tau_{4}}^{\overline{a_{3}a_{4}}}\right\rangle \nonumber\\
	&=-\frac{2}{\widetilde{\Delta}_{\mathcal{\epsilon}}^{2}}\left\langle \mathcal{Q}_{\vec{r}_{1}\vec{r}_{2},\tau_{1}\tau_{2}}^{\overline{a_{1}a_{2}}}\right\rangle \nonumber\\
	&=-\frac{2}{\widetilde{\Delta}_{\mathcal{\epsilon}}^{2}}\mathcal{V}^{\mathcal{Q}_{12}},
	\label{eq:V^Q - 3}
\end{align}
\noindent hence
\begin{equation}
\mathcal{V}^{\mathcal{Q}_{12}}=-\frac{1}{2} \widetilde{\Delta}_{\mathcal{\epsilon}}^{2} \check{G}_{\vec{r}\vec{r},\tau_{1}\tau_{2}}^{a_{1}a_{2},c}.\label{eq:V^Q - 4}
\end{equation}

Substituting Eqs.~\eqref{eq: g_0^psi^psi inv - 1},~\eqref{eq:g^psi^psi - 1}-\eqref{eq:g^psi^psi^psi^psi - 1},
\eqref{eq:V^z - 1},~\eqref{eq:V^z^z - 1}, and~\eqref{eq:V^Q - 4} into
Eq.~\eqref{eq:Dyson's equation - 3} gives

\begin{equation}
\check{G}_{\vec{r}_{1}\vec{r}_{2},\tau_{1}\tau_{2}}^{a_{1}a_{2},c}=\mathcal{G}_{\tau_{1}\tau_{2}}^{a_{1}a_{2},c}+\mathcal{G}_{\tau_{1}\tau_{3}}^{a_{1}a_{3},c}\left[\Sigma^{zz}\right]_{\vec{r}_{3}\vec{r}_{4},\tau_{3}\tau_{4}}^{\overline{a_{3}}\overline{a_{4}}}\check{G}_{\vec{r}_{4}\vec{r}_{2},\tau_{4}\tau_{2}}^{a_{4}a_{2},c},\label{eq:Dyson's equation - 4}
\end{equation}

\noindent where

\begin{align}
\left[\Sigma^{zz}\right]_{\vec{r}_{1}\vec{r}_{2},\tau_{1}\tau_{2}}^{a_{1}a_{2}} & =\left[\Sigma_{1}^{zz}\right]_{\vec{r}_{1}\vec{r}_{2},\tau_{1}\tau_{2}}^{a_{1}a_{2}}\nonumber \\
 & \quad+\delta_{\vec{r}_{1}\vec{r}_{2}}\left(\left[\Sigma_{2}^{zz}\right]_{\vec{r}_{1},\tau_{1}\tau_{2}}^{a_{1}a_{2}}+\left[\Sigma_{3}^{zz}\right]_{\vec{r}_{1},\tau_{1}\tau_{2}}^{a_{1}a_{2}}\right),\label{eq:Sigma^z^z - 1}
\end{align}

\begin{equation}
\left[\Sigma_{1}^{zz}\right]_{\vec{r}_{1}\vec{r}_{2},\tau_{1}\tau_{2}}^{a_{1}a_{2}}=-2J_{\vec{r}_{1}\vec{r}_{2},\tau_{1}\tau_{2}}^{a_{1}a_{2}},\label{eq:Sigma_1^z^z - 1}
\end{equation}
\noindent and the contributions to the self energy are
\begin{align}
\left[\Sigma_{2}^{zz}\right]_{\vec{r},\tau_{1}\tau_{2}}^{a_{1}a_{2}} & =-\tilde{u}_{\tau_{1}\tau_{2}}^{a_{1}a_{2}}-\frac{1}{2}u_{\tau_{1}\tau_{2}\tau_{3}\tau_{4}}^{a_{1}a_{2}a_{3}a_{4}}\left(i\check{G}_{\vec{r},\tau_{3}\tau_{4}}^{\overline{a_{3}}\overline{a_{4}},c}\right),\label{eq:Sigma_2^z^z - 1}
\end{align}
\noindent and
\begin{equation}
\left[\Sigma_{3}^{zz}\right]_{\vec{r},\tau_{1}\tau_{2}}^{a_{1}a_{2}}=-\frac{i}{6} \widetilde{\Delta}_{\mathcal{\epsilon}}^{2} u_{\tau_{1}\tau_{2}\left\llbracket \tau_{3}\tau_{4}\right\rrbracket }^{a_{1}a_{2}\left\llbracket a_{3}a_{4}\right\rrbracket }\left(i\check{G}_{\vec{r},\tau_{3}\tau_{4}}^{\overline{a_{3}}\overline{a_{4}},c}\right).\label{eq:Sigma_3^z^z - 1}
\end{equation}
\noindent The next step is to calculate the mean field $\mathcal{V}^{z_{1}}$. Using Eqs.~\eqref{eq: S_eff^dis - 2},~\eqref{eq:D^inv - 1} and~\eqref{eq:Sigma^1-loop - 1} we can rewrite Eqs.~\eqref{eq:Mean-field EOM-1} as

\begin{equation}
\begin{aligned}0 & =\left(2J^{z_{1}z_{2}}+\tilde{u}^{z_{1}z_{2}}+\left[\left(\mathcal{G}^{c}\right)^{-1}\right]^{z_{1}z_{2}}\right)\mathcal{V}^{z_{2}}\\
 &\quad +g^{z_{1}z_{2}Q_{34}}\mathcal{V}^{z_{2}}\mathcal{V}^{Q_{34}} + \frac{1}{3!}g^{z_{1}z_{2}z_{3}z_{4}}\mathcal{V}^{z_{2}}\mathcal{V}^{z_{3}}\mathcal{V}^{z_{4}}\\
 &\quad +i g^{z_{1}Q_{23}z_{4}}\mathcal{V}^{Q_{23}z_{4},c}+\frac{i}{2}g^{z_{1}z_{2}z_{3}z_{4}}\mathcal{V}^{z_{2}}\mathcal{V}^{z_{3}z_{4},c}, \label{eq:Mean-field EOM-2}
\end{aligned}
\end{equation}

\noindent where we used $\mathcal{V}^{Q_{12}z_{3},c}=\mathcal{V}^{z_{1}Q_{23},c}$. Next we need to calculate $\mathcal{V}^{Q_{12}z_{3},c}$. Using Eq.~\eqref{eq:Dyson's equation - 2} again, we get

\begin{equation}
\begin{aligned}\mathcal{V}^{Q_{12}z_{3},c} & = \left[g_{0}\right]^{Q_{12}z_{3}}+\left[g_{0}\right]^{Q_{12}Q_{45}}\Sigma^{Q_{45}z_{6}}\mathcal{V}^{z_{6}z_{3},c}\\
&\quad +\left[g_{0}\right]^{Q_{12}Q_{45}}\Sigma^{Q_{45}Q_{67}}\mathcal{V}^{Q_{67}z_{3},c}.\label{eq:VzQ - 1}
\end{aligned}
\end{equation}

\noindent In order to make progress we treat the disorder strength perturbatively. Our results here are based on keeping terms to $\mathcal{O}\left(\widetilde{\Delta}_{\mathcal{\epsilon}}^{2} \right)$. Since $\left[g_{0}\right]^{Q_{12}Q_{45}}$ is of order $\widetilde{\Delta}_{\mathcal{\epsilon}}^{2}$, it forces the last term in Eq.~\eqref{eq:VzQ - 1}
to be of order $\widetilde{\Delta}_{\mathcal{\epsilon}}^{4}$. Therefore we approximate Eq.~\eqref{eq:VzQ - 1} by
\begin{equation}
\mathcal{V}^{Q_{12}z_{3},c} =\left[g_{0}\right]^{Q_{12}Q_{45}}\Sigma^{Q_{45}z_{6}}\mathcal{V}^{z_{6}z_{3},c}+\mathcal{O}\left(\widetilde{\Delta}_{\mathcal{\epsilon}}^{4} \right),
\label{eq:VzQ - 2}
\end{equation}
where we also used the fact that $\left[g_{0}\right]^{Q_{12}z_{3}}=0$. Now, using Eqs.~\eqref{eq: g_0^Q^Q inv - 1},~\eqref{eq:g^psi^psi - 1},~\eqref{eq:Sigma^psi^psi - 1},~\eqref{eq:Sigma^psi^Q - 1},~\eqref{eq:V^z - 1},~\eqref{eq:V^z^z - 1},~\eqref{eq:V^Q - 4} and ~\eqref{eq:VzQ - 2} we can rewrite Eq.~\eqref{eq:Mean-field EOM-2} as follows
\begin{equation}
    \begin{aligned}
    0 & =\left(2J_{\vec{r}_{1}\vec{r}_{2}\tau_{1}\tau_{2}}^{a_{1}a_{2}}+\left[\left(\mathcal{G}^{c}\right)^{-1}\right]_{\vec{r}_{1}\vec{r}_{2},\tau_{1}\tau_{2}}^{a_{1}a_{2}}\right)\check{\phi}_{\vec{r}_{2}\tau_{2}}^{\overline{a_{2}}}\\
     & \quad+\frac{1}{3!}u_{\tau_{1}\tau_{2}\tau_{3}\tau_{4}}^{a_{1}a_{2}a_{3}a_{4}}\check{\phi}_{\vec{r}_{1}\tau_{2}}^{\overline{a_{2}}}\check{\phi}_{\vec{r}_{1}\tau_{3}}^{\overline{a_{3}}}\check{\phi}_{\vec{r}_{1}\tau_{4}}^{\overline{a_{4}}}\\
     & \quad+\frac{1}{2!}u_{\tau_{1}\tau_{2}\tau_{3}\tau_{4}}^{a_{1}a_{2}a_{3}a_{4}}\check{\phi}_{\vec{r}_{1}\tau_{2}}^{\overline{a_{2}}}\left(i\check{G}_{\vec{r}_{1}\vec{r}_{1},\tau_{3}\tau_{4}}^{\overline{a_{3}a_{4}},c}-i\mathcal{G}_{\vec{r}_{1}\vec{r}_{1},\tau_{3}\tau_{4}}^{\overline{a_{3}a_{4}},c}\right)\\
     & \quad-\frac{\widetilde{\Delta}_{\mathcal{\epsilon}}^{2}}{9}u_{\tau_{4}\tau_{1}\left\llbracket \tau_{2}\tau_{3}\right\rrbracket }^{a_{4}a_{1}\left\llbracket a_{2}a_{3}\right\rrbracket }u_{ \left\llbracket \tau_{2}\tau_{3}\right\rrbracket \tau_{5}\tau_{6}}^{\overline{\left\llbracket a_{2}a_{3}\right\rrbracket }a_{5}a_{6}}\check{\phi}_{\vec{r}_{1}\tau_{6}}^{\overline{a_{6}}}\check{G}_{\vec{r}_{1}\vec{r}_{1},\tau_{5}\tau_{4}}^{\overline{a_{5}a_{4}},c}\\
     & \quad-\frac{\widetilde{\Delta}_{\mathcal{\epsilon}}^{2}}{6}u_{\tau_{1}\tau_{2}\left\llbracket \tau_{3}\tau_{4}\right\rrbracket }^{a_{1}a_{2}\left\llbracket a_{3}a_{4}\right\rrbracket }\check{\phi}_{\vec{r}_{1}\tau_{2}}^{\overline{a_{2}}}\check{G}_{\vec{r}_{1}\vec{r}_{1},\tau_{3}\tau_{4}}^{\overline{a_{3}a_{4}},c}.
    \end{aligned}
    \label{eq:Mean-field EOM-3}
\end{equation}
\noindent We will also be interested in calculating particle number to obtain the Mott insulator phase boundary,
which can be calculated from $G$. To calculate $G$, we solve Eq.~\eqref{eq:Dyson's equation - 4}, however, the form shown here is still not particularly amenable to
solution. We now discuss simplifications that allow us to obtain more tractable equations of motion.

\subsection{Low-frequency approximation\label{subsec:low-frequency approximation}}

Equation~\eqref{eq:Dyson's equation - 4}, whilst having a compact form
in our notation, contains as many as four time-integrals, making it
computationally expensive to solve the equations numerically. This
suggests that some level of approximation beyond simply
truncating the self-energy is required in order to obtain physical insight from
the equations above. Following Refs.~\citep{Kennett2011,Fitzpatrick2018a},
we focus on the low-frequency components of the equations of motion,
specifically the self-energy terms $\Sigma_{2}^{zz}$ and $\Sigma_{3}^{zz}$. 

$\Sigma_{1}^{zz}+\Sigma_{2}^{zz}$ is almost identical in form to the self-energy
obtained in Ref.~\citep{Fitzpatrick2018b}, the only difference being
that the $u$-vertices have a trivial spatial dependency. Therefore
the low-frequency calculation of $\Sigma_{2}^{zz}$ is almost identical
to that in Ref.~\citep{Fitzpatrick2018b} and can be written as

\begin{align}
\left[\Sigma_{2}^{zz}\right]_{\vec{r},\tau_{1}\tau_{2}}^{a_{1}a_{2}} & \simeq 2\delta\left(\tau_{1},\tau_{2}\right)\sigma_{1}^{a_{1}a_{2}}u_{1}\nonumber \\
 & \quad\times\left\{ \check{n}_{\vec{r}}\left(\tau_{1}\right)-\check{n}_{\vec{r}}\left(\tau=0\right)\right\} ,\label{eq:Sigma_2^z^z - 2}
\end{align}

\noindent where $\check{n}_{\vec{r}}\left(\tau\right)$ is the disorder-averaged particle number at site $\vec{r}$ and contour time $\tau$, and $u_{1}$ comes from
taking the low-frequency approximation of $u_{\tau_{1}\tau_{2}\tau_{3}\tau_{4}}^{a_{1}a_{2}a_{3}a_{4}}$,
the expression for which is given in Appendix \ref{sec:Parameters in the self energy}. Provided $\mu/U$ is not close to an integer, then $u_1/U\ll 1 $ ~\citep{Fitzpatrick2018a, Kennett2011}; and we also focus on the small $\widetilde{\Delta}_{\mathcal{\epsilon}}^{2}$ limit, we keep terms of either order $u_1$ or $\widetilde{\Delta}_{\mathcal{\epsilon}}^{2}$ but not terms of order $u_1 \widetilde{\Delta}_{\mathcal{\epsilon}}^{2}$ or higher.

\noindent Now, to calculate the low-frequency approximation to $\left[\Sigma_{3}^{zz}\right]_{\vec{r},\tau_{1}\tau_{2}}^{a_{1}a_{2}}$
it is helpful to rewrite $u_{\tau_{1}\tau_{2}\left\llbracket \tau_{3}\tau_{4}\right\rrbracket }^{a_{1}a_{2}\left\llbracket a_{3}a_{4}\right\rrbracket }$
as
\begin{align}
 u_{\tau_{1}\tau_{2}\left\llbracket \tau_{3}\tau_{4}\right\rrbracket }^{a_{1}a_{2}\left\llbracket a_{3}a_{4}\right\rrbracket }& =-i\frac{3}{2}\prod_{m=3}^{4}\left\{ \mathcal{G}_{\tau_{m}\tau_{m}^{\prime}}^{a_{m}a_{m}^{\prime},c}\right\} u_{\tau_{1}\tau_{2}\tau_{3}^{\prime}\tau_{4}^{\prime}}^{a_{1}a_{2}\overline{a_{3}^{\prime}}\overline{a_{4}^{\prime}}}\nonumber \\
 & \quad\quad+\frac{3}{2}\left\{ \delta_{\tau_{1}\tau_{3}}\sigma_{1}^{a_{1}a_{3}}\right\} \left\{ \delta_{\tau_{2}\tau_{4}}\sigma_{1}^{a_{2}a_{4}}\right\} \nonumber \\
 & \quad\quad+\frac{3}{2}\left\{ \delta_{\tau_{1}\tau_{4}}\sigma_{1}^{a_{1}a_{4}}\right\} \left\{ \delta_{\tau_{2}\tau_{3}}\sigma_{1}^{a_{2}a_{3}}\right\} .\label{eq:rewrite u^z^z^Q - 1}
\end{align}
\noindent By casting $u_{\tau_{1}\tau_{2}\left\llbracket \tau_{3}\tau_{4}\right\rrbracket }^{a_{1}a_{2}\left\llbracket a_{3}a_{4}\right\rrbracket }$
in this form, one can then perform a similar calculation to that for
$\Sigma_{2}^{zz}$ to obtain

\begin{align}
& \left[\Sigma_{3}^{zz}\right]_{\vec{r}}^{a_{1}a_{2},\left(R,A,K\right)} \left(\tau_{1},\tau_{2}\right)\nonumber\\   &\quad\quad\quad\simeq-\frac{i}{2}\widetilde{\Delta}_{\mathcal{\epsilon}}^{2} \left\{i\check{G}_{\vec{r}}^{a_{1}a_{2},\left(R,A,K\right)}\left(\tau_{1},\tau_{2}\right) \right\}.
\end{align}

After applying the low-frequency approximation, we obtain
a self-energy that is identical in Keldysh structure to the
low-frequency self-energy obtained in Ref.~\citep{Fitzpatrick2018b} {(}for a discussion
about Keldysh structure, see Ref.~\citep{Fitzpatrick2018a}{)}. Therefore the equations of motion of the full propagator for the disordered system are
identical in structure to those obtained in Ref.~\citep{Fitzpatrick2018b}.
With this in mind, it is straightforward to show that the equations
of motion can be written as follows:
\begin{widetext}

\begin{equation}
\check{G}_{\vec{r}\vec{r}^{\prime}}^{\left(R,A\right)}\left(t_1,t_2\right)=\mathcal{G}_{\vec{r}}^{\left(R,A\right)}\left(t_1,t_2\right)+\sum_{\vec{r}}\int_{0}^{\infty}\int_{0}^{\infty}dt_3 dt_4\mathcal{G}_{\vec{r}}^{\left(R,A\right)}\left(t_1,t_3\right)\left[\Sigma^{zz}\right]_{\vec{r}\vec{r}^{\prime \prime}}^{ 
\left(R,A\right)}\left(t_3,t_4\right) \check{G}_{\vec{r}^{\prime \prime} \vec{r}^{\prime}}^{\left(R,A\right)}\left(t_4,t_2\right),\label{eq:A equation of motion - 1}
\end{equation}

\begin{align}
\check{G}_{\vec{r}\vec{r}^{\prime}}^{\left(K\right)}\left(t_1,t_2\right)
&=\mathcal{G}_{\vec{r}}^{\left(K\right)}\left(t_1,t_2\right)+\sum_{\vec{r}}\int_{0}^{\infty}\int_{0}^{\infty}dt_3 dt_4\mathcal{G}_{\vec{r}}^{\left(R\right)}\left(t_1,t_3\right)\left[\Sigma^{zz}\right]_{\vec{r}\vec{r}^{\prime \prime}}^{\left(R\right)}\left(t_3,t_4\right) \check{G}_{\vec{r}^{\prime \prime} \vec{r}^{\prime}}^{\left(K\right)}\left(t_4,t_2\right)\nonumber\\
 & \quad+\sum_{\vec{r}}\int_{0}^{\infty}\int_{0}^{\infty}dt_3 dt_4\mathcal{G}_{\vec{r}}^{\left(K\right)}\left(t_1,t_3\right)\left[\Sigma^{zz}\right]_{\vec{r}\vec{r}^{\prime \prime}}^{\left(A\right)}\left(t_3,t_4\right) \check{G}_{\vec{r}^{\prime \prime} \vec{r}^{\prime}}^{\left(A\right)}\left(t_4,t_2\right),
\label{eq:G^(K) equation of motion - 1}
\end{align}
\end{widetext}
\noindent where $\check{G}_{\vec{r}\vec{r}^{\prime}}^{\left(R\right)}\left(t,t^{\prime}\right)$ and $\check{G}_{\vec{r}\vec{r}^{\prime}}^{\left(A\right)}\left(t,t^{\prime}\right)$ are the disorder averaged full retarded and advanced Green's functions respectively obtained from performing a disorder average (as in Eq.~\eqref{eq:avg COGF defined - 1}) of the quantities
\begin{align}
G_{\vec{r}\vec{r}^{\prime}}^{\left(R\right)}\left(t,t^{\prime};\epsilon\right)&= -i \Theta\left( t-t^{\prime}\right)\nonumber\\
&\left\langle a_{\vec{r}}\left(t;\epsilon\right)a_{\vec{r}^{\prime}}^{\dagger}\left(t^{\prime};\epsilon\right)-a_{\vec{r}^{\prime}}^{\dagger}\left(t^{\prime};\epsilon\right)a_{\vec{r}}\left(t;\epsilon\right)\right\rangle_{S} ,\label{eq:retarded func def - 1}\\
G_{\vec{r}\vec{r}^{\prime}}^{\left(A\right)}\left(t,t^{\prime};\epsilon\right)&= i \Theta\left(t^{\prime}-t\right)\nonumber\\
&\left\langle a_{\vec{r}}\left(t;\epsilon\right)a_{\vec{r}^{\prime}}^{\dagger}\left(t^{\prime};\epsilon\right)-a_{\vec{r}^{\prime}}^{\dagger}\left(t^{\prime};\epsilon\right)a_{\vec{r}}\left(t;\epsilon\right)\right\rangle_{S} ,\label{eq:advanced func def - 1}
\end{align}

\noindent and $\check{G}_{\vec{r}}^{\left(K\right)}\left(t,t^{\prime}\right)$ is the disorder averaged full kinetic Green's function obtained from disorder averaging
\begin{equation}
G_{\vec{r}\vec{r}^{\prime}}^{\left(K\right)}\left(t,t^{\prime};\epsilon\right)=-i\left\langle a_{\vec{r}}\left(t;\epsilon\right)a_{\vec{r}^{\prime}}^{\dagger}\left(t^{\prime};\epsilon\right)+a_{\vec{r}^{\prime}}^{\dagger}\left(t^{\prime};\epsilon\right)a_{\vec{r}}\left(t;\epsilon\right)\right\rangle_{S}.
\label{eq:kinetic green's func def - 1}
\end{equation}
\noindent The quantities $\mathcal{G}_{\vec{r}}^{\left(R\right)}\left(t-t^{\prime}\right)$, $\mathcal{G}_{\vec{r}}^{\left(A\right)}\left(t-t^{\prime}\right)$, and  $\mathcal{G}_{\vec{r}}^{\left(K\right)}\left(t-t^{\prime}\right)$
that enter Eqs.~\eqref{eq:A equation of motion - 1} and \eqref{eq:G^(K) equation of motion - 1}
are the $W_{0}$-generated retarded, advanced and kinetic Green's
functions, respectively, which are all time-translational invariant
(expressions for each are presented in Appendix~\ref{sec:Propagator in the zero disorder and hopping limit}). It is important to note that $\check{n}_{\vec{r}}\left(t\right)$
can be obtained from $\check{G}_{\vec{r}\vec{r}^{\prime}}^{\left(K\right)}\left(t,t^{\prime}\right)$
as follows:

\begin{equation}
\check{n}_{\vec{r}}\left(t\right)=\frac{1}{2}\left\{ i\check{G}_{\vec{r}\vec{r}}^{\left(K\right)}\left(t,t\right)-1\right\} ,\label{eq:n_r from G^(K) - 1}
\end{equation}

\noindent which introduces an element of nonlinearity into the equations
of motion above. 

Finally, in the low-frequency limit, we can approximate Eq.~\eqref{eq:Mean-field EOM-3} as
\begin{equation}
\begin{aligned}
0 & =2 \sum_{\vec{r}'} J_{\vec{r}\vec{r}'}\left(t_1\right)\check{\phi}_{\vec{r}'}\left(t_1\right)+\left[\left(\mathcal{G}^{c}\right)^{-1}\right]_{\omega\rightarrow0}^{\left(R\right)}\check{\phi}_{\vec{r}}\left(t_1\right)\\
 & \quad-2u_{1}\check{\phi}_{\vec{r}}\left(t_1\right)\left[\check{n}_{\vec{r}}\left(t_1\right)-\check{n}_{\vec{r}}\left(t'_{1}=0\right)\right] \\
 & \quad-u_{1}\check{\phi}_{\vec{r}}\left(t_1\right)\left|\check{\phi}_{\vec{r}}\left(t_1\right)\right|^{2}-\widetilde{\Delta}_{\mathcal{\epsilon}}^{2}\check{\phi}_{\vec{r}}\left(t_1\right)\check{G}_{\vec{r}\vec{r}}^{12,\left(R\right)}\left(t_1 , t_2\right),
 \label{eq:Mean-field EOM-4}
\end{aligned}
\end{equation}
\noindent where $\left[\left(\mathcal{G}^{c}\right)^{-1}\right]_{\omega\to0}^{\left(R\right)}$
is the low-frequency approximation of the inverse retarded Green's function
obtained from $W_{0}$, the expression for which is given in Appendix \ref{sec:Parameters in the self energy}. 
\subsection{Equilibrium solution\label{subsec:Equilibrium solution}}
In studying the equilibrium solution to the equations of motion derived in Sec.~\ref{subsec:low-frequency approximation} we consider the system to be at zero temperature. We also work in $\vec{k}$-space rather than real space. Whilst disordered systems are not homogeneous due to the random potential in the Hamiltonian, the disorder-averaged COGFs respect translation invariance. Therefore we follow the same procedure as  Ref.~\citep{Fitzpatrick2018a} in order to obtain the Mott insulator phase boundary in the presence of disorder. The only difference being that now the COGFs are replaced by disorder-averaged COGFs. In $\vec{k}$-space Eq.~\eqref{eq:Dyson's equation - 4} becomes
\begin{align}
     \check{G}_{\vec{k}}^{a_{1}a_{2},\left(R\right)}\left(\omega\right) & =\mathcal{G}^{a_{1}a_{2},\left(R\right)}\left(\omega\right)\nonumber\\
     & \quad+\mathcal{G}^{a_{1}a_{3},\left(R\right)}\left(\omega\right)\left[\Sigma^{zz}\right]_{\vec{k}}^{\overline{a_{3}a_{4}},\left(R\right)}\left(\omega\right)\check{G}_{\vec{k}}^{a_{4}a_{2},\left(R\right)}\left(\omega\right),
     \label{eq:Dyson's equation - k-space}
\end{align}
\noindent where $\left[\Sigma^{zz}\right]_{\vec{k}}^{\overline{a_{3}a_{4}},\left(R\right)}\left(\omega\right)$ is the Fourier transform of $\left[\Sigma^{zz}\right]_{\vec{r}_{1}\vec{r}_{2},\tau_{1}\tau_{2}}^{a_{1}a_{2}}$. Now we can write
\begin{align}
    \left[\Sigma^{zz}\right]_{\vec{k}}^{12,\left(R\right)}\left(\omega\right) & =-2J\sum_{i=1}^{d}\cos(k_{i}a)+\frac{1}{2}\widetilde{\Delta}_{\mathcal{\epsilon}}^{2}\check{G}_{\vec{k}}^{12,\left(R\right)}\left(\omega\right)\nonumber\\
     & \quad+2 u_{1}\left[\left|\check{\phi}\right|^{2}+\left(\check{n}-\check{n}_{0}\right)\right]+\mathcal{O}\left(u_1\widetilde{\Delta}_{\mathcal{\epsilon}}^{2} \right),
\label{eq:Sigma-12 k-space-1}
\end{align}
\begin{align}
    \left[\Sigma^{zz}\right]_{\vec{k}}^{11,\left(R\right)}\left(\omega\right)	&=\frac{1}{2}u_{1}\left[2\left(\check{\phi}^{1}\right)^{2}+i\check{G}_{\vec{r'}=0}^{11,\left(K\right)}\left(s=0\right)\right]\nonumber\\
	&\quad+\frac{1}{2}\widetilde{\Delta}_{\mathcal{\epsilon}}^{2}\check{G}_{\vec{k}}^{11,\left(R\right)}\left(\omega\right),
	\label{eq:Sigma-11 k-space-1}
\end{align}

\begin{align}
    \left[\Sigma^{zz}\right]_{\vec{k}}^{22,\left(R\right)}\left(\omega\right)	&=\frac{1}{2}u_{1}\left[2 \left(\check{\phi}^{2}\right)^{2}+i\check{G}_{\vec{r'}=0}^{22,\left(K\right)}\left(s=0\right)\right]\nonumber\\
	&\quad+\frac{1}{2}\widetilde{\Delta}_{\mathcal{\epsilon}}^{2}\check{G}_{\vec{k}}^{22,\left(R\right)}\left(\omega\right),
	\label{eq:Sigma-22 k-space-1}
\end{align}
\noindent where $\check{n}$ is the average local particle number for $J \neq 0$
\begin{equation}
    \check{n}=\left\langle \check{n}_{\vec{k}}\right\rangle =\frac{1}{N_{\text{site}}}\sum_{\vec{k}}\check{n}_{\vec{k}}.
\end{equation}
\noindent Now, from Eq.~\eqref{eq:Dyson's equation - k-space} we can write
\begin{widetext}
\begin{align}
    \check{G}_{\vec{k}}^{12,\left(R\right)}\left(\omega\right)&=\frac{\left[\left\{ \mathcal{G}^{21,\left(R\right)}\left(\omega\right)\right\} ^{-1}-\left[\Sigma^{zz}\right]_{\vec{k}}^{21,\left(R\right)}\left(\omega\right)\right]}{\left[\left\{ \mathcal{G}^{21,\left(R\right)}\left(\omega\right)\right\} ^{-1}-\left[\Sigma^{zz}\right]_{\vec{k}}^{21,\left(R\right)}\left(\omega\right)\right]\left[\left\{ \mathcal{G}^{12,\left(R\right)}\left(\omega\right)\right\} ^{-1}-\left[\Sigma^{zz}\right]_{\vec{k}}^{12,\left(R\right)}\left(\omega\right)\right]-\left|\left[\Sigma^{zz}\right]_{\vec{k}}^{22,\left(R\right)}\left(\omega\right)\right|^{2}},\label{eq:G12R - 1}\\
    \check{G}_{\vec{k}}^{22,\left(R\right)}\left(\omega\right)&=\frac{\left[\Sigma^{zz}\right]_{\vec{k}}^{22,\left(R\right)}\left(\omega\right)}{\left[\left\{ \mathcal{G}^{21,\left(R\right)}\left(\omega\right)\right\} ^{-1}-\left[\Sigma^{zz}\right]_{\vec{k}}^{21,\left(R\right)}\left(\omega\right)\right]\left[\left\{ \mathcal{G}^{12,\left(R\right)}\left(\omega\right)\right\} ^{-1}-\left[\Sigma^{zz}\right]_{\vec{k}}^{12,\left(R\right)}\left(\omega\right)\right]-\left|\left[\Sigma^{zz}\right]_{\vec{k}}^{22,\left(R\right)}\left(\omega\right)\right|^{2}}.\label{eq:G22R - 1}
\end{align}
\end{widetext}
In the following, we will discuss equilibrium solutions for the Mott insulator and superfluid phases. For simplicity we only consider the static limit, i.e., $\omega = 0$.  As we demonstrate later, this assumption is acceptable at least in equilibrium. 
\subsubsection{Mott Insulator}\label{subsec: Mott-Insulator}
In the Mott insulator phase $\check{\phi}$ and $\left[\Sigma^{zz}\right]_{\vec{k}}^{22,\left(R\right)}\left(\omega\right)$ are zero and $ \left[\Sigma^{zz}\right]_{\vec{k}}^{12,\left(R\right)}\left(\omega\right)$ is
\begin{align}
    \left[\Sigma^{zz}\right]_{\vec{k}}^{12,\left(R\right)} &=-2J\sum_{i=1}^{d}\cos(\vec{k}_{i}a)+2u_{1}\left(\check{n}-\check{n}_{0}\right)\nonumber\\
    &\quad+ \frac{1}{2}\widetilde{\Delta}_{\mathcal{\epsilon}}^{2} \check{G}_{\vec{k}}^{12,\left(R\right)}\left(\omega=0\right).
    \label{eq:Sigma-12 k-space-2}
\end{align}
\noindent Therefore Eq.~\eqref{eq:G12R - 1} reduces to
\begin{equation}
    \check{G}_{\vec{k}}^{12,\left(R\right)}\left(\omega\right)=\frac{1}{\left[\left\{ \mathcal{G}^{\left(R\right)}\left(\omega\right)\right\} ^{-1}-\left[\Sigma^{zz}\right]_{\vec{k}}^{12,\left(R\right)}\right]},
    \label{eq: G12R - 2}
\end{equation}
which we can rewrite as~\citep{Fitzpatrick2018a}
\begin{align}
    \check{G}_{\vec{k}}^{12,\left(R\right)}\left(\omega\right)	&=\tilde{z}_{\mathrm{MI},\vec{k}}^{\left(+\right)}\frac{1}{\left\{ \omega-\Delta\tilde{E}_{\mathrm{MI},\vec{k}}^{\left(+\right)}\right\} +i0^{+}}\nonumber\\
	&\quad-\tilde{z}_{\mathrm{MI},\vec{k}}^{\left(-\right)}\frac{1}{\left\{ \omega+\Delta\tilde{E}_{\mathrm{MI},\vec{k}}^{\left(-\right)}\right\} +i0^{+}},
	\label{eq: G12R - 3}
\end{align}
\noindent where
\begin{equation}
    \Delta\tilde{E}_{\mathrm{MI},\vec{k}}^{\left(\pm\right)}=\frac{\mp B_{\vec{k}}+\sqrt{\left(B_{\vec{k}}\right)^{2}-4C_{\vec{k}}}}{2},
    \label{eq: Delta E +/- - 1}
\end{equation}
\begin{equation}
    B_{\vec{k}}=-\left\{ \Delta\mathcal{E}^{\left(+\right)}-\Delta\mathcal{E}^{\left(-\right)}\right\} -\Sigma_{\vec{k}}^{12,\left(R\right)},
    \label{eq: B vector - 1}
\end{equation}
\begin{equation}
    C_{\vec{k}}=-\left(\mu+U\right)\left[\Sigma_{\vec{k}}^{12,\left(R\right)}-\left\{ \mathcal{G}^{\left(R\right)}\left(\omega'=0\right)\right\} ^{-1}\right],
    \label{eq: C vector - 1}
\end{equation}
\begin{equation}
    \tilde{z}_{\mathrm{MI},\vec{k}}^{\left(\pm\right)}=\frac{\left(\mu+U\right)\pm\Delta\tilde{E}_{\mathrm{MI},\vec{k}}^{\left(\pm\right)}}{\Delta\tilde{E}_{\mathrm{MI},\vec{k}}^{\left(+\right)}+\Delta\tilde{E}_{\mathrm{MI},\vec{k}}^{\left(-\right)}},
    \label{eq: Delta Z +/- - 1}
\end{equation}
and $\Delta\mathcal{E}^{\left(\pm\right)}$ are the excitation energies in the atomic limit (i.e $J = 0$)
\begin{align}
    \Delta\mathcal{E}^{\left(+\right)}&=\mathcal{E}_{n_{\mathrm{MI}}+1}-\mathcal{E}_{n_{\mathrm{MI}}},\\
    \Delta\mathcal{E}^{\left(-\right)}&=\mathcal{E}_{n_{\mathrm{MI}}-1}-\mathcal{E}_{n_{\mathrm{MI}}}.
\end{align}
\noindent As explained in Ref.~\citep{Fitzpatrick2018a}, the local particle number can be expressed as
\begin{equation}
    \check{n}_{\vec{k}}=\frac{1}{2}\left\{ \tilde{z}_{\mathrm{MI},\vec{k}}^{\left(+\right)}+\tilde{z}_{\mathrm{MI},\vec{k}}^{\left(-\right)}-1\right\}.
    \label{eq: nk MI - 1}
\end{equation}

\noindent In Fig.~\ref{fig:fig2} we show the excitation energies $\Delta\tilde{E}_{\mathrm{MI},\vec{k}}^{\left(\pm\right)}$, and the spectral weights $\tilde{z}_{\mathrm{MI},\vec{k}}^{\left(\pm\right)}$ for different disorder strengths $\Delta_{\epsilon}$ in the Mott insulator phase where the relation between $\Delta_{\epsilon}$ and $\tilde{\Delta}_{\epsilon}$ is given in Eq.~\eqref{eq:Delta defined - 1} and from now on we drop the $\epsilon$ index in $\Delta_{\epsilon}$ to avoid confusion. In addition, we compare the quasi-momentum distribution $\check{n}_{\vec{k}}$ for two different disorder strengths:  $\Delta/U=0.1$ and $\Delta/U=0.3$. Here we have a $1000 \times 1000$ square lattice, the chemical potential is $\mu/U = 0.42$, the chosen hopping term is $J/U = 0.02$, and $\beta U = \infty$. As can be seen from Fig.~\ref{fig:fig2}, in the Mott phase, the excitation energies decrease with increasing disorder strength. In addition, with increasing disorder strength, the quasi-momentum distribution $\check{n}_{\vec{k}}$ becomes more localized, which implies that for the same hopping term, the system is closer to the transition point.

\begin{widetext}

\begin{figure}
\subfigure{
  \includegraphics[width=0.4\textwidth]{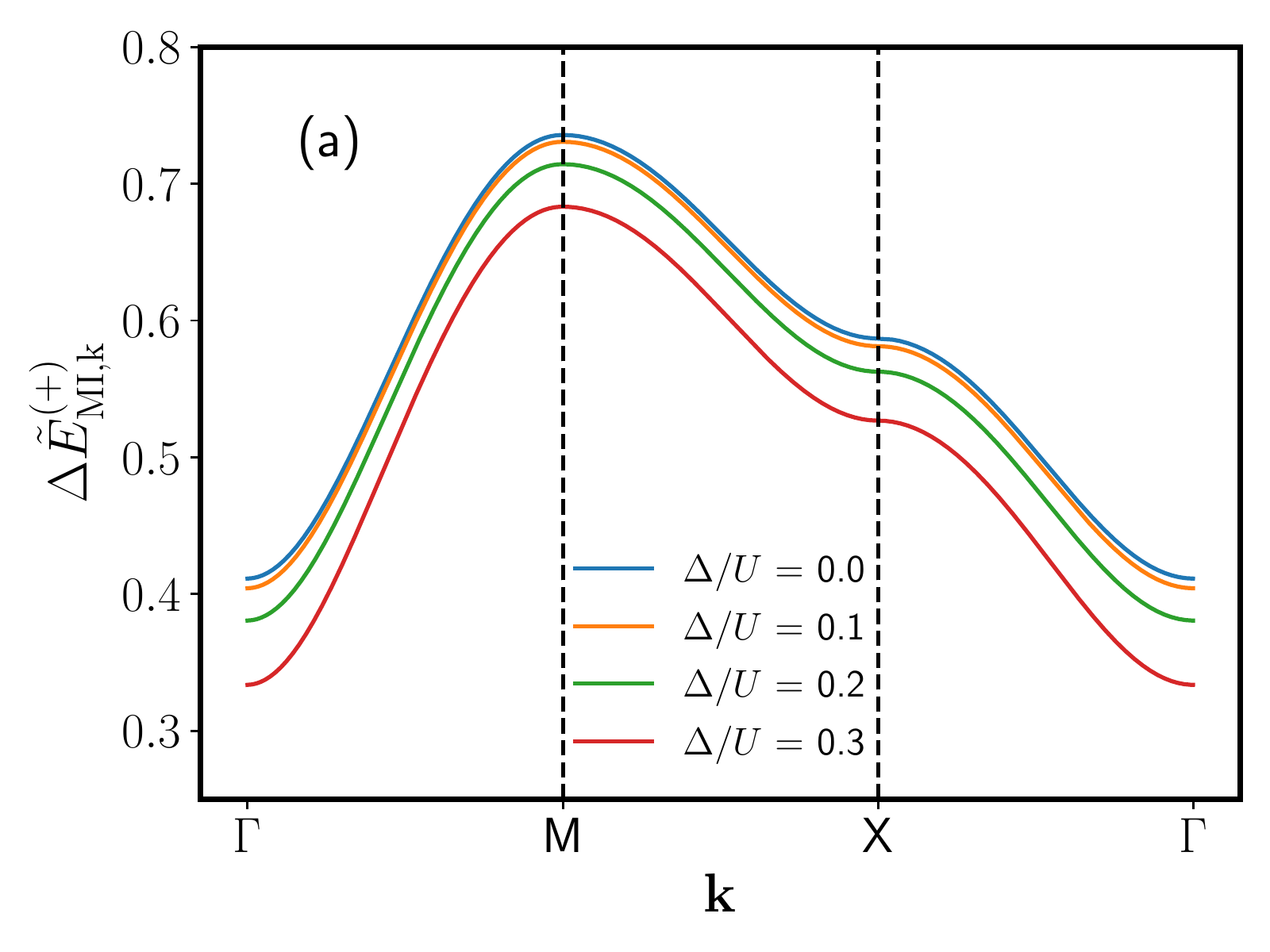}
  }
 \subfigure{
  \includegraphics[width=0.4\textwidth]{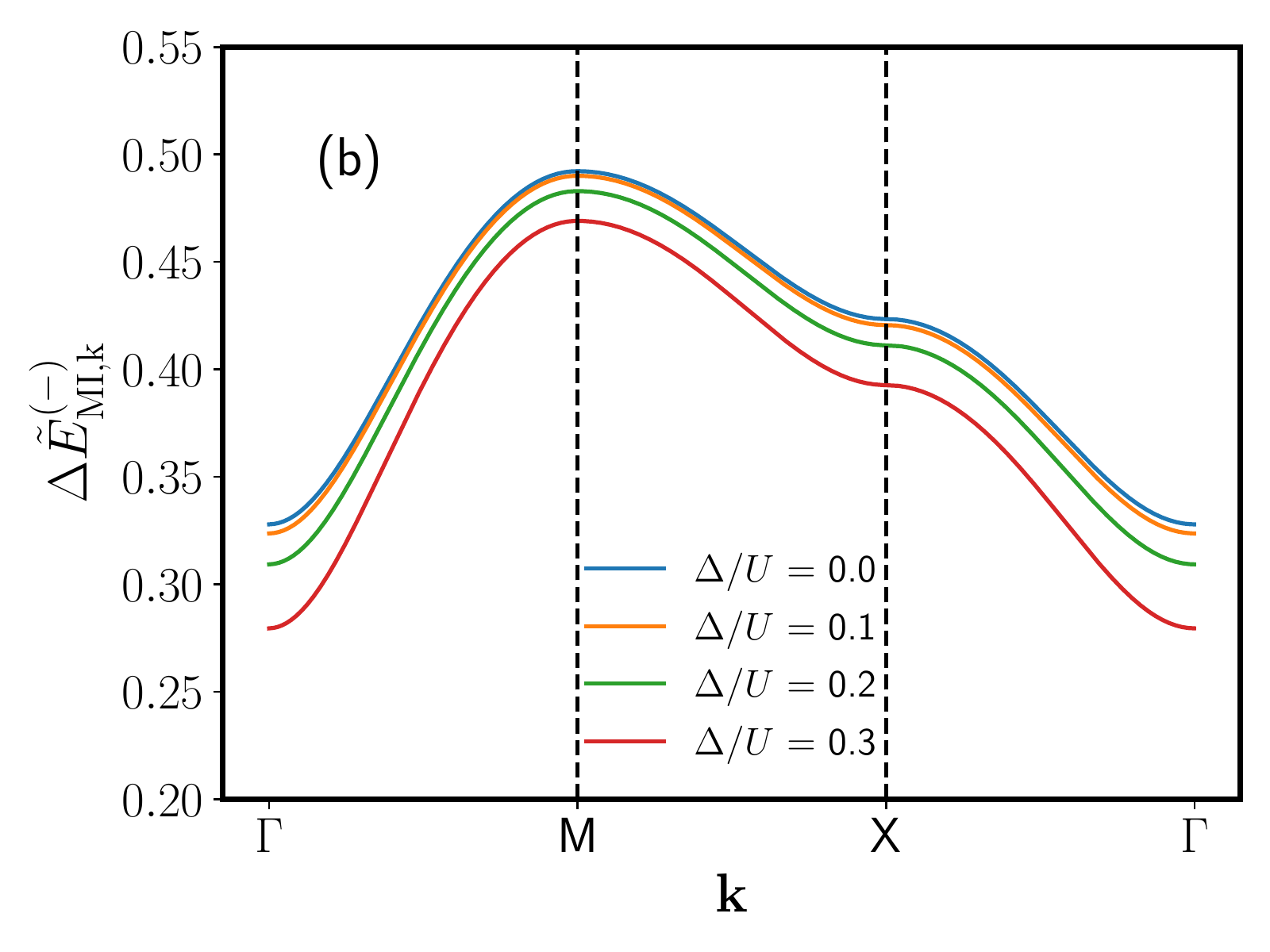}
  }
 \subfigure{
  \includegraphics[width=0.4\textwidth]{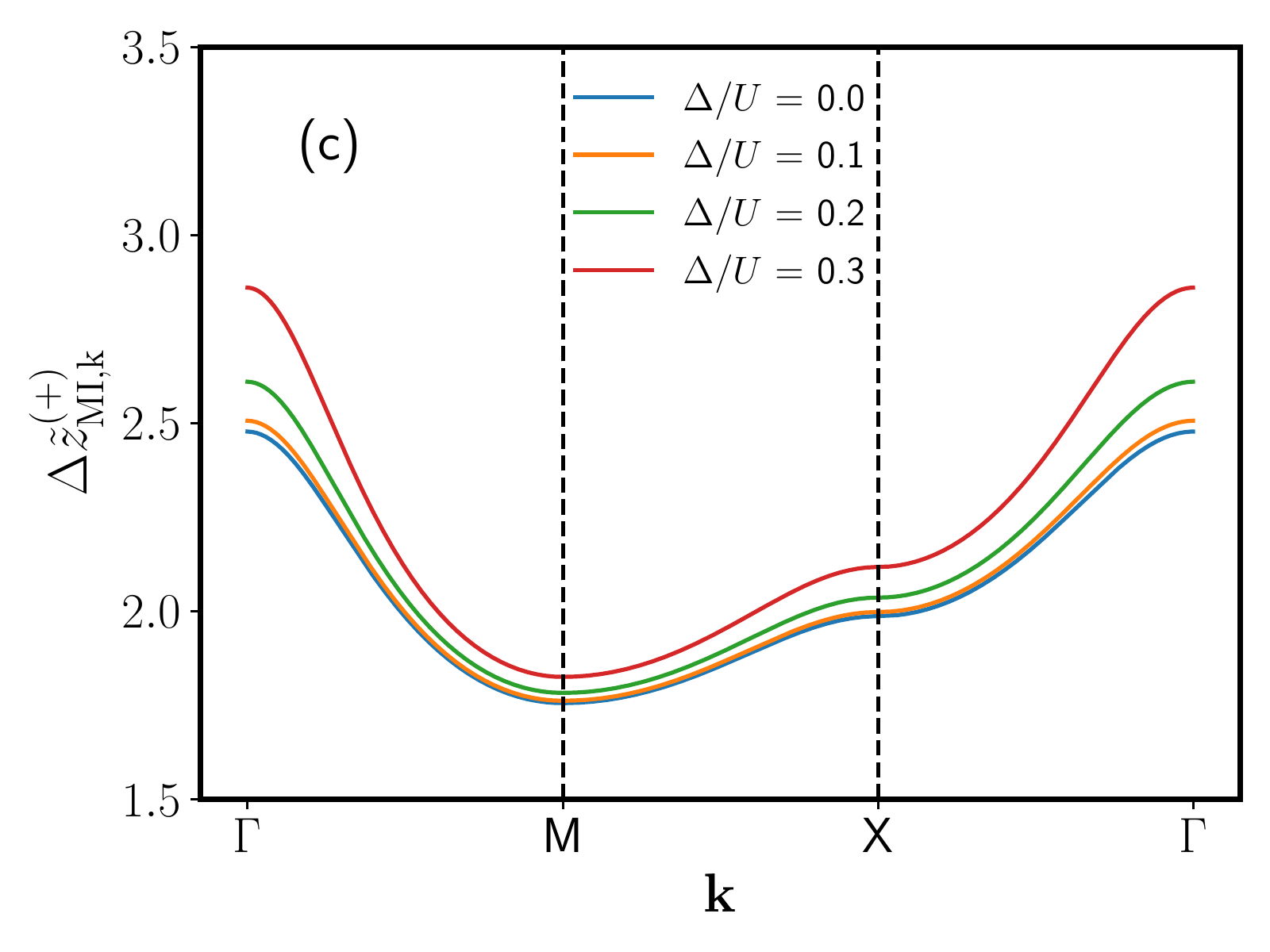}
  }
\subfigure{
  \includegraphics[width=0.4\textwidth]{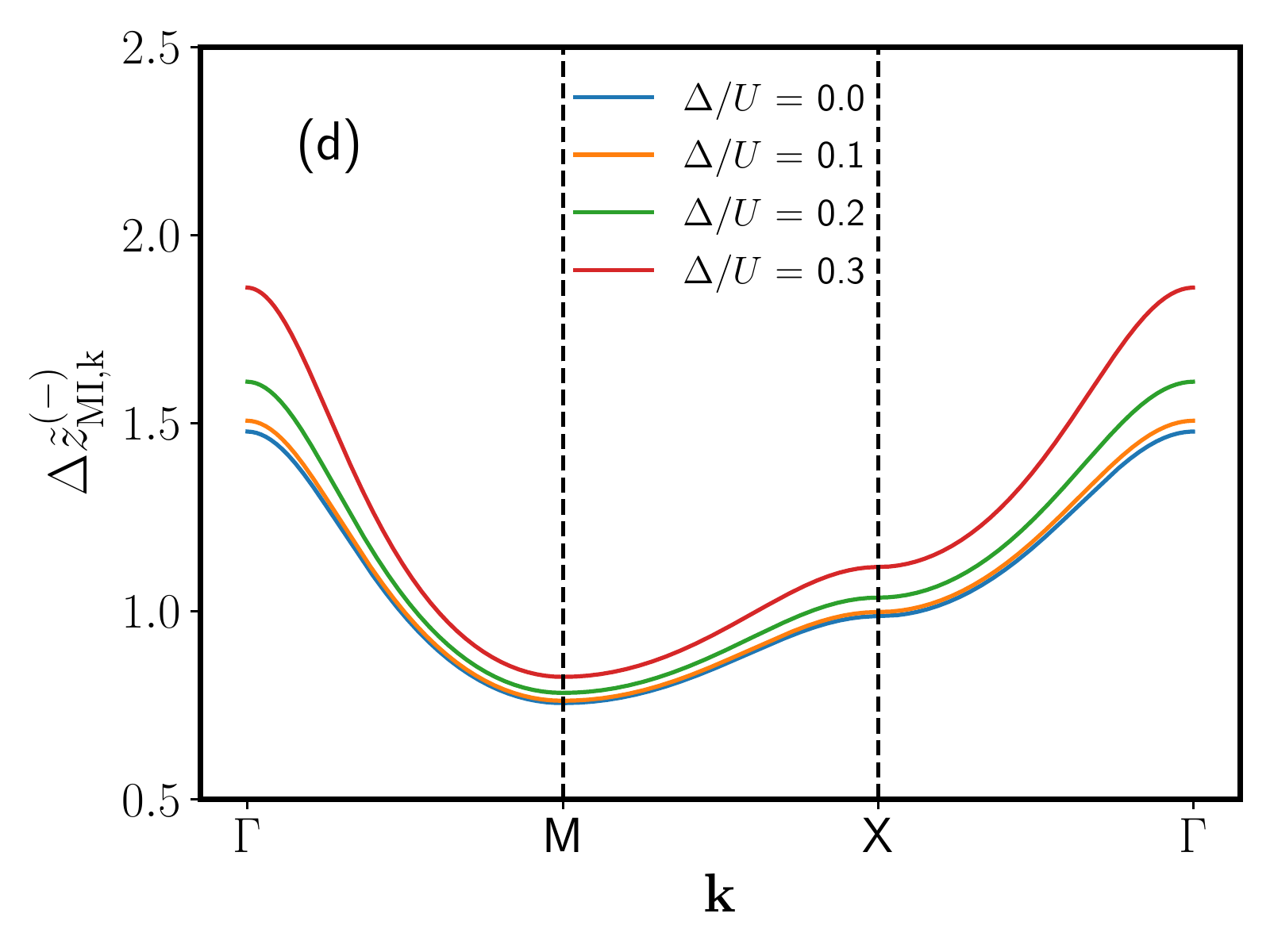}
  }
 \subfigure{
  \includegraphics[width=0.44\textwidth]{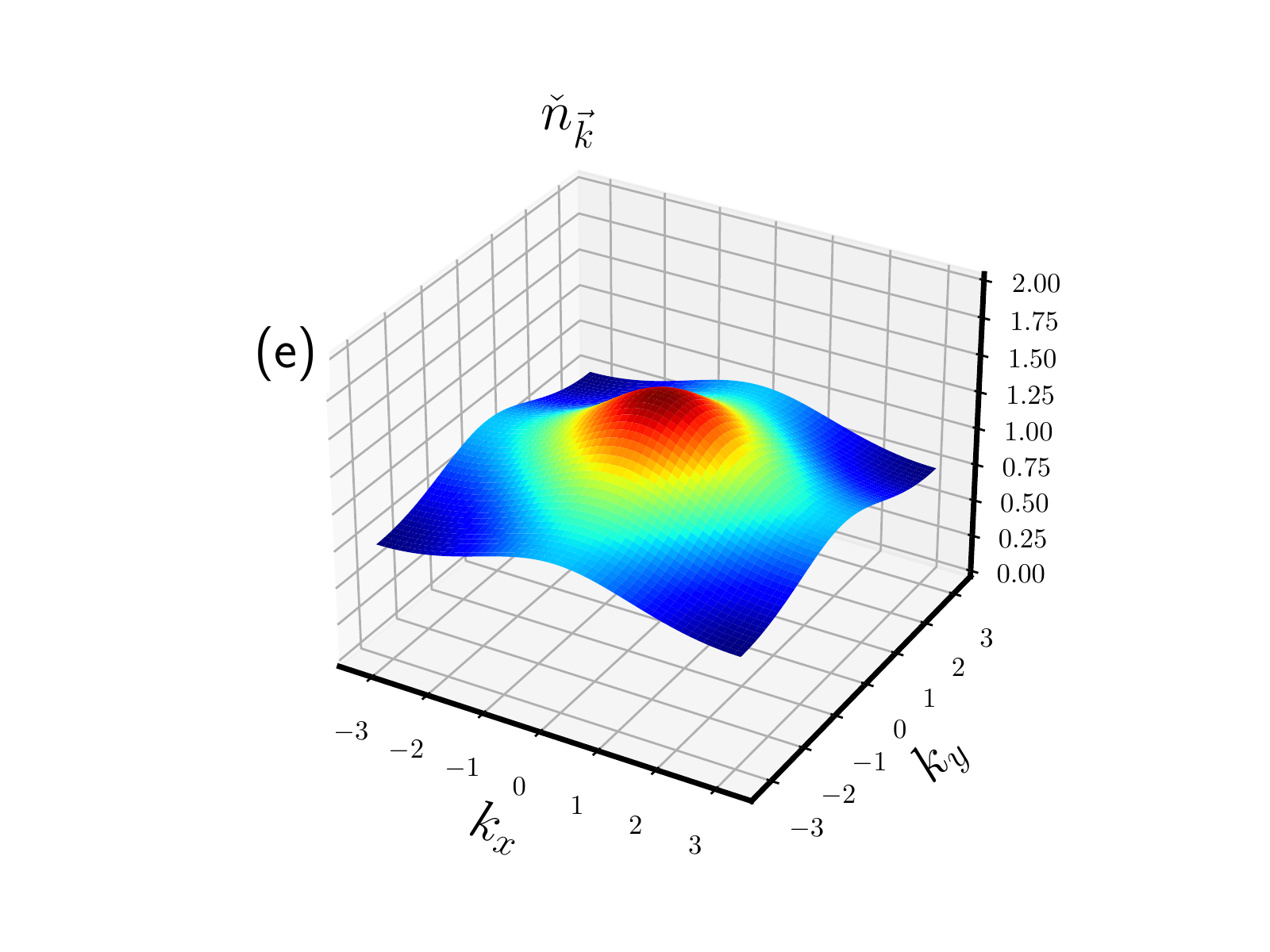}
  }
 \subfigure{
  \includegraphics[width=0.44\textwidth]{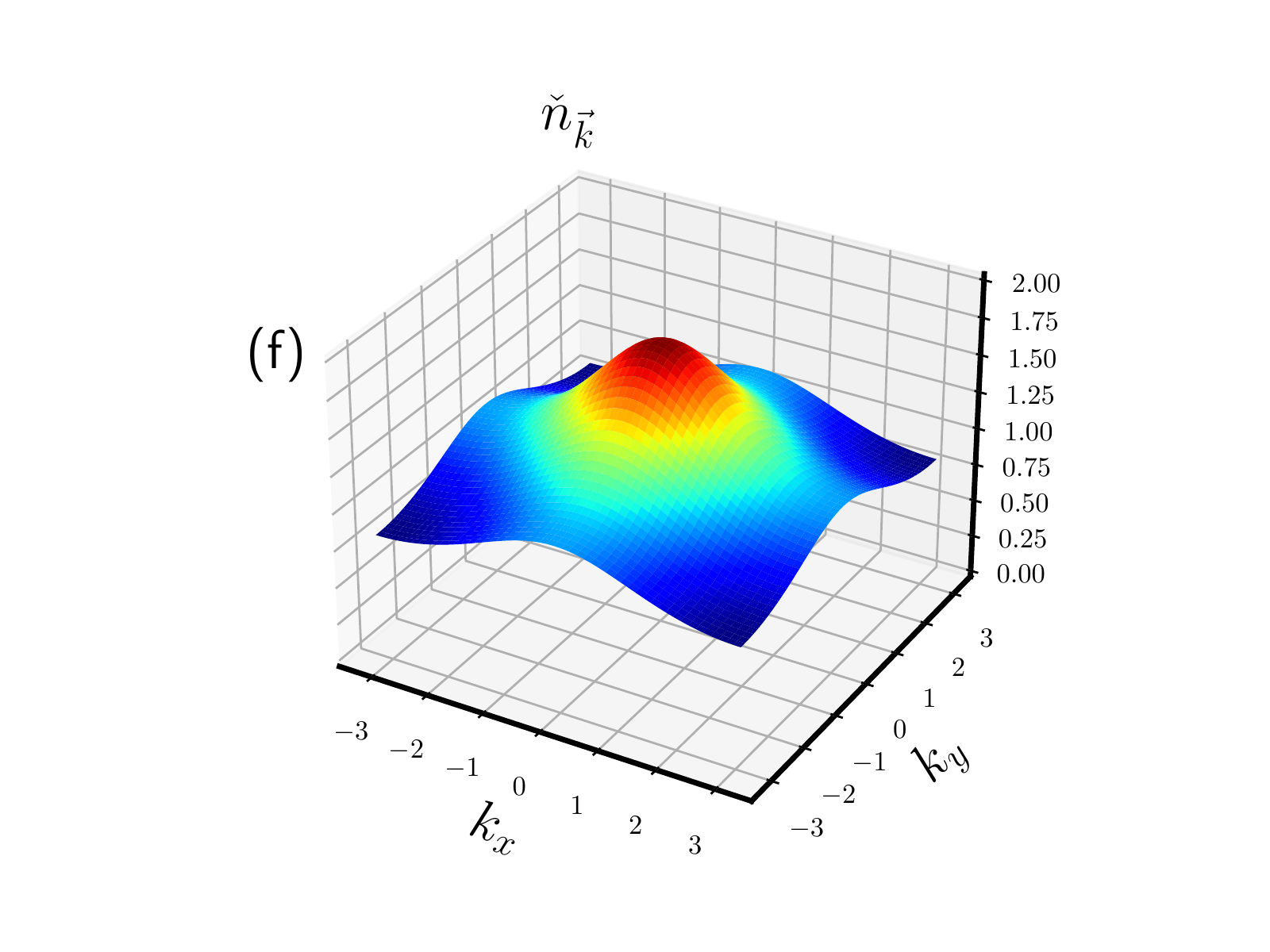}
  }
\caption{Collective excitations of the Mott phase of the 2-dimensional disordered-BHM (a) Quasi-particle excitation energies $\Delta\tilde{E}_{\mathrm{MI},\vec{k}}^{\left(+\right)}$, (b) Quasi-hole excitation energies $\Delta\tilde{E}_{\mathrm{MI},\vec{k}}^{\left(-\right)}$, (c) Quasi-particle spectral weights $\tilde{z}_{\mathrm{MI},\vec{k}}^{\left(+\right)}$, (d) Quasi-hole spectral weights $\tilde{z}_{\mathrm{MI},\vec{k}}^{\left(-\right)}$, for different disorder strengths. Panels (e) and (f) show quasi-momentum distribution $\check{n}_{\vec{k}}$ for $\Delta/U=0.1$, and $\Delta/U=0.3$ respectively. The parameters used were $N_{s}=1000^2$, $\mu/U = 0.42$, $J/U = 0.02$, and $\beta U = \infty$. Note that $\Gamma = \left( 0,0\right)$, $M = \left( \pi,\pi \right)$, and $X = \left( \pi,0 \right)$.}
\label{fig:fig2}
\end{figure}

\subsubsection{Superfluid}\label{subsec:Super-Fluid}
In the superfluid phase, $\check{\phi}$ and $\left[\Sigma^{zz}\right]_{\vec{k}}^{22,\left(R\right)}$
are non-zero, hence we must use the full forms of Eqs.~\eqref{eq:G12R - 1} and~\eqref{eq:G22R - 1}. We begin by calculating $\check{\phi}$ from Eq.~\eqref{eq:Mean-field EOM-4}. Note that in writing Eq.~\eqref{eq:Mean-field EOM-4}, in order to obtain a gapless energy spectrum, we used the HFB-Popov~\citep{Popov} (HFBP) approximation as explained in detail in Ref.~\citep{Fitzpatrick2018a}. Here, we extend the HFB-Popov approximation to the disordered case (see Appendix~\ref{sec:HFBP}). Therefore we have

\begin{equation}
    \check{\phi}=\sqrt{\frac{\left\{ \mathcal{G}^{\left(R\right)}\left(\omega'=0\right)\right\} ^{-1}+2dJ- \widetilde{\Delta}_{\mathcal{\epsilon}}^{2}\left\{ \check{G}^{12,\left(R\right)}_{\vec{k}}\left(\omega=0\right)\right\}}{u_{1}}-2\left(\check{n}-\check{n}_{0}\right)+\frac{ \widetilde{\Delta}_{\mathcal{\epsilon}}^{2}}{4 u_1}\left[ \check{G}_{\vec{k}=0}^{12,\left(R\right)}\left(\omega=0\right)+ \:\check{G}_{\vec{k}=0}^{22,\left(R\right)}\left(\omega=0\right)\right]} \: \:.
\label{eq:Mean-field EOM-5}
\end{equation}

\noindent Now, in the superfluid phase, using Eqs.~\eqref{eq:Sigma-12 k-space-1},~\eqref{eq:Sigma-11 k-space-1},~\eqref{eq:Sigma-22 k-space-1} and~\eqref{eq:Mean-field EOM-5} we have that the self-energy is

\begin{align}
    \left[\Sigma^{zz}\right]_{\vec{k}}^{12,\left(R\right)}\left(\omega\right)  =-2J\sum_{i=1}^{d}\cos(k_{i}a)+2 u_{1}\left[\left|\check{\phi}\right|^{2}+\left(\check{n}-\check{n}_{0}\right)\right]+\frac{1}{2}\widetilde{\Delta}_{\mathcal{\epsilon}}^{2}\check{G}_{\vec{k}}^{12,\left(R\right)}\left(\omega=0\right),
\label{eq:Sigma-12 k-space-3}
\end{align}
\begin{align}
\left[\Sigma^{zz}\right]_{\vec{k}}^{11,\left(R\right)}\left(\omega\right) = u_{1}\check{\phi}^{2}+\frac{1}{2}\widetilde{\Delta}_{\mathcal{\epsilon}}^{2}\check{G}_{\vec{k}}^{11,\left(R\right)}\left(\omega\right) -\frac{\widetilde{\Delta}_{\mathcal{\epsilon}}^{2}}{4}\left( \check{G}_{\vec{k}=0}^{12,\left(R\right)}\left(\omega=0\right)+ \:\check{G}_{\vec{k}=0}^{11,\left(R\right)}\left(\omega=0\right)\right),
	\label{eq:Sigma-11 k-space-3}
\end{align}
\noindent and
\begin{align}
 \left[\Sigma^{zz}\right]_{\vec{k}}^{22,\left(R\right)}\left(\omega\right) = u_{1}\check{\phi}^{2}+\frac{1}{2}\widetilde{\Delta}_{\mathcal{\epsilon}}^{2}\check{G}_{\vec{k}}^{22,\left(R\right)}\left(\omega\right)-\frac{\widetilde{\Delta}_{\mathcal{\epsilon}}^{2}}{4}\left( \check{G}_{\vec{k}=0}^{12,\left(R\right)}\left(\omega=0\right)+ \:\check{G}_{\vec{k}=0}^{22,\left(R\right)}\left(\omega=0\right)\right).
	\label{eq:Sigma-22 k-space-3}
\end{align}

\noindent Next, we calculate $\check{G}^{\left(R \right)}$: starting from Eq.~\eqref{eq:G22R - 1}, one can show that

\begin{equation}
    \check{G}_{\vec{k}}^{12,\left(R\right)}\left(\omega\right)=\frac{\left\{ \omega^{+}+\Delta\tilde{E}_{\mathrm{MI},\vec{k}}^{\left(+\right)}\right\} \left\{ \omega^{+}-\Delta\tilde{E}_{\mathrm{MI},\vec{k}}^{\left(-\right)}\right\} \left\{ \omega^{+}+\left(\mu+U\right)\right\} }{\left\{ \omega^{+}-\Delta\tilde{E}_{\mathrm{SF},\vec{k}}^{\left(1\right)}\right\} \left\{ \omega^{+}+\Delta\tilde{E}_{\mathrm{SF},\vec{k}}^{\left(1\right)}\right\} \left\{ \omega^{+}-\Delta\tilde{E}_{\mathrm{SF},\vec{k}}^{\left(2\right)}\right\} \left\{ \omega^{+}+\Delta\tilde{E}_{\mathrm{SF},\vec{k}}^{\left(2\right)}\right\} },
    \label{eq:G12R - 4}
\end{equation}

\noindent where

\begin{equation}
    \Delta\tilde{E}_{\mathrm{SF},\vec{k}}^{\left(s\right)}=\sqrt{\frac{-\tilde{B}_{\vec{k}}-\left(-1\right)^{s}\sqrt{\left(\tilde{B}_{\vec{k}}\right)^{2}-4\tilde{C}_{\vec{k}}}}{2}},
    \label{eq: Delta E SF - 1}
\end{equation}

\begin{equation}
    \tilde{B}_{\vec{k}}=\left|\left[\Sigma^{zz}\right]_{\vec{k}}^{22,\left(R\right)}\right|^{2}-\left\{ \Delta\tilde{E}_{\mathrm{MI},\vec{k}}^{\left(+\right)}\right\} ^{2}-\left\{ \Delta\tilde{E}_{\mathrm{MI},\vec{k}}^{\left(-\right)}\right\} ^{2},
    \label{eq: B-vector - 2}
\end{equation}
\noindent and
\begin{equation}
    \tilde{C}_{\vec{k}}=\left\{ \Delta\tilde{E}_{\mathrm{MI},\vec{k}}^{\left(+\right)}\Delta\tilde{E}_{\mathrm{MI},\vec{k}}^{\left(-\right)}\right\} ^{2}-\left(\mu+U\right)\left|\left[\Sigma^{zz}\right]_{\vec{k}}^{22,\left(R\right)}\right|^{2}.
    \label{eq: C-vector - 2}
\end{equation}
\noindent It is important to note that the expressions for B and C differ from those for the Mott insulator in that B has units of $\textit{energy}$ for the MI but units of $\left[\textit{energy} \right]^2$ for the SF. Following the same approach as Ref.~\citep{Fitzpatrick2018a}, the  quasi-momentum $\check{n}_{\vec{k}}$ for $\vec{k}\neq0$ is
\begin{equation}
    \check{n}_{\vec{k}} = \begin{cases}
        \frac{1}{2}\left\{ \tilde{z}_{\mathrm{SF},\vec{k}}^{\left(1,+\right)}+\tilde{z}_{\mathrm{SF},\vec{k}}^{\left(1,-\right)}+\tilde{z}_{\mathrm{SF},\vec{k}}^{\left(2,+\right)}+\tilde{z}_{\mathrm{SF},\vec{k}}^{\left(2,-\right)}-1\right\}, & \text{if } \vec{k}\neq0\\[0.3cm]
        \frac{1}{2}\left\{ \tilde{z}_{\mathrm{SF},\vec{k}}^{\left(1,+\right)}+\tilde{z}_{\mathrm{SF},\vec{k}}^{\left(1,-\right)}+2N_{sites}\left|\check{\phi}\right|^{2}-1\right\}, & \text{if } \vec{k}= 0
        \end{cases}
    \: \:,\label{eq: nk-SF}
\end{equation}

\noindent where

\begin{equation}
    \tilde{z}_{\mathrm{SF},\vec{k}}^{\left(s,\pm\right)}=\left(-1\right)^{s+1}\frac{\left\{ \Delta\tilde{E}_{\mathrm{SF},\vec{k}}^{\left(s\right)}\pm\Delta\tilde{E}_{\mathrm{MI},\vec{k}}^{\left(+\right)}\right\} \left\{ \Delta\tilde{E}_{\mathrm{SF},\vec{k}}^{\left(s\right)}\mp\Delta\tilde{E}_{\mathrm{MI},\vec{k}}^{\left(-\right)}\right\} \left\{ \left(\mu+U\right)\pm\Delta\tilde{E}_{\mathrm{SF},\vec{k}}^{\left(s\right)}\right\} }{2\Delta\tilde{E}_{\mathrm{SF},\vec{k}}^{\left(s\right)}\left[\left\{ \Delta\tilde{E}_{\mathrm{SF},\vec{k}}^{\left(1\right)}\right\} ^{2}+\left\{ \Delta\tilde{E}_{\mathrm{SF},\vec{k}}^{\left(2\right)}\right\} ^{2}\right]}.
    \label{eq: spectral function}
\end{equation}

\noindent In Fig.~\ref{fig:fig3} we show the collective mode spectra and quasi-particle spectral weight in the superfluid phase for different disorder strengths as calculated from Eqs.~\eqref{eq: Delta E SF - 1} and~\eqref{eq: spectral function}. To perform the numerical calculations we used a $1000 \times 1000$ square lattice, and set the chemical potential $\mu/U = 0.36$, the hopping to $J/U = 0.03$, and $\beta U = \infty$.

\begin{figure}
\subfigure{
  \includegraphics[width=0.4\textwidth]{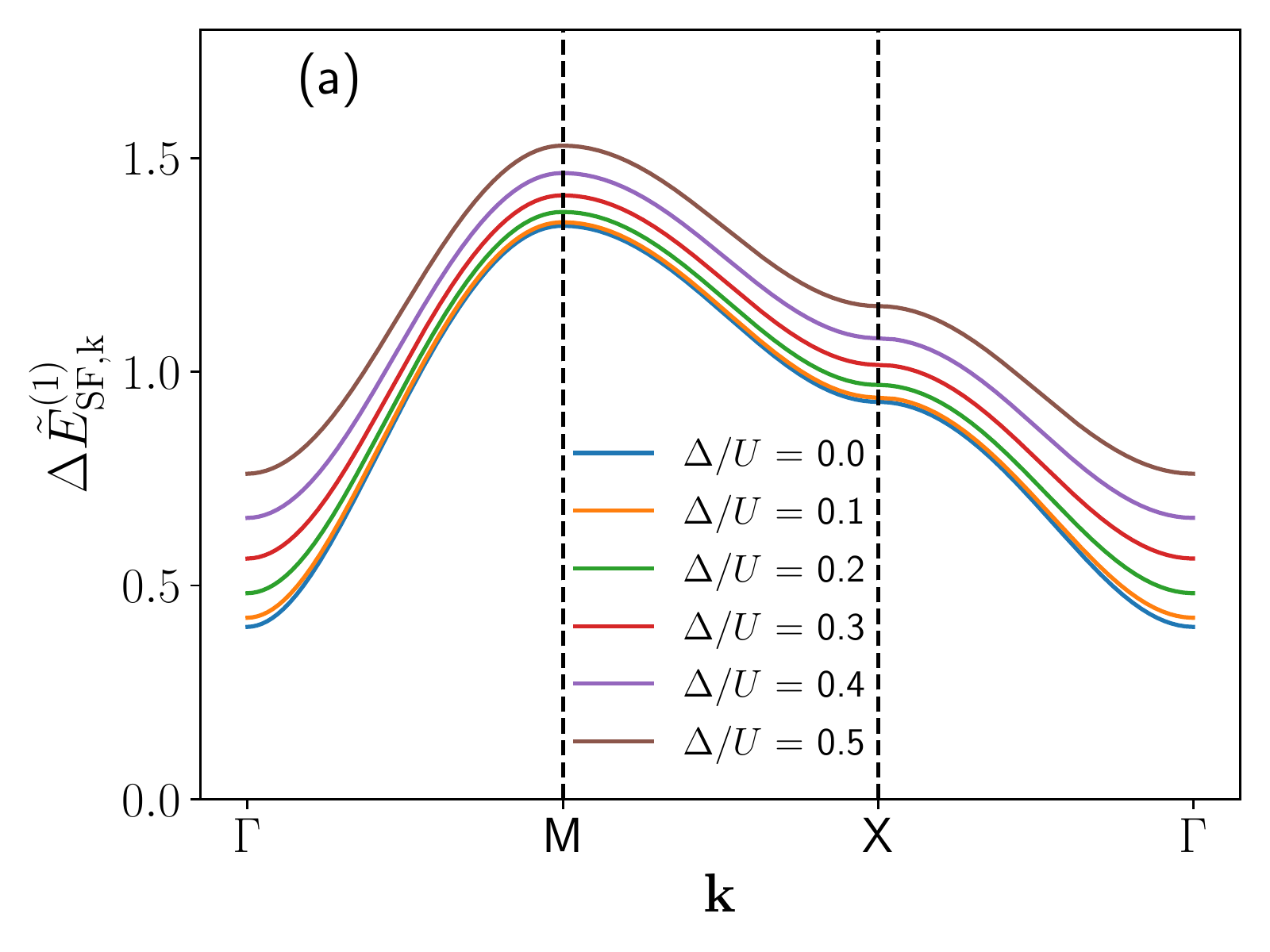}
  }
 \subfigure{
  \includegraphics[width=0.4\textwidth]{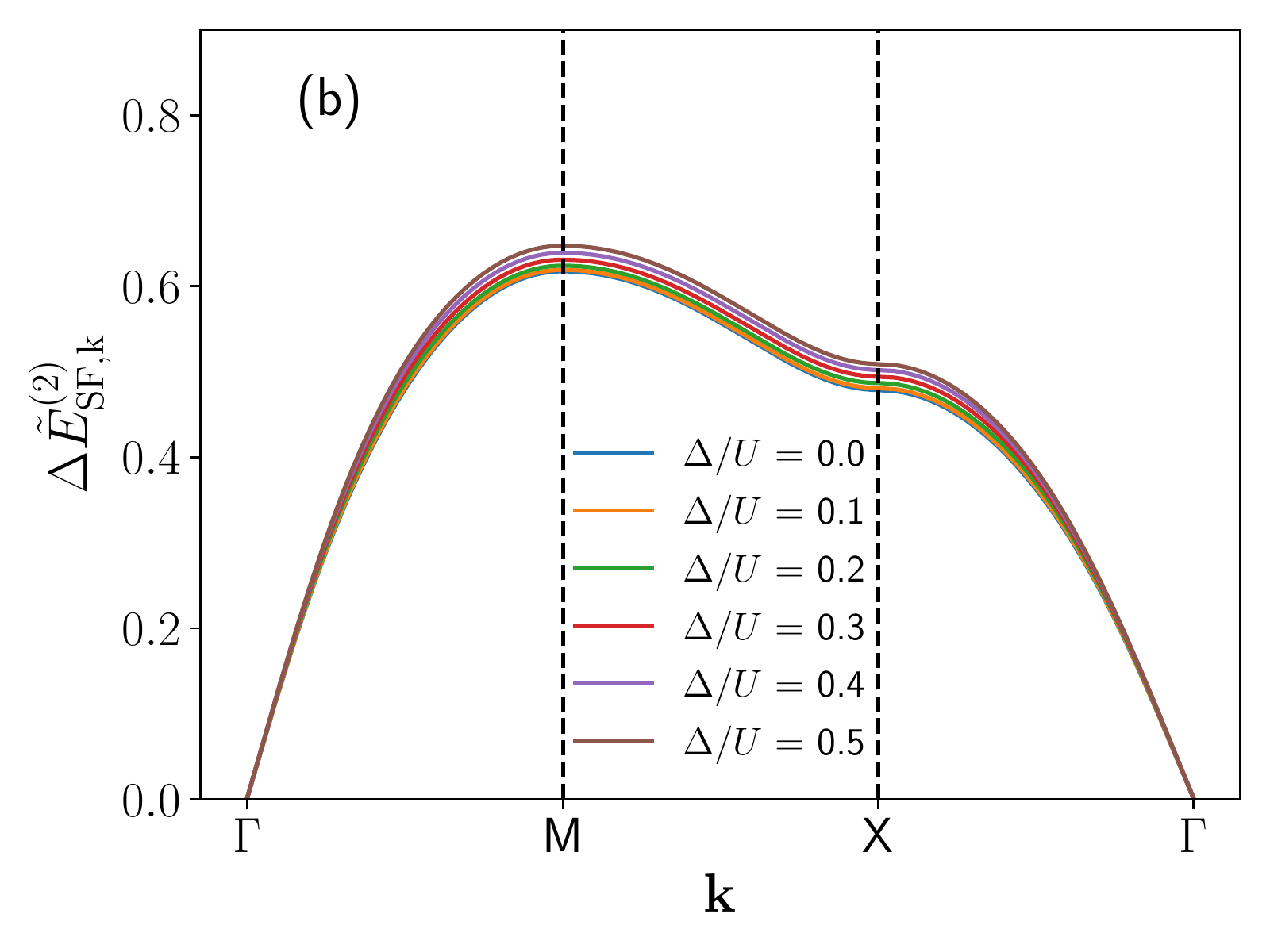}
  }
 \subfigure{
  \includegraphics[width=0.4\textwidth]{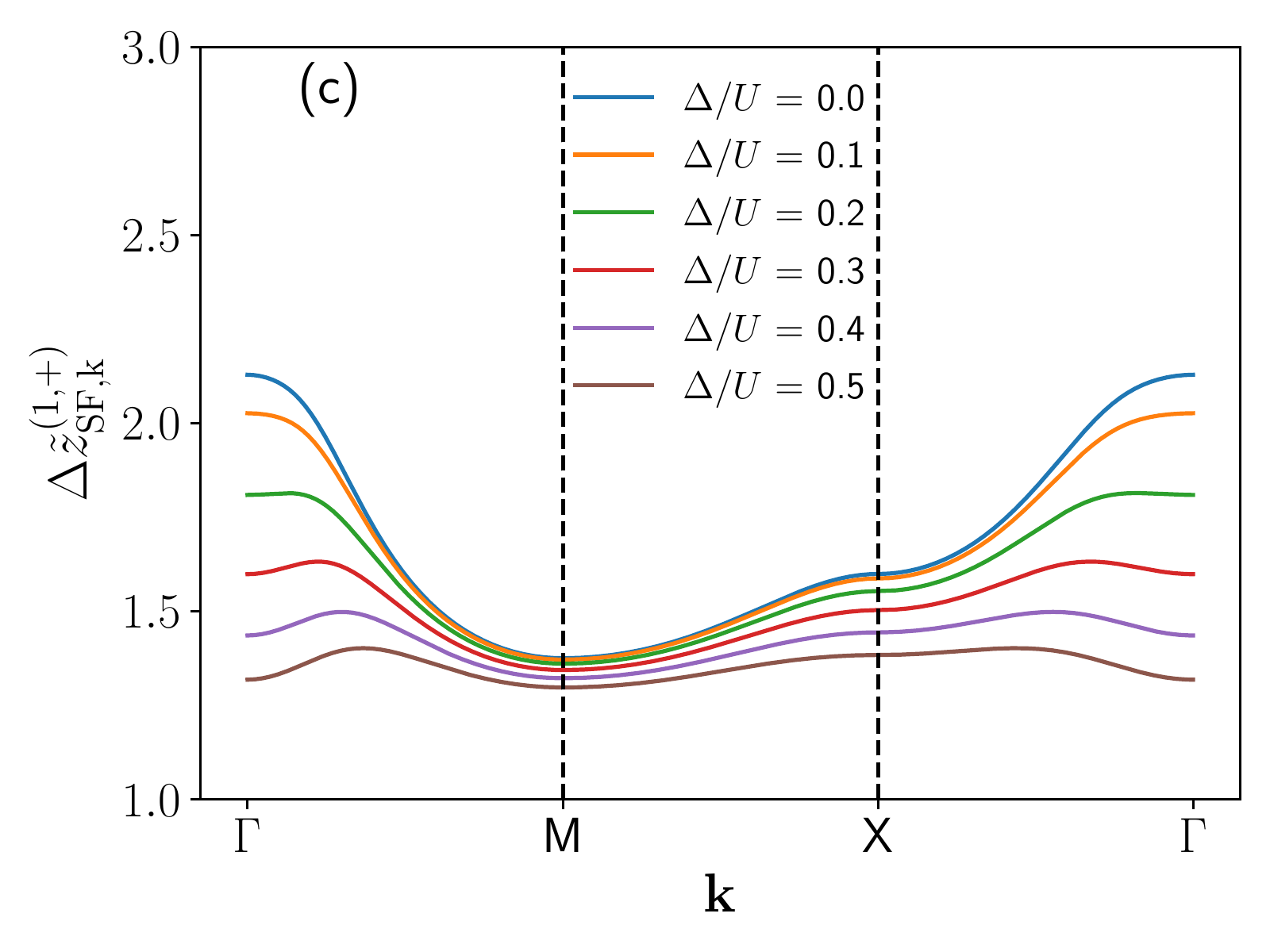}
  }
\subfigure{
  \includegraphics[width=0.4\textwidth]{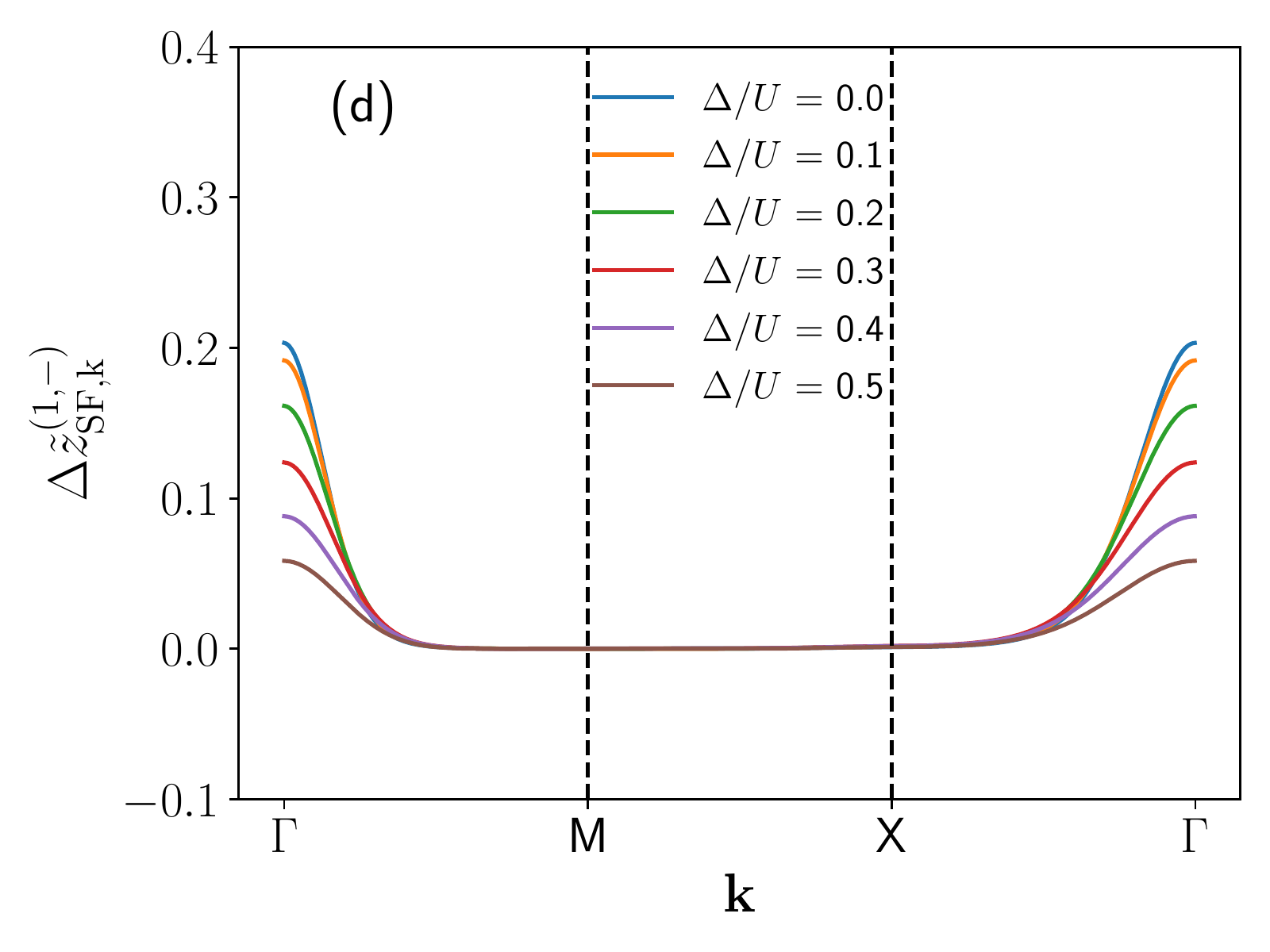}
  }
 \subfigure{
  \includegraphics[width=0.4\textwidth]{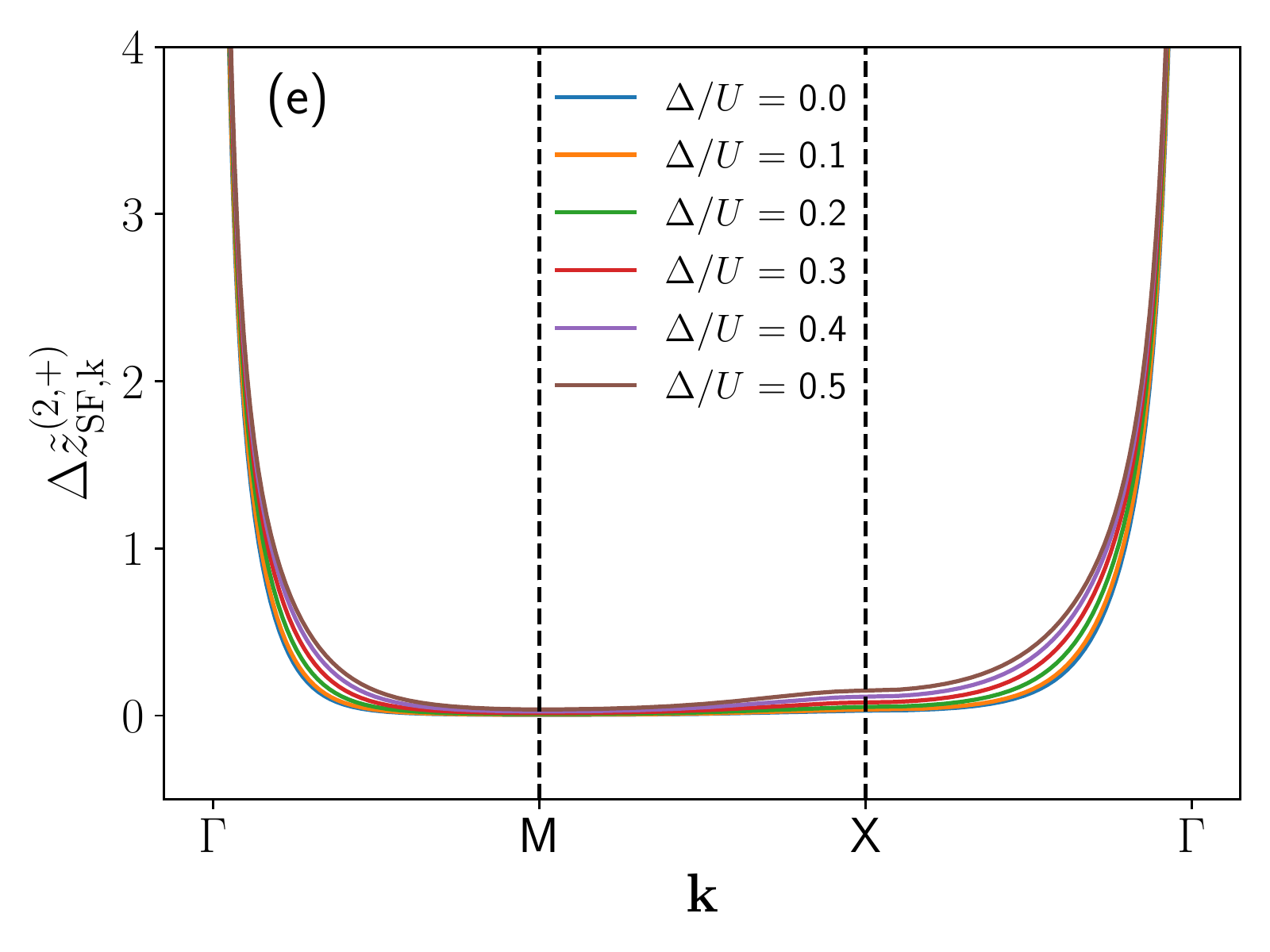}
  }
 \subfigure{
  \includegraphics[width=0.4\textwidth]{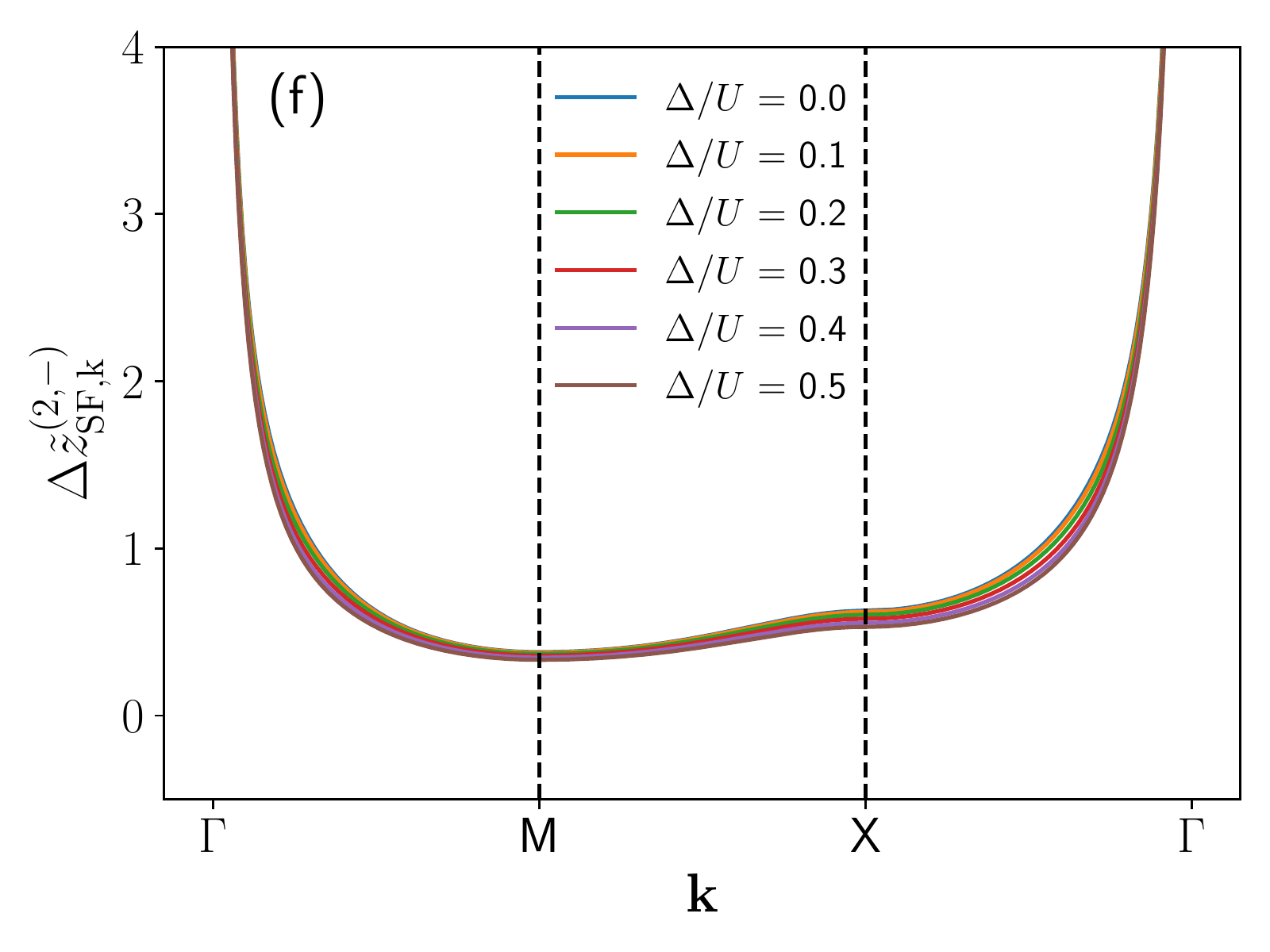}
  }
\caption{Collective excitations for the superfluid phase of the two-dimensional disordered-BHM (a) first quasi-particle/hole excitation energy branch $\Delta\tilde{E}_{\mathrm{SF},\vec{k}}^{\left(1\right)}$, (b) second quasi-particle/hole excitation energy branch $\Delta\tilde{E}_{\mathrm{SF},\vec{k}}^{\left(2\right)}$, (c) quasi-particle spectral weights $\tilde{z}_{\mathrm{SF},\vec{k}}^{\left(1,+\right)}$ for the first branch, (d) quasi-hole spectral weights $\tilde{z}_{\mathrm{SF},\vec{k}}^{\left(1,-\right)}$ for the first branch, (e) quasi-particle spectral weights $\tilde{z}_{\mathrm{SF},\vec{k}}^{\left(2,+\right)}$ for the second branch, (f) quasi-hole spectral weights $ \tilde{z}_{\mathrm{SF},\vec{k}}^{\left(2,-\right)}$ for the second branch for different disorder strengths. The parameters used were  $N_{s}=1000^2$, $\mu/U = 0.36$, $J/U = 0.07$, and $\beta U = \infty$. Note that $\Gamma = \left( 0,0\right)$, $M = \left( \pi,\pi \right)$, and $X = \left( \pi,0 \right)$.}
\label{fig:fig3}
\end{figure}

\end{widetext}
\subsubsection{Mott insulator  phase boundary}\label{subsec:Phase boundary}
\begin{figure}
\subfigure{
  \includegraphics[width=0.43\textwidth]{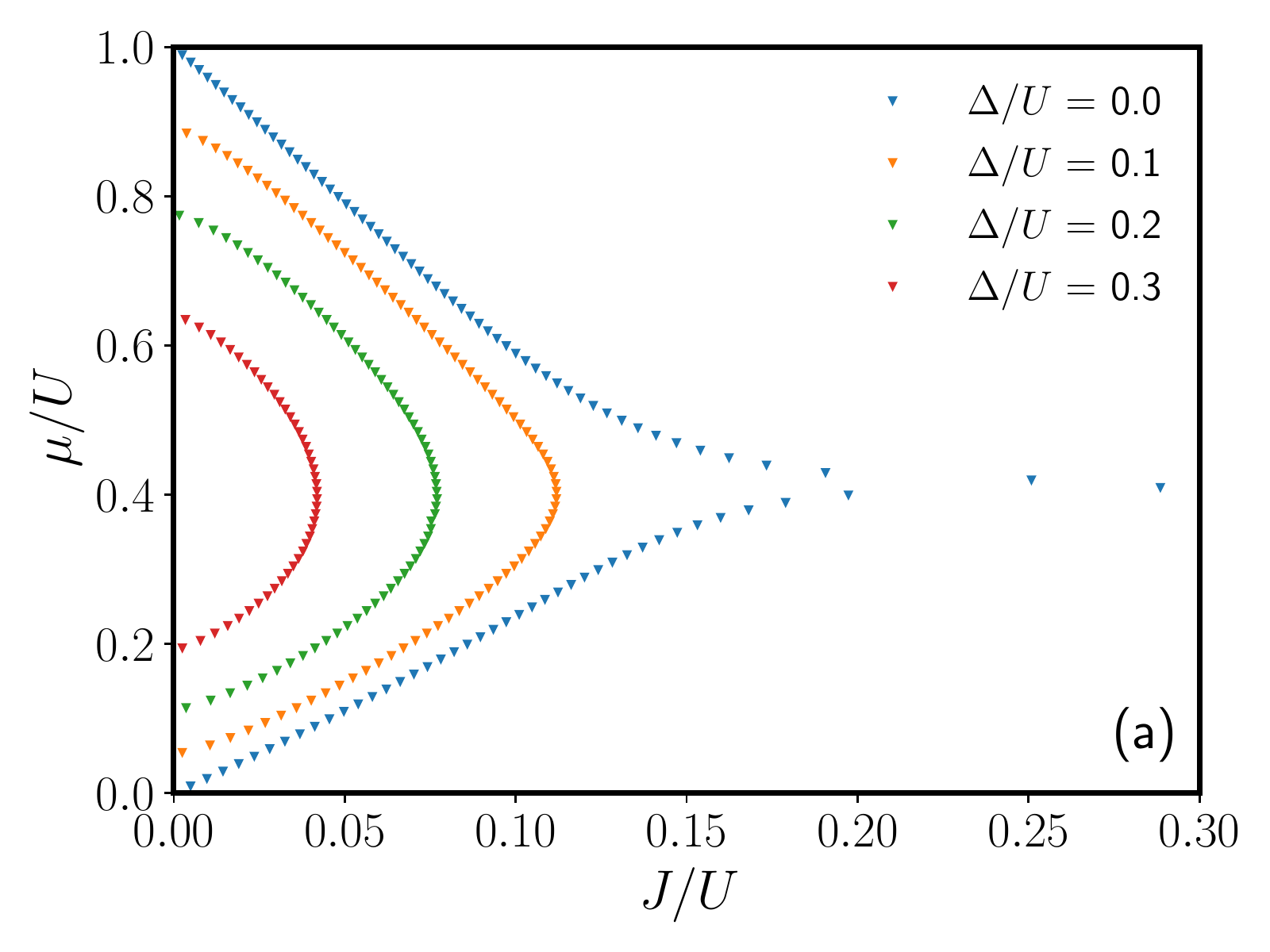}
  }
 \subfigure{
  \includegraphics[width=0.43\textwidth]{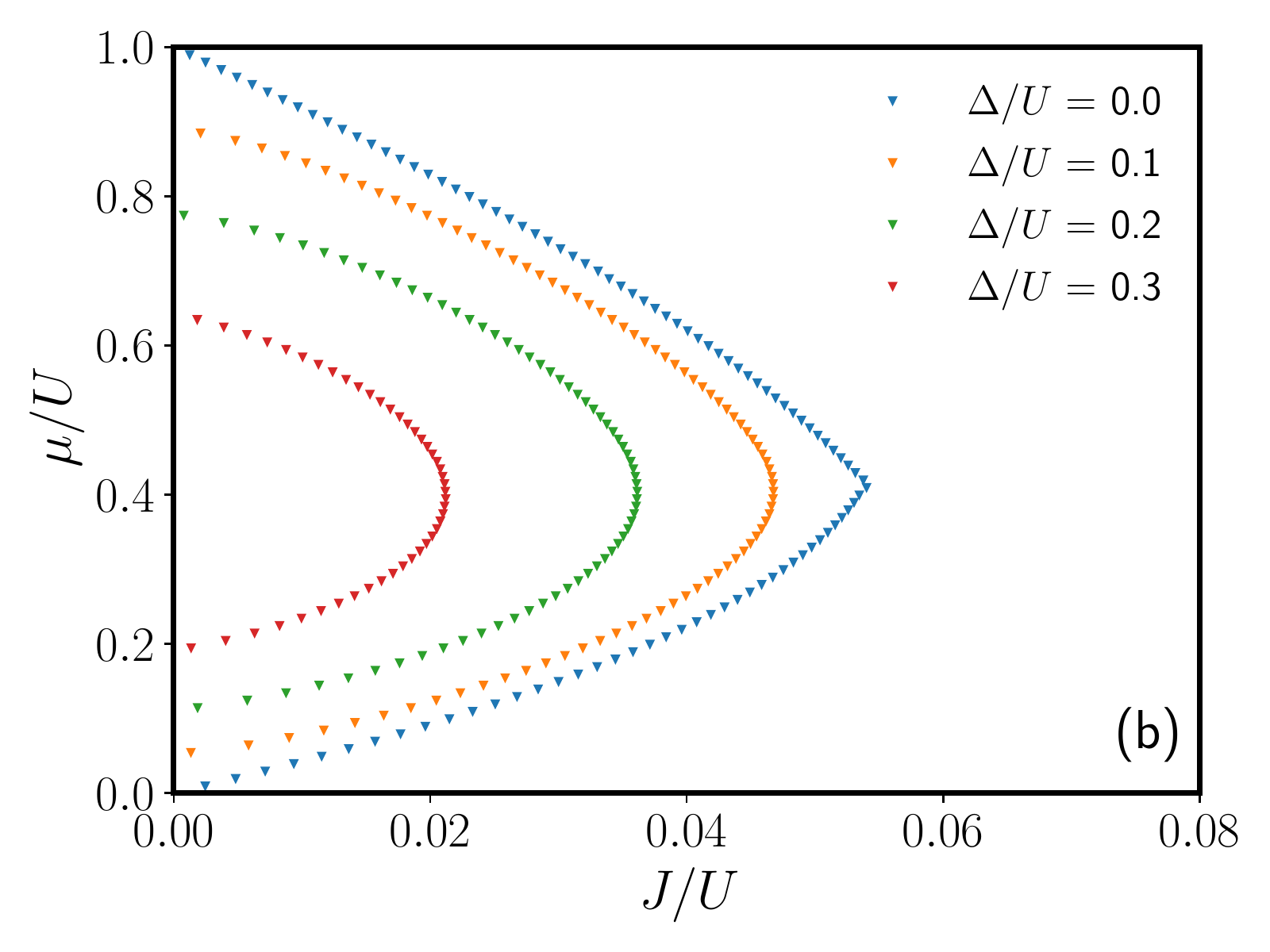}
  }
 \subfigure{
  \includegraphics[width=0.43\textwidth]{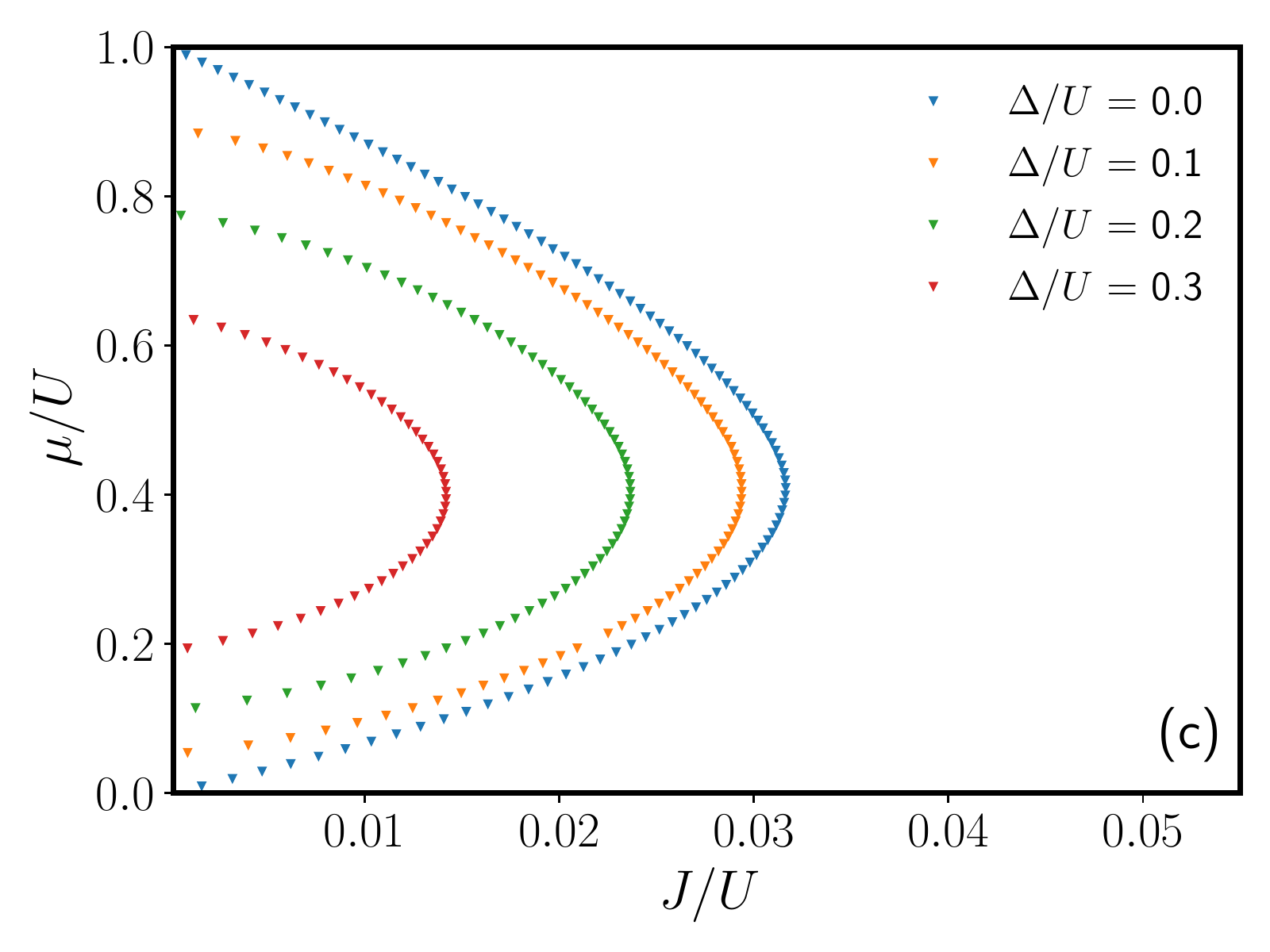}
  }
\caption{Mott insulator phase boundaries for different disorder strengths for (a) d=1, (b) d=2, and (c) d=3. Calculations performed keeping disorder terms to $\mathcal{O}\left( \widetilde{\Delta}_{\mathcal{\epsilon}}^{2}\right)$.}
\label{fig:fig4}
\end{figure}

To obtain the Mott Insulator -- Bose Glass (MI-BG) phase boundary, we calculate the critical hopping $J_c$ at which  $\check{\phi} =0$. This can be done numerically using Eq.~\eqref{eq:Mean-field EOM-5}. In Fig.~\ref{fig:fig4} we show the phase boundary for different disorder strengths for dimensions one, two, and three.

\begin{figure}
\subfigure{
  \includegraphics[width=0.43\textwidth]{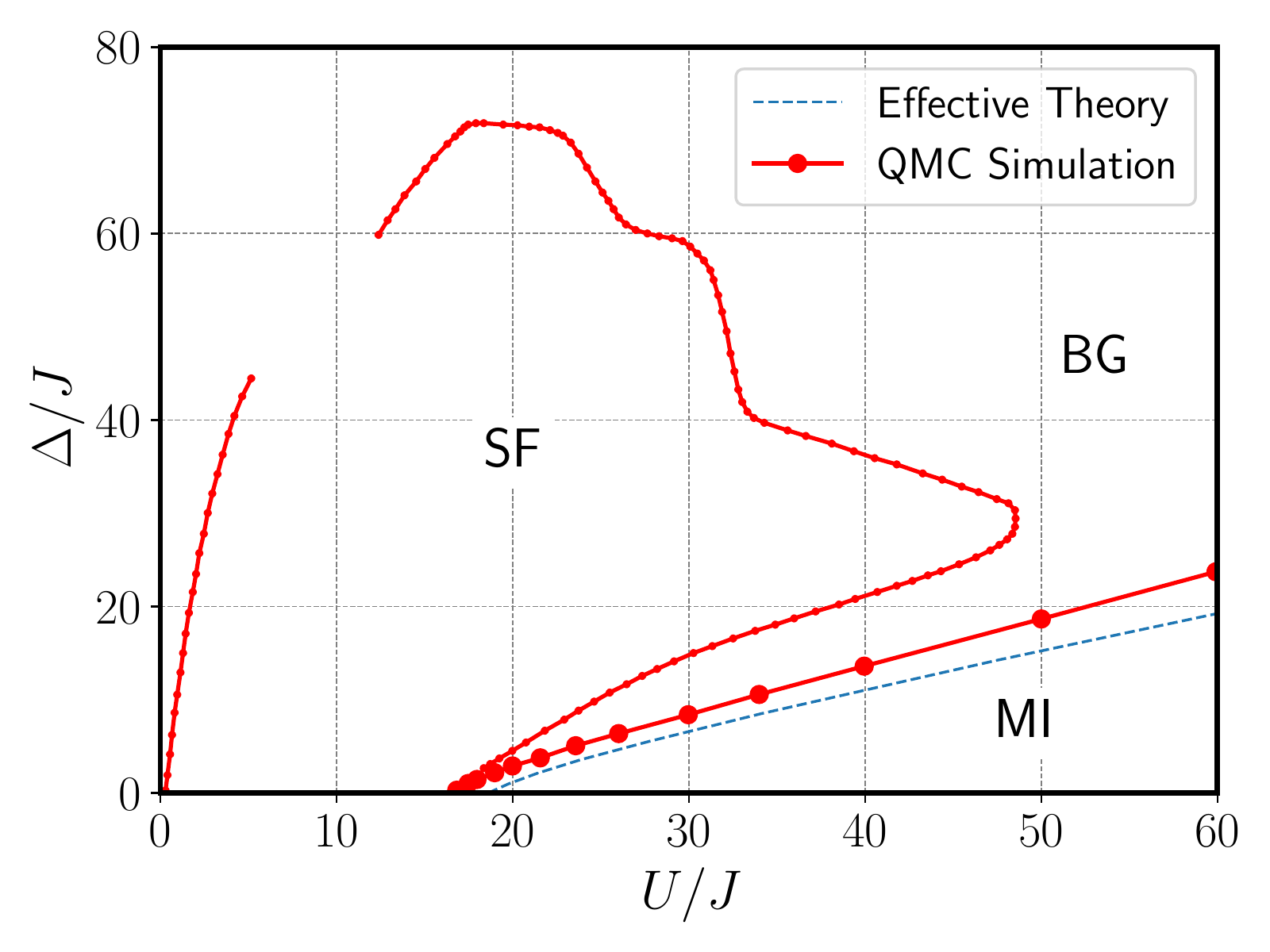}
  }
 \subfigure{
  \includegraphics[width=0.43\textwidth]{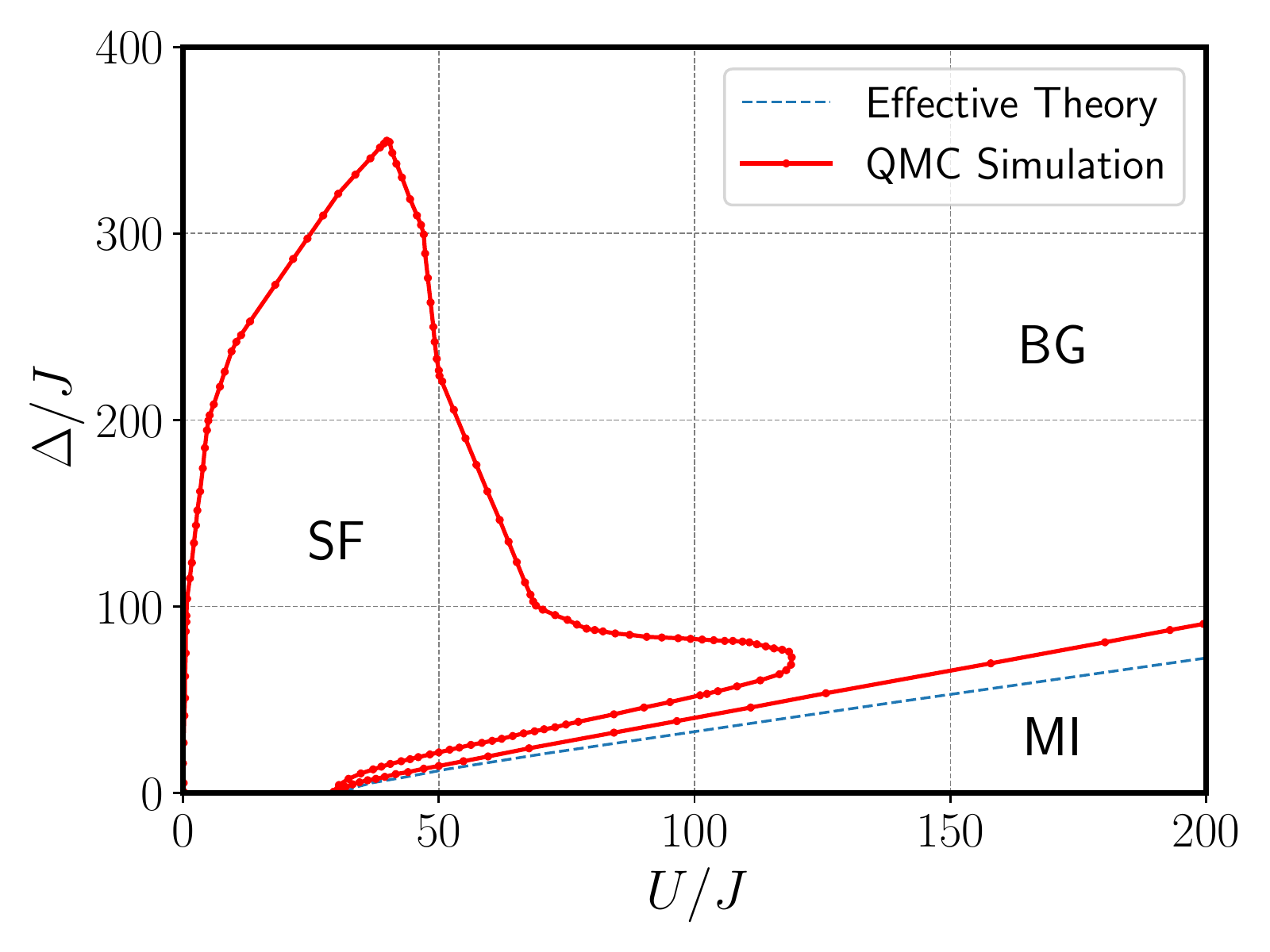}
  }
\caption{Comparison of the results from effective theory to $\mathcal{O}\left( \widetilde{\Delta}_{\mathcal{\epsilon}}^{2}\right)$ and QMC for the Mott insulator -- Bose glass transition for (a) $d=2$ and (b) $d=3$. QMC data for both the Mott insulator -- Bose glass and Bose glass -- superfluid transition taken for $d=2$ from Ref.~\citep{Soyler2011} and for $d=3$ from Ref.~\citep{Gurarie2009}.}
\label{fig:fig5}
\end{figure}

In Refs.~\citep{Soyler2011, Gurarie2009}, the MI-BG transition for the chemical potential at the tip of the Mott lobe was calculated for two and three dimensional cubic lattices with random disorder uniformly distributed on the interval $\left[ -\Delta, \Delta \right] $. In Fig.~\ref{fig:fig5}, using the disordered BHM in the strong coupling regime, we calculated the MI-BG transition and compared our results keeping terms to $\mathcal{O}\left(\widetilde{\Delta}_{\mathcal{\epsilon}}^{2} \right)$ with those obtained using QMC simulations~\citep{Soyler2011, Gurarie2009}. We find good agreement between our results and QMC simulations. It should be noted that we consider  a Gaussian distribution of disorder, while Refs.~\citep{Soyler2011, Gurarie2009} used a box distribution. 

\begin{figure}
\subfigure{
  \includegraphics[width=0.43\textwidth]{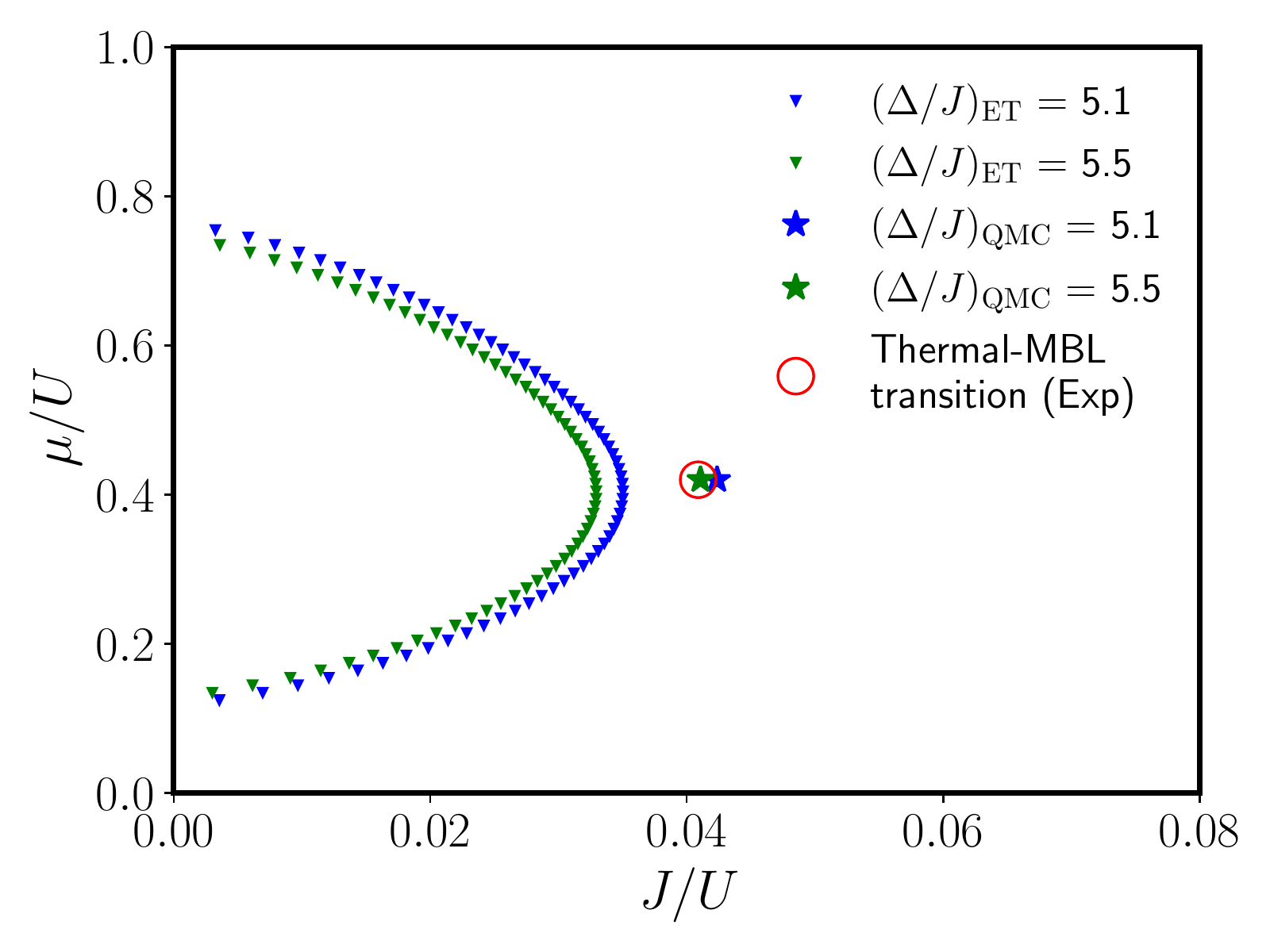}
  }
\caption{Comparison of the experimentally identified thermal-MBL transition point (Red dot) at $U/J=24.4$ for unit filling with the Mott insulator -- Bose glass transition curves for $\Delta /J = 5.1$ and $\Delta /J = 5.5$ (corresponding to $\Delta /J = 5.3(2)$) obtained using the effective theory to $\mathcal{O} \left( \tilde{\Delta}_{\mathcal{\epsilon}}^{2} \right)$. Blue and green stars are the corresponding QMC Mott insulator -- Bose glass transition points for $\Delta /J = 5.1$ and $\Delta /J = 5.5$, respectively taken from Ref.~\citep{Soyler2011}.}
\label{fig:fig6}
\end{figure}

Finally, we consider the same set of parameters reported in Ref.~\citep{Choi2016} in which thermalization-MBL transition occurs. As is demonstrated in Fig.~\ref{fig:fig6} we note that the reported critical point in Ref.~\citep{Choi2016} sits on top of the QMC determined phase transition.

\section{Discussion and conclusions}
\label{sec:Conc}
In this work, we extended the 2PISC approach to the BHM to include the effects of the disorder. We obtained a disorder-averaged effective theory from which we obtained the 2PI equations of motion for the superfluid order parameter and two, three and four-point correlations. These equations apply both in and out of equilibrium.
A strength of the 2PISC is that it is applicable in one, two, and three spatial dimensions. This is particularly advantageous for out-of-equilibrium dynamics, where numerical methods that are essentially exact, such as exact diagonalization or DMRG, are limited to one dimension or very small system sizes.

A weakness of the 2PISC method is that in order to make progress, one needs to truncate the effective action which was done at quartic order, and this is an uncontrolled approximation. However, previous results in the clean case for both the phase boundary in equilibrium~\citep{Fitzpatrick2018a} and agreement with exact diagonalization for out-of-equilibrium dynamics~\citep{Fitzpatrick2018b} give confidence in its usefulness. It should be noted that the accuracy appears to be greatest for larger $U/J$~\citep{Mokhtari-Jazi2021}. 
The main result of this paper is the derivation of the effective theory and the 2PI equations of motion, but as a check on the theory, we solved the disorder-averaged equations of motion in the equilibrium limit. We obtained the collective excitation spectra for the disordered BHM in  both the Mott and superfluid phases and also obtained the Mott insulator phase boundary at a variety of disorder strengths in one, two, and three dimensions. We compared our results with QMC simulations performed in Refs.~\citep{Soyler2011, Gurarie2009} and found very good agreement with the exact phase boundary.

Previous comparison of Mott insulator phase boundaries with QMC calculations in the clean case~\citep{Fitzpatrick2018a} found the 2PISC method gave a big improvement over the mean-field theory but was not in complete quantitative agreement with QMC calculation. We find a similar situation in the disordered case. However, there are caveats, in that we consider a Gaussian distribution of disorder rather than the box distribution used in Refs.~\citep{Soyler2011} and~\citep{Gurarie2009}. We also treat the disorder perturbatively and keep the highest-order term only (calculations to order $\tilde{\Delta}_{\mathcal{\epsilon}}^{2}$). Given that there are uncontrolled approximations in the 2PISC method, these results give confidence in the results here and future applications to the out-of-equilibrium dynamics of the disordered Bose-Hubbard model. One limitation of our method is that we have not been able to determine the Bose glass -- Superfuid phase boundary, which corresponds to the vanishing of the superfluid stiffness.

We noted that a motivation for our work was the experiments by Choi {\it et al.}~\citep{Choi2016} on thermalization in the disordered two-dimensional BHM. In that work, there was an identification of a thermal to MBL transition at a critical disorder value of $\Delta/U = 5.3(2)$ when $U/J=24.4$. Using these same parameter values, in Fig.~\ref{fig:fig6}, we show that this point appears to lie essentially at the Mott insulator -- Bose glass phase transition identified in QMC simulations. While our calculations and the QMC calculations in Ref.~\citep{Soyler2011} do not include a trap, this result adds further to the questions raised in Ref.~\citep{Yan2017} as to whether the experiments in Ref.~\citep{Choi2016} probe an MBL transition or a glass transition. We intend to explore this question further in future work on out-of-equilibrium dynamics of the disordered BHM.\\
\begin{acknowledgments}
The authors thank NSERC for support of this work. 
\end{acknowledgments}
\begin{widetext}
\appendix

\section{Propagator in the zero disorder and hopping limit\label{sec:Propagator in the zero disorder and hopping limit}}

In the zero disorder and hopping limit, for an initial state $\hat{\rho}_{i}$
of the form given in Eq.~\eqref{eq:rho-density - 1}, the
spectral function $\mathcal{A}_{\vec{r}}\left(t\right)$ and the kinetic
Green's function $\mathcal{G}_{\vec{r}}^{\left(K\right)}\left(t\right)$
can be expressed as follows:
\begin{align}
\mathcal{G}_{\vec{r}}^{\left(R\right)}\left(t\right) & =-i \Theta \left(t\right)\left\{\left(n_{\vec{r}}+1\right)e^{-i\left\{ \mathcal{E}\left(\vec{r},n_{\vec{r}}+1\right)-\mathcal{E}\left(\vec{r},n_{\vec{r}}\right)\right\} t}-n_{\vec{r}}e^{i\left\{ \mathcal{E}\left(\vec{r},n_{\vec{r}}-1\right)-\mathcal{E}\left(\vec{r},n_{\vec{r}}\right)\right\} t}\right\},\label{eq:G0^(R) - 1}\\
\mathcal{G}_{\vec{r}}^{\left(A\right)}\left(t\right) & =i\Theta\left(-t\right)\left\{\left(n_{\vec{r}}+1\right)e^{-i\left\{ \mathcal{E}\left(\vec{r},n_{\vec{r}}+1\right)-\mathcal{E}\left(\vec{r},n_{\vec{r}}\right)\right\} t}-n_{\vec{r}}e^{i\left\{ \mathcal{E}\left(\vec{r},n_{\vec{r}}-1\right)-\mathcal{E}\left(\vec{r},n_{\vec{r}}\right)\right\} t}\right\},\label{eq:G0^(A) - 1}\\
\mathcal{G}_{\vec{r}}^{\left(K\right)}\left(t\right) & =-i\left\{ \left(n_{\vec{r}}+1\right)e^{-i\left\{ \mathcal{E}\left(\vec{r},n_{\vec{r}}+1\right)-\mathcal{E}\left(\vec{r},n_{\vec{r}}\right)\right\} t}+n_{\vec{r}}e^{i\left\{ \mathcal{E}\left(\vec{r},n_{\vec{r}}-1\right)-\mathcal{E}\left(\vec{r},n_{\vec{r}}\right)\right\} t}\right\} ,\label{eq:G0^(K) - 1}
\end{align}
\noindent where $n_{\vec{r}}$ is particle density profile of the initial
state, and
\begin{equation}
    \mathcal{E}\left(\vec{r},n_{\vec{r}}\right)=\frac{U}{2}\sum_{\vec{r}}n_{\vec{r}}\left(n_{\vec{r}}-1\right)+\sum_{\vec{r}}\left(V_{\vec{r}}-\mu\right)n_{\vec{r}}.\label{eq:local energy - 1}
\end{equation}
\section{Local quantities in the self-energy $\Sigma^{zz}$\label{sec:Parameters in the self energy}}

In obtaining the effective self-energy $\Sigma_{\delta}^{zz}$, we
introduced two local quantities that are non-trivial functions of
the initial particle density profile $n_{\vec{r}}$ and the chemical
potential $\mu$:
\begin{align}
    \mathcal{G}_{\vec{r},\omega\to0}^{\left(R\right)} & =-\frac{n_{\vec{r}}+1}{\mathcal{E}\left(\vec{r},n_{\vec{r}}+1\right)-\mathcal{E}\left(\vec{r},n_{\vec{r}}\right)}-\frac{n_{\vec{r}}}{\mathcal{E}\left(\vec{r},n_{\vec{r}}-1\right)-\mathcal{E}\left(\vec{r},n_{\vec{r}}\right)},\label{eq:static limit of G0^(R) - 1}
\end{align}
\noindent and
\begin{align}
    \left[u_{1}\right]_{\vec{r}} & =-2\left\{ \mathcal{G}_{\vec{r},\omega\to0}^{\left(R\right)}\right\} ^{-4}\nonumber \\
     & \quad\times\left\{ \frac{\left(n_{\vec{r}}+1\right)\left(n_{\vec{r}}+2\right)}{\left\{ \mathcal{E}\left(\vec{r},n_{\vec{r}}+2\right)-\mathcal{E}\left(\vec{r},n_{\vec{r}}\right)\right\} \left\{ \mathcal{E}\left(\vec{r},n_{\vec{r}}+1\right)-\mathcal{E}\left(\vec{r},n_{\vec{r}}\right)\right\} ^{2}}\right.\nonumber \\
     & \quad\phantom{\times}\left.+\frac{n_{\vec{r}}\left(n_{\vec{r}}-1\right)}{\left\{ \mathcal{E}\left(\vec{r},n_{\vec{r}}-2\right)-\mathcal{E}\left(\vec{r},n_{\vec{r}}\right)\right\} \left\{ \mathcal{E}\left(\vec{r},n_{\vec{r}}-1\right)-\mathcal{E}\left(\vec{r},n_{\vec{r}}\right)\right\} ^{2}}\right.\nonumber \\
     & \quad\phantom{\times}\left.-\frac{\left(n_{\vec{r}}+1\right)^{2}}{\left\{ \mathcal{E}\left(\vec{r},n_{\vec{r}}+1\right)-\mathcal{E}\left(\vec{r},n_{\vec{r}}\right)\right\} ^{3}}-\frac{n_{\vec{r}}^{2}}{\left\{ \mathcal{E}\left(\vec{r},n_{\vec{r}}-1\right)-\mathcal{E}\left(\vec{r},n_{\vec{r}}\right)\right\} ^{3}}\right.\nonumber \\
     & \quad\phantom{\times}\left.-\frac{n_{\vec{r}}\left(n_{\vec{r}}+1\right)}{\left\{ \mathcal{E}\left(\vec{r},n_{\vec{r}}+1\right)-\mathcal{E}\left(\vec{r},n_{\vec{r}}\right)\right\} \left\{ \mathcal{E}\left(\vec{r},n_{\vec{r}}-1\right)-\mathcal{E}\left(\vec{r},n_{\vec{r}}\right)\right\} ^{2}}\right.\nonumber \\
     & \quad\phantom{\times}\left.-\frac{n_{\vec{r}}\left(n_{\vec{r}}+1\right)}{\left\{ \mathcal{E}\left(\vec{r},n_{\vec{r}}+1\right)-\mathcal{E}\left(\vec{r},n_{\vec{r}}\right)\right\} ^{2}\left\{ \mathcal{E}\left(\vec{r},n_{\vec{r}}-1\right)-\mathcal{E}\left(\vec{r},n_{\vec{r}}\right)\right\} }\right\} ,\label{eq:static limit of u^(4) - 1}
\end{align}
\noindent where $\mathcal{E}\left(\vec{r},n_{\vec{r}}\right)$ is
given by Eq.~\eqref{eq:local energy - 1}. Note that the expressions
for $\mathcal{G}_{\vec{r},\omega\to0}^{\left(R\right)}$ and $\left[u_{1}\right]_{\vec{r}}$
are very similar to those introduced in Ref.~\citep{Fitzpatrick2018a}
for $\mathcal{G}^{12,\left(R\right)}\left(\omega\to0\right)$ and
$u_{1}$ respectively, with the biggest difference being the spatial
dependence in the present case. In our numerical work, we do not consider a trap, and so $u_1$ can be taken to be independent of $\vec{r}$. We present the more general expression here for completeness.
 
 \section{Gapless spectrum in the HFBP approximation for the disordered-BHM \label{sec:HFBP}}
 In this appendix, we extend the HFBP approximation presented in Ref.~\citep{Fitzpatrick2018b} to the disordered-BHM. In the SF phase, in order for the excitation spectrum to be gapless, we require that 
 \begin{equation}
      \tilde{C}_{\vec{k}=0} = 0,
      \label{eq: HFBP condition}
 \end{equation}
 where $ \tilde{C}_{\vec{k}}$ was defined in Eq.~\eqref{eq: C-vector - 2}. Following the same approach as presented in Ref.~\citep{Fitzpatrick2018b}, for $      \tilde{C}_{\vec{k}=0}$ we obtain
\begin{align}
          \tilde{C}_{\vec{k}=0} &= \left( U+\mu \right)^2 \left\{\frac{ \widetilde{\Delta}_{\mathcal{\epsilon}}^{2} \left( \check{G}_{\vec{k}=0}^{12,\left(R\right)}\left(\omega=0\right)- \:\check{G}_{\vec{k}=0}^{22,\left(R\right)}\left(\omega=0\right)\right)}{2}-2 u_1 \phi^2 \right\}\nonumber\\
          &\quad\quad\quad\quad \times \left\{ \frac{\widetilde{\Delta}_{\mathcal{\epsilon}}^{2} \left(\check{G}_{\vec{k}=0}^{12,\left(R\right)}\left(\omega=0\right)+\check{G}_{\vec{k}=0}^{22,\left(R\right)}\left(\omega=0\right)\right)}{2} +u_1 \left\{i\check{G}_{\vec{r'}=0}^{22,\left(K\right)}\left(s=0\right)\right\} \right\}.
          \label{eq: C_k 3}
\end{align}
From Eq.~\eqref{eq: C_k 3} it is clear that to have Eq.~\eqref{eq: HFBP condition} satisfied we should have
\begin{equation}
\check{G}_{\vec{r'}=0}^{22,\left(K\right)}\left(s=0\right) = \frac{i \widetilde{\Delta}_{\mathcal{\epsilon}}^{2}}{2 u_1}\left( \check{G}_{\vec{k}=0}^{12,\left(R\right)}\left(\omega=0\right)+ \:\check{G}_{\vec{k}=0}^{22,\left(R\right)}\left(\omega=0\right)\right).
\label{eq: G22K}
\end{equation}
Using the same procedure for $\check{G}_{\vec{r'}=0}^{11,\left(K\right)}\left(s=0\right)$ we can write
\begin{equation}
\check{G}_{\vec{r'}=0}^{11,\left(K\right)}\left(s=0\right) = \frac{i\widetilde{\Delta}_{\mathcal{\epsilon}}^{2}}{2 u_1}\left( \check{G}_{\vec{k}=0}^{12,\left(R\right)}\left(\omega=0\right)+ \:\check{G}_{\vec{k}=0}^{11,\left(R\right)}\left(\omega=0\right)\right).
\label{eq: G11K}
\end{equation}
\end{widetext}
\clearpage
\bibliographystyle{apsrev4-1}
\bibliography{arxiv}

\end{document}